  \documentclass{article}

\usepackage{a4wide,natbib}  

\usepackage{timet}
\usepackage{amssymb,amsmath}
\usepackage{graphicx,subcaption}
\usepackage{hyperref}

\title
{Rotating double-diffusive convection in stably stratified planetary cores}
\author
{R. Monville$^1$, J. Vidal$^{1,2}$, D. C\'ebron$^1$\thanks{david.cebron@univ-grenoble-alpes.fr}, N. Schaeffer$^1$ \\
  \small $^1$Universit\'e Grenoble Alpes, CNRS, ISTerre, Grenoble, France \\
  \small $^2$Department of Applied Mathematics, School of Mathematics, University of Leeds, Leeds, LS2 9JT, UK
  }
 
\date{Accepted 2019 July 25 for publication in GJI}

\begin{document}

\label{firstpage}
\maketitle

\begin{abstract}
\normalsize
In planetary fluid cores, the density depends on temperature and chemical composition, which diffuse at very different rates. This leads to various instabilities, bearing the name of double-diffusive convection. We investigate rotating double-diffusive convection (RDDC) in fluid spheres. We use the Boussinesq approximation with homogeneous internal thermal and compositional source terms. We focus on the finger regime, in which the thermal gradient is stabilising whereas the compositional one is destabilising. 
First, we perform a global linear stability analysis in spheres. 
The critical Rayleigh numbers drastically drop for stably stratified fluids, yielding large-scale convective motions where local analyses predict stability. 
We evidence the inviscid nature of this large-scale double-diffusive instability, enabling the determination of the marginal stability curve at realistic planetary regimes. 
In particular, we show that in stably stratified spheres, the Rayleigh numbers $Ra$ at the onset 
evolve like $Ra \sim Ek^{-1}$, where $Ek$ is the Ekman number. 
This differs from rotating convection in unstably stratified spheres, for which $Ra \sim Ek^{-4/3}$. The domain of existence of inviscid convection thus increases as $Ek^{-1/3}$. 
Second, we perform nonlinear simulations. 
We find a transition between two regimes of RDDC, controlled by the strength of the stratification.
Furthermore, far from the RDDC onset, we find a dominating equatorially anti-symmetric, large-scale zonal flow slightly above the associated linear onset.
Unexpectedly, a purely linear mechanism can explain this phenomenon, even far from the instability onset, yielding a symmetry breaking of the nonlinear flow at saturation.
For even stronger stable stratification, the flow becomes mainly equatorially-symmetric and intense zonal jets develop.
Finally, we apply our results to the early Earth core. 
Double diffusion can reduce the critical Rayleigh number by four decades for realistic core conditions. 
We suggest that the early Earth core was prone to turbulent RDDC, with large-scale zonal flows.
\end{abstract}


\section{Introduction}
\subsection{Geophysical context}
Thermo-compositional convection stirs motions in the Earth's core \citep{jones2015tog}, that sustain large-scale magnetic fields via dynamo action.
The thermal part is generated by the super-adiabatic thermal gradient.
It mainly comes from the secular cooling of the core, driven by the heat extracted at the core-mantle boundary (CMB).
Additionally, because of this cooling, latent heat is released by the crystallisation of the inner core \citep{verhoogen1961heat}.
Radioactive heat sources can also participate, although their contribution is debated \citep[e.g.][]{hirao2006partitioning,bouhifd2007potassium,chidester2017metal}.
The compositional part is sustained by the ejection of light elements into the fluid core, mainly due to the solidification of the inner core \citep[e.g.][]{fearn1981compositional}. 
Currently, compositional buoyancy is expected to dominate over thermal buoyancy \citep{braginsky1995equations,lister1995strength,buffett1996core}. Few models have considered individual contributions of thermal and compositional buoyancies for the present dynamics of the core, by using experiments \citep{cardin1992experimental}, asymptotic models \citep{busse2002low,simitev2011double} or numerical simulations 
\citep[e.g.][]{glatzmaier1996anelastic,kutzner2000effects,hori2012influence,bouffard2017double}.

The crystallisation of the inner core is a rather recent geophysical feature, initiated 1 Ga or 2 Ga ago \citep{labrosse2015thermal}.
However, the geodynamo is active since at least 3.45 Ga \citep{usui2009evidence,tarduno2010geodynamo}, despite the absence of the main buoyancy source (crystallization of the inner core).
Moreover, driving the early geodynamo by thermal buoyancy alone requires large secular cooling rates \citep{gubbins2003can}.
Such fast cooling rates are problematic for most thermal histories \citep{labrosse2015thermal}, although allowed by the large remaining uncertainties \citep[e.g.][]{williams2018thermal}.
Prior the inner core crystallization, a large fraction of the core is expected to present a sub-adiabatic temperature \citep{nimmo2015tog,labrosse2015thermal}, inhibiting (thermal) convective motions.
Therefore, determining the origin of the fluid motions sustaining the early geodynamo is elusive.

It has been suggested that light elements, dissolved during the core formation \citep[e.g.][]{badro2015core}, may have been exsolved due to the secular cooling \citep{buffett2000sediments}.
The exsolution of buoyant magnesium oxide would provide compositional buoyancy, notably prior to the nucleation of the inner core \citep{o2016powering,badro2016early}.
This mechanism has been criticised, e.g. because the magnesium solubility in the core depends not only on the temperature but also strongly on the oxygen content \citep{du2017insufficient}.
Moreover, this scenario requires a core formation at extremely high temperature to incorporate a sufficient amount of magnesium. 
Instead, \citet{hirose2017crystallization} advocated for top-down crystallisation of silicon oxides, incorporated in the core via the metal-segregation processes in a deep magma ocean at moderate temperatures. These non-standard mechanisms put forward the possibility to drive the early geodynamo by double-diffusive convection.

\subsection{Double-diffusive convection}
Double-diffusive convection (DDC) refers to various buoyancy-driven instabilities, generated by two different components of buoyancy. For planetary cores, we refer to thermal and chemical buoyancies.
The two sources diffuse at different rates, with the thermal (fast) diffusivity $\kappa_T$ and the chemical (slow) one $\kappa_C$.
Their ratio defines the dimensionless Lewis number $L = \kappa_T/\kappa_C$, which is expected to be at least $10^3$ \citep{braginsky1995equations} in planetary cores (see table \ref{table:nondim}).

DDC takes different forms, depending on the value of $L$ and on the sign of the mean gradients of each individual component of the density.
Classical convection occurs when both thermal and compositional gradients are destabilising.
Then, we distinguish (i) the finger regime \citep{stern1960salt}, when the chemical gradient is unstable and the thermal one stable, and (ii) the semi-convection quadrant \citep{spiegel1969semiconvection} with a stabilising compositional gradient and a destabilising thermal one.
Recently, double-diffusive effects have been evidenced even with slightly stabilising thermal gradients, leading to finger convection for unstable stratification \citep[e.g.][]{kellner2014transition}.

DDC has been mainly studied for oceanographic purposes \citep[e.g.][]{schmitt1994double,radko2013double}. Applications has become also apparent in astrophysics \citep[e.g.][]{garaud2018double} or mantle physics \citep{hansen1988numerical,hansen1989subcritical,hansen1990nonlinear}. Rotational effects have been largely neglected in these works. Only a few studies investigated rotating double-diffusive convection (RDDC), usually by considering rotational effects in local Cartesian models. Under this assumption, rotation has essentially a stabilising effect \citep{acheson1980stable,pearlstein1981effect,moll2017double,sengupta2018effect}.
Yet, the relevance of these local models remains elusive for rapidly rotating planetary cores.
Indeed, a subtle interplay between the rapid rotation and the bounded spherical geometry is expected for RDDC. Notably, \citet{busse2002low} predicted asymptotically the existence of double-diffusive convection at low Rayleigh numbers in rapidly rotating fluids cores, by extending his reduced annulus model \citep{busse1970thermal}. 
\citet{simitev2011double} did confirm these predictions numerically in the annulus geometry.
Finally, only few studies tackled RDDC in spherical geometries with both unstable buoyancies \citep{glatzmaier1996anelastic,breuer2010thermochemically,trumper2012numerical,takahashi2014double}, and even fewer with antagonist gradients \citep{manglik2010dynamo,net2012numerical}.

\begin{table}
 \centering
 \begin{tabular}{ccccc} 
 \hline
 Symbol & Name & Definition & Earth (current) & Stars \\ [0.5ex] 
 \hline
 $L$ & Lewis & $\kappa_T/\kappa_C$ & $10^4$ & $10^3 - 10^7$ \\ [0.5ex] 
 $Pr$ & Prandtl  & ${\nu}/{\kappa_T}$ & $0.01 - 0.1$ & $10^{-6}$ \\[0.5ex]
 $Sc$ & Schmidt & ${\nu}/{\kappa_C}$ & $10^2 - 10^3$ & $10^{-3} - 10^{1}$ \\[0.5ex]
 $Ek$ & Ekman & ${\nu}/({\Omega_s R^2})$ & $10^{-15}$ & $10^{-18}$ \\[0.5ex]
 \hline
\end{tabular}
\caption{Dimensionless numbers characterising diffusive effects and typical values in the Earth's liquid core \citep{braginsky1995equations,labrosse2015thermal} and stably stratified stellar envelopes \citep{garaud2015excitation}. Kinematic viscosity $\nu$, thermal diffusivity $\kappa_T$, compositional diffusivity $\kappa_C$, planetary angular velocity $\Omega_s$ and radius $R$.}
\label{table:nondim}
\end{table}

\subsection{Computational methods}
Simulations of RDDC in spherical geometry are computationally challenging. 
A major difficulty is to use small enough values of $\kappa_C$ for fixed values of $\kappa_T$, to probe the regime $L \gg 1$.
This means that the spatial resolution must be adequate, for simulating both the fine-scale compositional structures and the thermal ones.
In addition, planetary cores are generally rapidly rotating, as measured by the dimensionless Ekman number $Ek \ll 1$ (table \ref{table:nondim}). Thus, RDDC must be investigated in the regime $Ek \ll 1$ simultaneously with $L \gg 1$. 
Eulerian numerical methods cannot presently encompass this broad range of length (and time) scales properly. Hence, computations are always performed for dimensionless parameters orders of magnitude away from core values.

To circumvent these issues, a "particle-in-cell" (PIC) method has been developed \citep{bouffard2017double,bouffard2019chemical}.
It models the compositional field in the limit $L \gg 1$ as a collection of advected particles, while keeping an Eulerian description for velocity and temperature fields.
While PIC methods excel in the diffusionless limit $\kappa_C = 0$, they suffer from several drawbacks at finite values of $L$.
For instance, \citet{bouffard2017particle} showed that the PIC approach currently does not compare well with proposed benchmarks of RDDC in spherical geometry \citep{breuer2010thermochemically}, obtained at finite values of $L$. 
Finally, even if mixing Eulerian and PIC methods may be desirable for initial value problems, this approach prevents from efficiently finding the instability onset.
In contrast, the determination of the onset with Eulerian methods reduces to eigenvalue problems, which can be solved efficiently \citep[e.g. for convection][]{net2012numerical,kaplan2017subcritical}.

\subsection{Outline}
In this study, we aim at investigating numerically RDDC in spherical bodies. We are motivated by explaining the origin of the early geodynamo and by the potential importance of the double-diffusive effects highlighted by \citet{busse2002low} and \citet{simitev2011double}.
We will focus on rotating full spheres, without inner cores.
Beyond the geophysical motivation, a full sphere geometry is the simpler configuration to illustrate the intricate influence of rotation and global geometry on RDDC. Moreover, we will employ the classical Eulerian description, for which efficient codes are available.

The paper is organised as follows. The formulation of the problem is described in \S\ref{description_pb}, together with our numerical method of choice.
In \S\ref{sec:wkb}, we draw physical insights from existing local stability analyses. Then, we conduct a global stability analysis in spheres in \S\ref{sec:linearstab}, and we compare it with the asymptotic theory of RDDC in cylindrical geometry of \cite{busse2002low}
In \S\ref{sec:nonlinear}, we perform nonlinear simulations to study the rotating finger convection (i.e. for a destabilising compositional gradient and a stabilising thermal one).
In \S\ref{sec:applications}, we predict the onset of RDDC for core conditions and discuss the geophysical implications.
Finally, we end the paper in \S\ref{sec:ccl} with a conclusion and outline several perspectives for geophysical and astrophysical bodies.

\section{Description of the problem}
\label{description_pb}
\subsection{Dimensional background state}
We model RDDC in planetary cores by studying thermal and compositional Boussinesq convection in a rotating sphere. We consider a full sphere of radius $R$, filled with an homogeneous incompressible Newtonian fluid of density $\rho$, molecular kinematic viscosity $\nu$, thermal diffusivity $\kappa_T$ and compositional diffusivity $\kappa_C \ll \kappa_T$. 
The fluid is co-rotating with the sphere at the angular velocity $\boldsymbol{\Omega} =\Omega_s \boldsymbol{1}_z$ in the inertial frame. The fluid is also stratified in density under the (dimensional) imposed gravitational field $\boldsymbol{g} = -g_0 r \, \boldsymbol{1}_r$, where $g_0 R$ is the dimensional value of the gravity field at the outer spherical boundary $r=R$ and $\boldsymbol{1}_r$ is the unit radial vector in spherical coordinates $(r,\theta,\phi)$. 

Within the Boussinesq approximation \citep{spiegel1960boussinesq}, variations of the density $\rho^*$ due to the (dimensional) temperature $T^*$ and concentration of light elements $C^*$ are only taken in the buoyancy force. We use the following linear equation of state
\begin{equation}
\rho^*/\rho_m  = 1-\alpha_T(T^*-T_m)-\alpha_C(C^*-C_m)
\label{eq:EoS}
\end{equation}
by assuming $|\rho^* - \rho_m| / \rho_m \ll 1$, where $(T_m,C_m,\rho_m)$ are the mean reference values at $r=R$ and $(\alpha_T,\alpha_C)$ are the thermal and compositional expansion coefficients.
In equation of state (\ref{eq:EoS}), $T^*$ is actually the departure from the adiabatic reference temperature profile. 
Similarly, $C^*$ is the departure from the compositional reference barodiffusive profile \citep{davies2011buoyancy}, which is rather small compared to the adiabatic density profile \citep{gubbins1979thermal,gubbins2004gross}.

We work in the co-rotating reference frame. We study slight departures from a motionless, hydrostatic background state for the  temperature $T_0^*$ and composition $C_0^*$. The latter profiles are governed by the dimensional temperature and composition equations in the Boussinesq approximation
\begin{equation}
\kappa_T \nabla^2 T_0^*  = - \mathcal{Q} _T, \ \, \ 
\kappa_C \nabla^2 C_0^* =  - \mathcal{Q}_C,
\label{eq:T0C0}
\end{equation}
with $\mathcal{Q}_T$ and $\mathcal{Q}_C$ the thermal and compositional source (or sink) terms. 

Thermo-compositional convection is sustained by the thermal and compositional gradients ($\nabla T_0^*, \nabla C_0^*$). They can be maintained by (i) non-zero internal sources/sinks $(\mathcal{Q} _T,\mathcal{Q} _C)$, (ii) thermal or compositional fields externally imposed at the boundary or (iii) flux conditions. In the Earth's core, the thermal gradient is mainly imposed by heat extracted at core-mantle boundary (CMB), yielding flux conditions. The compositional gradient is presently mainly driven by the crystallisation of the solid inner core \citep[e.g.][]{loper1981study}, while, in the early Earth, it may have been driven by the precipitation of light elements at the top of the core \citep{o2016powering,badro2016early}. Hence, flux-type conditions are more relevant for compositional effects. 
Actually, the proper boundary condition ties the heat flux and the compositional flux to the local core dynamics \citep{braginsky1995equations}. This intricate condition has only been implemented in the anelastic simulations of \citet{glatzmaier1996anelastic}, who also treated separately thermal and chemical buoyancies. Yet, they assumed identical turbulent diffusivities, which discards double-diffusive effects. 

However, the choice of the boundary conditions is less crucial for the dynamics in the full sphere geometry (investigated here) than in spherical shells \citep{kutzner2000effects,hori2012influence}.
To ensure stationary solutions, we assume that thermal and compositional background profiles are sustained by spatially homogeneous sources $(\mathcal{Q} _T,\mathcal{Q} _C)$.
Hence, the dimensional solutions of equations (\ref{eq:T0C0}) are
\begin{subequations}
\label{eq:T0C0sol}
	\begin{align}
    	T_0^* (r) &= T_m + \frac{\mathcal{Q} _T}{6 \kappa_T} (R^2 - r^2), \\
        C_0^* (r) &= C_m + \frac{\mathcal{Q} _C}{6 \kappa_C} (R^2 - r^2).
    \end{align}
\end{subequations}
Without loss of generality, we set $(T_m,C_m)=(0,0)$, because they do not play any dynamical role (only the gradients do have a role).

\subsection{Dimensionless governing equations}
For numerical convenience, we work with dimensionless quantities. We use the length scale $R$, the viscous time scale ${R^2}/{\nu}$, the temperature scale $(\nu \mathcal{Q}_T R^2)/(6 {\kappa_T}^2)$ and the composition scale $(\nu  \mathcal{Q}_C R^2)/(6 {\kappa_C} ^2)$. Note that temperature and composition scales can be either positive or negative, depending on the signs of $(\mathcal{Q}_T,\mathcal{Q}_C)$.
In the following, we write the dimensionless velocity, temperature and composition without asterisk to differentiate them from their dimensional counterparts. In dimensionless form, dimensional background state (\ref{eq:T0C0sol}) yields 
\begin{equation}
	T_0 (r) = \frac{1- r^2}{Pr}, \ \, \ C_0 (r) = \frac{1- r^2}{Sc},
	\label{eq:prof-tc}
\end{equation}
with
\begin{equation}
Pr = \frac{\nu}{\kappa_T}, \ \, \ Sc = \frac{\nu}{\kappa_C}, \ \, \ L = \frac{Sc}{Pr},
\label{eq:PrSc}
\end{equation}
the Prandtl number, the Schmidt number and the Lewis number.

In the co-rotating frame, we assume centrifugal effects to be small compared to the self-gravity of the fluid sphere $\boldsymbol{g}$. This condition is typically met in planetary cores, such that we neglect the centrifugal buoyancy in the Boussinesq equations \citep{lopez2013boussinesq}. 
We denote $\boldsymbol{u}$ the dimensionless velocity field, $\Theta$ the dimensionless temperature and $\xi$ the dimensionless concentration departing from motionless background state (\ref{eq:T0C0}). The governing dimensionless equations are
\begin{subequations}
\label{eq:dimensionlessEQN}
\begin{align}
\frac{\partial \boldsymbol{u}}{\partial t} + (\boldsymbol{u} \boldsymbol{\cdot} \boldsymbol{\nabla}) \boldsymbol{u} &= -\frac{2}{Ek}\boldsymbol{1}_z \times \boldsymbol{u} - \nabla p +  \boldsymbol{\nabla}^2 \boldsymbol{u} \label{eq:nd-NS}  \\
&+ \left( Ra_{T} \, \Theta + Ra_{C}\,  \xi \right) \, r \boldsymbol{1}_r, \nonumber \\
\frac{\partial \Theta}{\partial t} + (\boldsymbol{u} \boldsymbol{\cdot} \boldsymbol{\nabla}) \Theta &= \frac{1}{Pr} \left ( 2 \, \boldsymbol{r} \cdot\boldsymbol{u} + \nabla^2 \Theta\right ), \label{eq:nd-Tp} \\
\frac{\partial \xi}{\partial t}  + (\boldsymbol{u} \boldsymbol{\cdot} \boldsymbol{\nabla}) \xi &= \frac{1}{Sc} \left ( 2 \, \boldsymbol{r} \cdot\boldsymbol{u} + \nabla^2 \xi \right ), \label{eq:nd-Ch}\\
\boldsymbol{\nabla} \boldsymbol{\cdot} \boldsymbol{u} &= 0 \label{eq:nd-div},
\end{align}
\end{subequations}
with $\boldsymbol{u}$ the dimensionless velocity field, $p$ the dimensionless reduced pressure (including the centrifugal force). In equations (\ref{eq:dimensionlessEQN}), we have introduced the Ekman number
\begin{equation}
    Ek = \frac{\nu}{\Omega_s R^2},
\end{equation}
the thermal and compositional Rayleigh numbers
\begin{equation}
	 Ra_{T} = \frac{\alpha_T g_0 \mathcal{Q}_T R^6}{6 \nu  {\kappa_T}^2}, \ \, \ Ra_{C} = \frac{\alpha_C g_0 \mathcal{Q}_C R^6}{6 \nu {\kappa_C} ^2}
    \label{eq:RaTRaC}
\end{equation}
which can be either positive or negative, depending on the signs of $(\mathcal{Q}_T,\mathcal{Q}_C)$, 
Typical values of numbers $[Pr,Sc,L,Ek]$ are given in table \ref{table:nondim}.

Equations (\ref{eq:dimensionlessEQN}) are supplemented by boundary conditions (BC). At the outer spherical boundary modeling the CMB, the velocity field satisfies the non-penetration and no-slip boundary conditions in the co-rotating frame, i.e.
\begin{equation}
\boldsymbol{u} \boldsymbol{\cdot} \boldsymbol{1}_r = 0, \ \, \ \boldsymbol{u} \times \boldsymbol{1}_r = \boldsymbol{0} \ \, \ \text{at} \ \, \ r=1.
\label{eq:nsBC}
\end{equation}
For the thermal and compositional perturbations $(\Theta, \xi)$, we impose zero radial fluxes
\begin{equation}
\frac{\partial \Theta}{\partial r} = \frac{\partial \xi}{\partial r} = 0 \ \, \ \text{at} \ \, \ r=1.
\label{eq:nofluxBC}
\end{equation}
The above boundary conditions (\ref{eq:nofluxBC}) are relevant for a planetary core, in which the CMB controls heat and compositional fluxes through the profiles $T_0(r)$ and $C_0(r)$.
The fixed temperature flux models the heat flux exctracted by the mantle, while the fixed compositional flux models exsolution of light elements by the mantle \citep[e.g.][]{o2016powering}.

\subsection{Brunt-V\"ais\"al\"a frequency}
\begin{figure}
\centering
\includegraphics[width=0.5\textwidth]{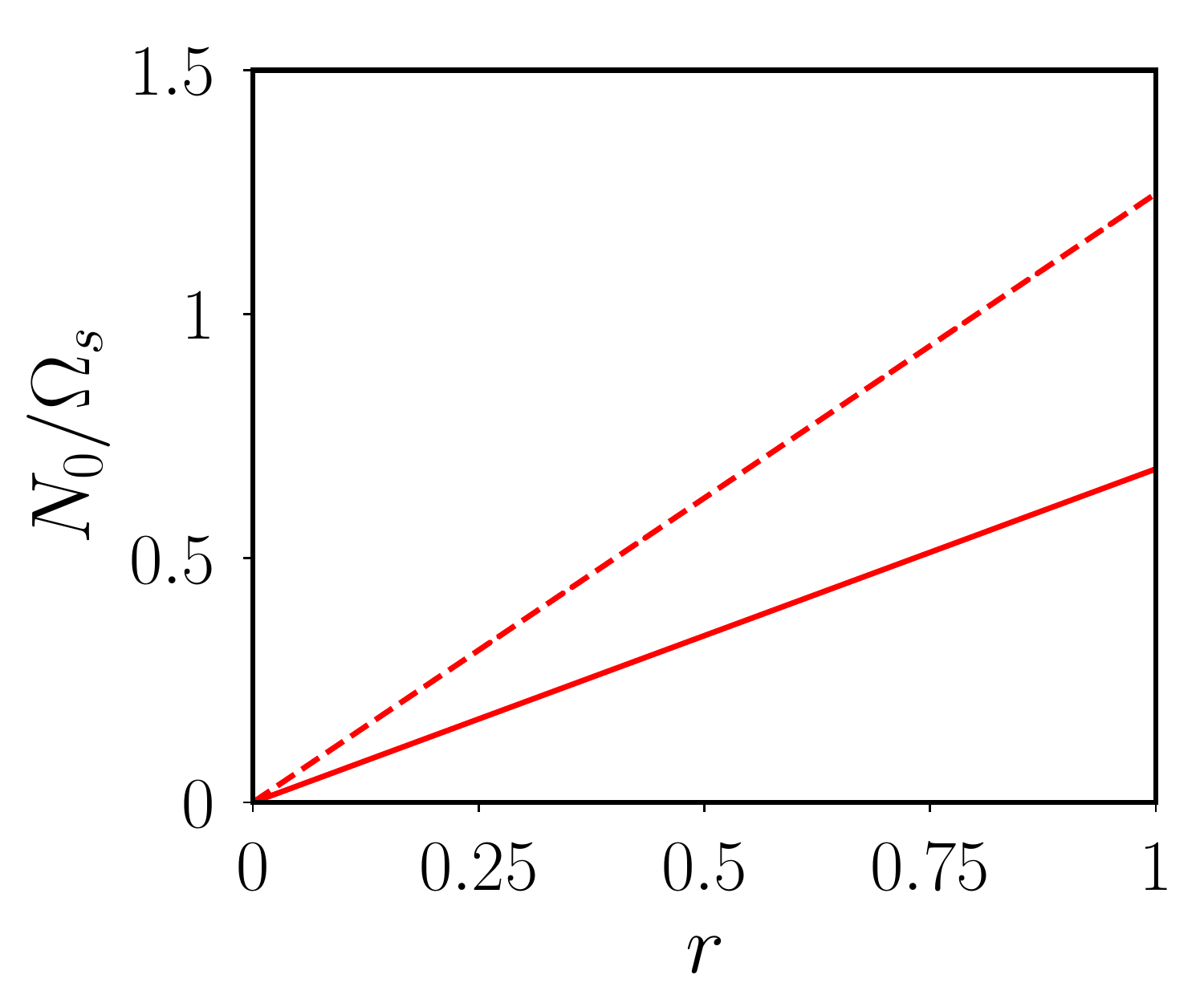}
\caption{Dimensionless background Brunt-V\"ais\"al\"a frequency for stably stratified background states $(N^2_0 / \Omega_s^2 >0)$. Parameters: $Pr=0.3, Sc = 3$,  $Ek = 10^{-5}$ and $Ra_T= -Ra_C/3$, with $Ra_C=3 \times 10^{9}$ (solid line) and $Ra_C=10^{10}$ (dashed line).}
\label{fig:BV}
\end{figure}

To compare heat and composition gradients, we introduce the total dimensional Brunt-V\"ais\"al\"a frequency $N$. The latter is defined in the Boussinesq approximation by \citep[e.g.][]{bullen1975earth}
\begin{equation}
N^2 = \boldsymbol{g} \boldsymbol{\cdot} \nabla \left ( \frac{\rho}{\rho_m} \right ).
\label{eq:N2Boussi}
\end{equation}
The fluid is stably stratified in density when $N^2 >0$, neutral when $N^2=0$ and unstably stratified when $N^2 < 0$. 
The total dimensional Brunt-V\"ais\"al\"a frequency characterising the background state, denoted $N_0$ in the following, is such that $N^2_0 = N^2_{0,T} + N^2_{0,C}$ where
\begin{equation}
N^2_{0,T} = - \alpha_T \, \boldsymbol{g} \boldsymbol{\cdot} \nabla T_0^* \ \, \ \text{and} \ \, \ N^2_{0,C} = - \alpha_C \,  \boldsymbol{g} \boldsymbol{\cdot} \nabla C_0^*
\end{equation}
are the thermal and compositional contributions. 
Solutions (\ref{eq:T0C0sol}) show that positive values of $(\mathcal{Q}_T,\mathcal{Q}_C)$ (respectively negative) give destabilising (respectively stabilising) thermal and compositional gradients. 

To compare the rotational effects with the stratification, a relevant quantity is the square of the Brunt-V\"ais\"al\"a frequency normalised by the fluid angular velocity $\Omega_s$. In dimensionless variables, it reads for the background state
\begin{equation}
	\frac{N_0^2}{\Omega_s^2}  (r) = - 2 r^2 \, Ek^2 \left( \frac{Ra_T}{Pr}+\frac{Ra_C}{Sc}\right) = -2 r^2 Ro_c^2,
	\label{eq:N2/Ws2}
\end{equation}
where $Ro_c = Ek \, (Ra_T/Pr + Ra_C/Sc)^{1/2}$ is the double-diffusive convective Rossby number. Formula (\ref{eq:N2/Ws2}) is illustrated in figure \ref{fig:BV}. Because of the quadratic radial dependence in the background state (\ref{eq:T0C0sol}),  the background Brunt-V\"ais\"al\"a frequency is linear in $r$ in our model. 
In pure thermal convection ($Ra_C=0, Ra_T>0$), $Ro_c$ is often employed as a proxy of the ratio between buoyancy and Coriolis forces \citep{gastine2016scaling}. In the strongly stratified regime, characterised by $Ro_c\gg1$, the scaling properties become reminiscent to non-rotating convection, whereas turbulent rotating convection is expected when $Ro_c\ll1$. Hence, we can expect a similar distinction between a strongly stratified regime of RDDC, when $|{N_0^2}/{\Omega_s^2}| \gg 1$ (i.e. $|Ro_c| \gg 1$), and a weakly stratified regime when $|{N_0^2}/{\Omega_s^2}| \ll 1$ (i.e. $|Ro_c| \ll 1$).

\subsection{Numerics in spheres}

We will employ the classical Eulerian description in spherical geometry to solve equations (\ref{eq:dimensionlessEQN}). So far, most Eulerian simulations of RDDC have neglected the distinction between thermal and compositional buoyancies. This lead to the co-density approach, first proposed by \citet{lister1995strength} and \citet{braginsky1995equations}, in which the two components have the same diffusivities $\kappa_T=\kappa_C$.
This assumption is widely used \citep[e.g.][]{schaeffer2017turbulent} and is mostly motivated by simplicity and numerical convenience, reducing by one both the number of parameters and equations.
The proposed justification is that these molecular diffusivities should be replaced by a turbulent one, accounting for the mixing by unresolved small-scale eddies.
However, this assumption is highly questionable and only possibly valid for highly turbulent flows, as found for overturning convection \citep{nataf2015turbulence}. Additionally, it filters out double-diffusive effects.

Only few Eulerian codes have treated separately the two buoyant components in spherical geometry, by using pseudo-spectral methods \citep{glatzmaier1996anelastic,manglik2010dynamo,net2012numerical,takahashi2014double} or finite volumes \citep{breuer2010thermochemically}.
Here, we use the linear code SINGE (\url{https://bitbucket.org/vidalje/singe}) and the nonlinear code XSHELLS (\url{https://nschaeff.bitbucket.io/xshells/}), which are both open-source codes. 
We have implemented in both codes the composition equation (\ref{eq:nd-Ch}) to account for double-diffusive effects. 
The SINGE code has been used for linear computations of waves \citep{vidal2015quasi} and convection onsets \citep{gastine2016scaling,kaplan2017subcritical} in spherical geometry.
On the other hand, XSHELLS 
can simulate turbulent flows in several contexts \citep{schaeffer2017turbulent,kaplan2017subcritical,kaplan2018dts}, scaling on thousands of cores, by using a domain decomposition in the radial direction (MPI and OpenMP standards).
XSHELLS solves the dynamical equations with a second order time-stepping scheme and treats the diffusive terms implicitly, while the nonlinear terms are handled explicitly. 

Both codes use a pseudo-spectral method, by describing the velocity field $\boldsymbol{u}$ with poloidal and toroidal scalars \citep[e.g.][]{backus1986poloidal}.
Then, poloidal and toroidal scalars are expanded onto spherical harmonics $Y_l^m (\theta,\phi)$ of degree $l$ and azimuthal wave number $m$, truncated at $(l_{\max}, m_{\max})$ in the simulations.
Similarly, temperature $\Theta$ and composition $\xi$ are also expanded onto spherical harmonics.
The two codes use second order finite differences in radius with $N_r$ points and spherical harmonic expansions provided by the fast SHTns library \citep{schaeffer2013efficient}.
At the origin ($r=0$), geometric conditions are applied: scalar fields ($\Theta$, $\xi$) can have a non-zero value at the origin. Since it must be independent of $\theta$ and $\phi$, only shperical harmonic $l=0$ is allowed.
Similary for vector fields that have a non-zero vector at the origin, only $l=1$ is allowed.
These all translate into appropriate boundary conditions, that distinguish $l=0$, $l=1$ and $l>1$ and which are used in both codes.
For nonlinear simulations, numerical instabilites can arise because of the clustering of points near the origin. In the XSHELLS code, these instabilities are suppressed by truncating the spherical harmonic degree at $l_{tr}(r) = 1 + \sqrt{r/r_s} \, l_{max}$ with $r_s=0.5$.
The XSHELLS code passes benchmarks designed to highlight issues arising at the origin \citep{marti2014full}.

\begin{figure}              
	\centering
    \begin{tabular}{c}
	\includegraphics[width=0.6\textwidth]{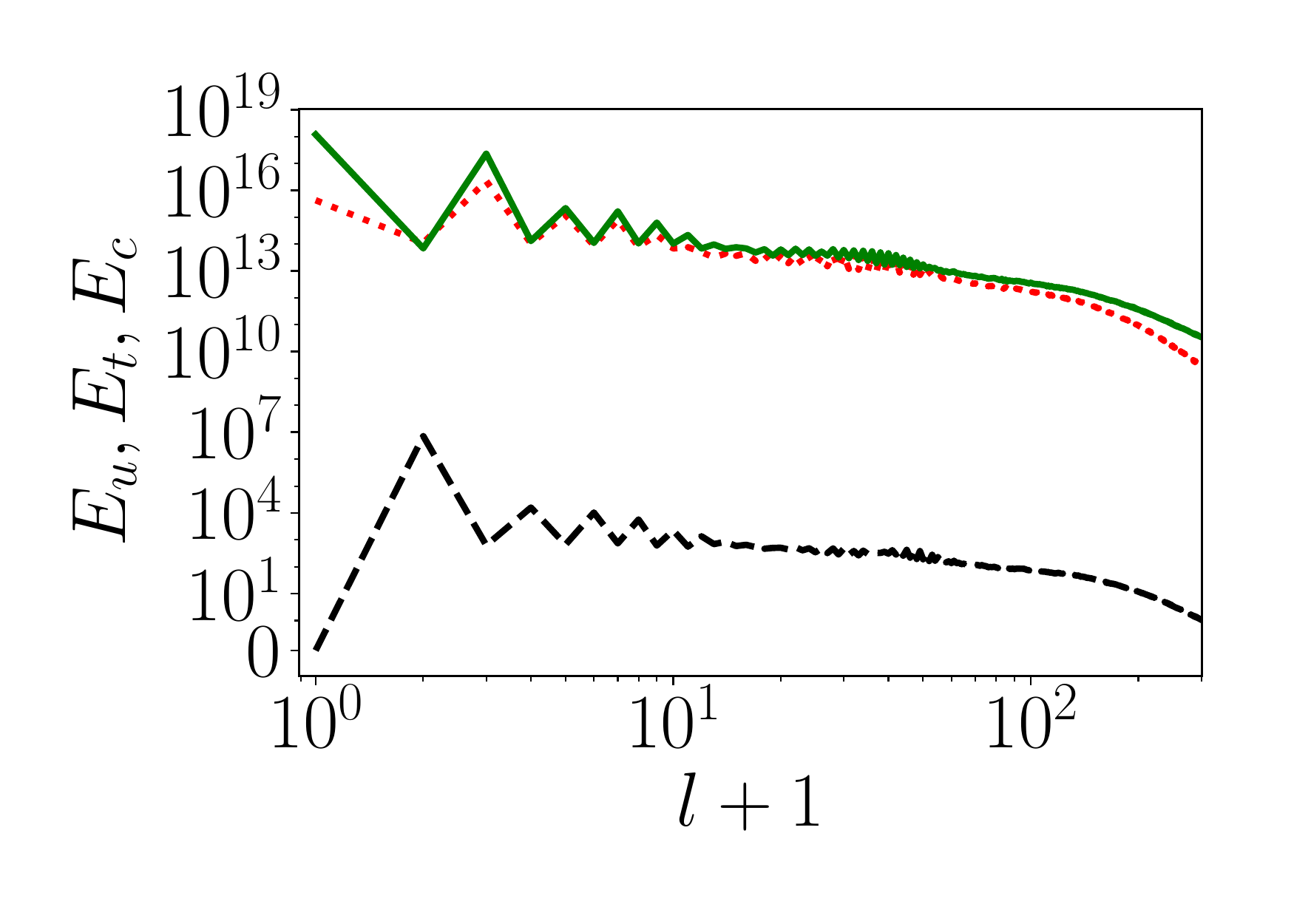} \\
    (a) \\
    \includegraphics[width=0.6\textwidth]{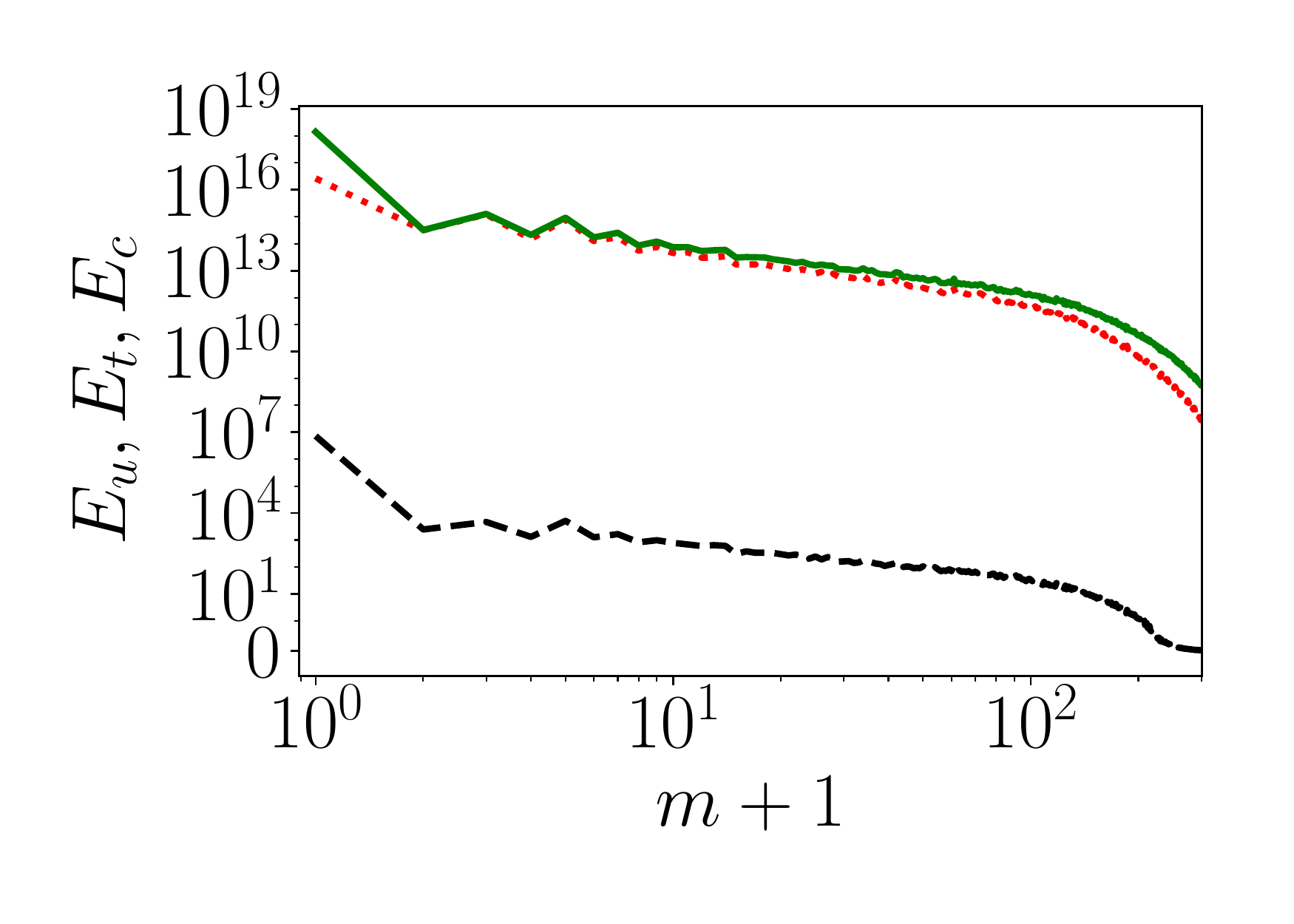} \\
    (b) \\
    \end{tabular}
	\caption{Instantaneous radial average of kinetic $E_u$ (dashed black), thermal $E_t$ (dotted red) and compositional $E_c$ (solid green) energy spectra as a function of the spherical degree $l+1$ (a) and azimuthal order $m+1$ (b) at $t = 2.7$ (viscous time). Nonlinear simulations at $Pr=0.3,Sc=3,Ek=10^{-5}$, $Ra_C=10^{10}$ and $Ra_T=-Ra_C/3$. The azimuthal spectrum (b) is dominated by the $m=0$ component due to the presence of zonal flows for large enough $Ra_C$, see figure \ref{fig:Uphizon} below.}
	\label{fig:spectre}
\end{figure}

The typical spatial resolution at $Ek=10^{-5}$ is $N_r=192$, $l_{\max}=120$, $m_{\max}=110$. For the most demanding nonlinear simulations (at large $Ra_C,Ra_T$), the numerical resolution is $N_r=384$, $l_{\max}=320$ and $m_{\max}=300$.
For such simulations, we show in figure \ref{fig:spectre} typical instantaneous spectra of the volume average of kinetic, thermal and compositional energies defined by
\begin{equation}
    E_{\{u, t, c\}} = \frac{1}{2} \int \left \{ |\boldsymbol{u}|^2, \Theta^2, \xi^2 \right \} \, \mathrm{d} V.
    \label{eq:KETECE}
\end{equation}
Spectra are numerically well converged.
We have also integrated the dynamics over several viscous time units (to skip any possible transient) to ensure reliable numerical results.

\section{Insights from local stability analyses}
\label{sec:wkb}

Composition and heat do not play a symmetrical role when $L \neq 1$. Several canonical situations occur and various local stability criteria have been devised for non-rotating fluids \citep[e.g.][]{garaud2018double}. Although the spherical geometry is natural for planetary cores, ruling out boundary effects yields physical insights for the stability. We briefly apply them for the background state (\ref{eq:T0C0sol}).  

Pioneering stability criteria have been inferred for non-rotating, diffusionless stellar interiors. 
\citet{ledoux1947stellar} obtained the stability criterion (in dimensional and dimensionless forms)
\begin{equation}
   N_0^2 = N^2_{0,T} + N^2_{0,C} > 0, \ \, \ \text{i.e.} \ \, \ Ra_T \, L <- Ra_C. 
   \label{eq:ledoux}
\end{equation}
Note that in the absence of compositional effects, Ledoux criteria (\ref{eq:ledoux}) reduces to the Schwarzschild criterion 
\citep{schwarzschild1959maximum} 
\begin{equation}
	N^2_{0,T} > 0, \ \, \ \text{i.e.} \ \, \ Ra_T>0.
    \label{eq:schwarchild}
\end{equation}
When the background state is both Schwarzschild ($N_{0,T}^2 < 0$) and Ledoux unstable ($N_0^2 < 0$), the fluid is prone to overturning convection driven by thermal and compositional buoyancies. 

However, Ledoux and Schwarzschild criteria (\ref{eq:ledoux})-(\ref{eq:schwarchild}) are not sufficient when heat (rapid diffuser) and composition (slow diffuser) have opposite destabilising/stabilising effects. Actually, the stability of the system depends on the density ratio $R_0$ \citep{stern1960salt}, given by
\begin{equation}
	R_0 = \frac{\alpha_T}{\alpha_C} \cdot \frac{|\nabla T_0^*|}{|\nabla C_0^*|} \sim \left | \frac{Ra_T}{Ra_C} \right | L,
    \label{eq:criteriafinger}
\end{equation}
in which the last estimate holds for our background state (\ref{eq:T0C0sol}) at the outer boundary. 
When the fluid is Ledoux unstable, i.e. $R_0 \leq 1$, the system is usually prone to overturning convection, but also sometimes to finger convection \citep{schmitt2011thermohaline}.
When the fluid is stable according to Ledoux criterion (\ref{eq:ledoux}) the situation depends on the values of $(Pr,Sc)$.
On the one hand, the situation $N^2_{0,T}>0$ (i.e. $Ra_T < 0$) and $N^2_{0,C} <0$ (i.e. $Ra_C > 0$) refers to the finger regime. In addition to overturning convection for $R_0 \leq 1$, the finger configuration is prone to double-diffusive instabilities when \citep{baines1969thermohaline}
\begin{equation}
1 \leq R_0 < L, \ \, \ \text{i.e.} \ \, \ |Ra_T| \leq Ra_C.
\label{eq:fingeringNRDDC}
\end{equation}
In that case, several finger DDC patterns can develop. On the other hand, the situation $N^2_{T,0}<0$ and $N^2_{C,0} >0$ refers to the semi-convection regime \citep{spiegel1969semiconvection}. The fluid is prone to double-diffusive instabilities when \citep[e.g.][]{radko2013double}
\begin{equation}
1 \leq R_0^{-1} \leq \frac{Pr + 1}{Pr + 1/L}. 
\label{eq:criteriaODDC}
\end{equation}
Based on typical values of dimensionless Lewis and Prandtl numbers given in table \ref{table:nondim}, we can expect many celestial fluid bodies to be unstable against double-diffusive convection according to criteria (\ref{eq:criteriafinger})-(\ref{eq:criteriaODDC}).

The aforementioned  local criteria do not account for rotational effects. 
Because the background state (\ref{eq:T0C0sol}) is spatially varying, we cannot directly use plane-wave perturbations usually employed in local analyses \citep[e.g.][]{cebron2013elliptical}. However, the spatial extent of a local model is much smaller than the size of the global domain. Hence, we can linearise (\ref{eq:T0C0sol}) in first approximation around a given position $\boldsymbol{r}_0$ to use WKB-type perturbations of the form
\begin{equation}
	\left [ \boldsymbol{u}, \Theta, \xi \right ] \propto \exp \left ( \mathrm{i}\boldsymbol{k} \boldsymbol{\cdot} \boldsymbol{r}_0 + \lambda t \right ),
    \label{eq:KelvinDDC}
\end{equation}
with $\boldsymbol{k}$ the local wave vector and $\lambda = \sigma + \mathrm{i} \omega$ the eigenvalue, where $\sigma \geq 0$ is the growth rate (or damping rate if $\sigma \leq 0$) and $\omega$ is the angular frequency. 
Note that perturbations (\ref{eq:KelvinDDC}) differ from WKB-type perturbations considered by \citet{yano1992asymptotic} and \citet{jones2000onset} for thermal convection. Indeed, the latter perturbations are exponentially decaying in the cylindrical direction around a given cylindrical radius (to fulfill the boundary conditions).
After some algebra, this yields a polynomial equation for the eigenvalue $\lambda$, similar to the one obtained by \citet{sengupta2018effect}, valid at the local position $\boldsymbol{r}_0$ of colatitude angle $\theta$. 
Note that \citet{braginsky2006formation} obtained a similar polynomial but considered a truncated version of the Coriolis force. 
As first obtained by \citet{sengupta2018effect}, the local analysis shows that the aforementioned non-rotating criteria are asymptotically valid for weakly rotating RDDC. Moreover, it indubitably shows that fastest-growing unstable waves for local rotating finger convection are largely unaffected by rotation. 
The unstable waves span the height of the local domain, with typical wave numbers $\boldsymbol{k} \boldsymbol{\cdot} \boldsymbol{g}=0$, called elevator waves \citep[e.g.][]{sengupta2018effect}. All other waves are merely stabilised by rotation. 
Moreover, the range of density ratios $R_0$ for which RDDC takes place is unchanged compared to non-rotating DDC, given by (\ref{eq:fingeringNRDDC}) in the finger regime.

However, this local behaviour may be misleading. Indeed, it is known that WKB-type local solutions do not necessarily provide approximations to the complete three-dimensional global solutions. For instance, they can severely differ for thermal convection in the limit $Ek \ll 1$ \citep{busse1970thermal,soward1977finite,yano1992asymptotic,jones2000onset}. 
Therefore, the local analysis, predicting unavoidably elevator modes as the fastest-growing modes, is likely inaccurate to describe the onset of RDDC for rapidly rotating cores, and we turn to a global stability analysis.

\begin{figure}
	\centering
	\includegraphics[width=0.8\textwidth]{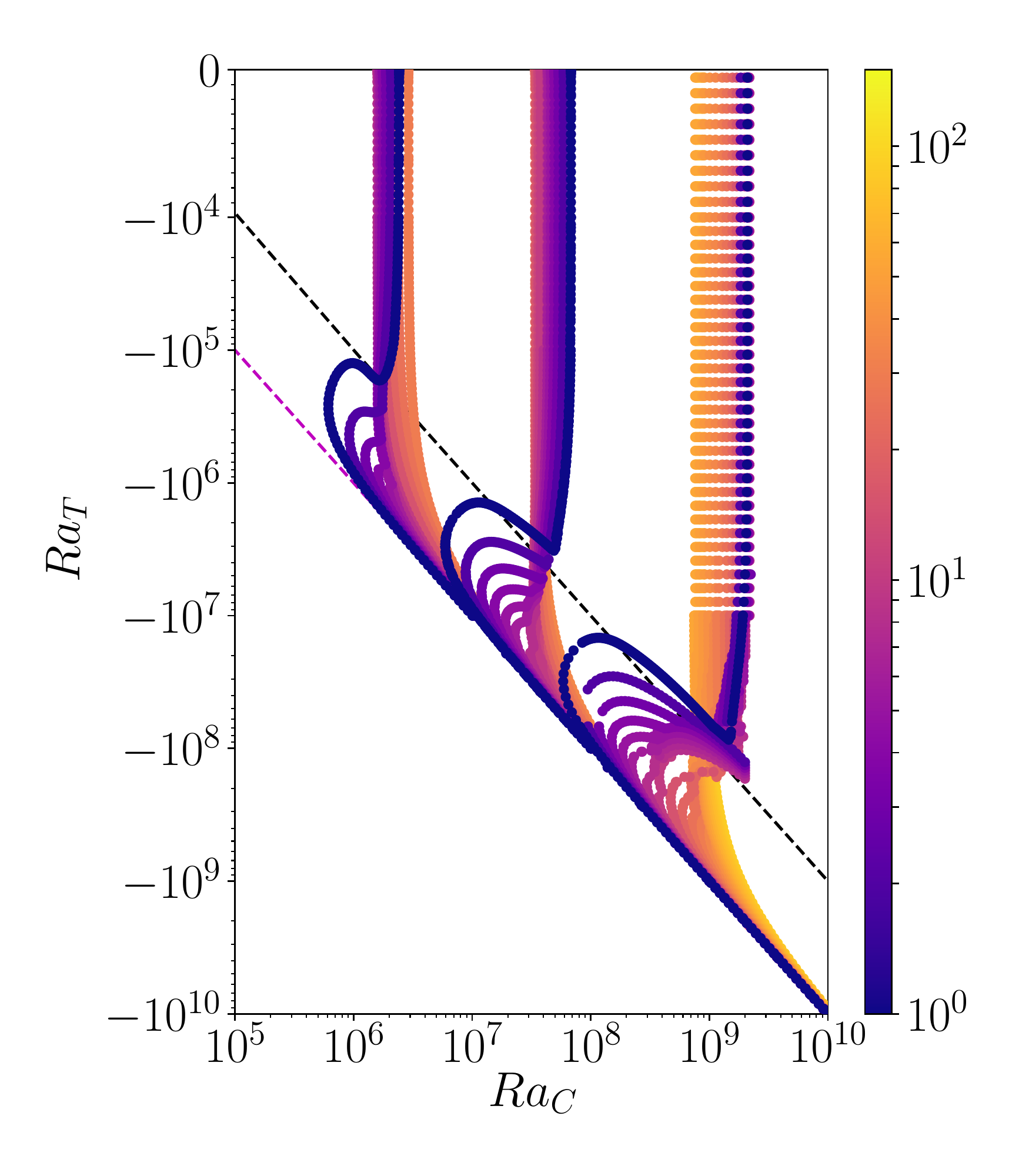}
    \caption{Linear onset of RDDC in the finger regime $(Ra_T<0, Ra_C>0)$ computed with SINGE. Computations at $Pr=0.3, Sc=3$ and for $Ek=10^{-4}$ (left), $Ek=10^{-5}$ (middle) and $Ek=10^{-6}$ (right). Colour shows the considered critical wave number $m$ at the onset. The upper (black) dashed  line is the neutral stability curve $N^2_0=0$. The lower (magenta) dashed line is the upper bound (\ref{eq:fingeringNRDDC}) for non-rotating finger convection, i.e. $Ra_T = -Ra_C$.}
    \label{fig:SINGE_Busse_bas}
\end{figure}

\section{Global stability analysis}
\label{sec:linearstab}
\subsection{Generalised eigenvalue problem}
In this section, we perform a global linear stability analysis of background state (\ref{eq:T0C0sol}). 
To do so, we discard the nonlinear terms ($\boldsymbol{u} \boldsymbol{\cdot} \boldsymbol{\nabla} \boldsymbol{u}, \boldsymbol{u} \boldsymbol{\cdot} \boldsymbol{\nabla} \Theta, \boldsymbol{u} \boldsymbol{\cdot} \boldsymbol {\nabla} \xi)$ in equations (\ref{eq:dimensionlessEQN}).
The symmetries of the background state and the linearised equations leads to uncoupled families of modes.
The axisymmetry implies all azimuthal wave numbers $m$ are uncoupled and can be considered separately.
Similarly, the reflexion symmetry about the equatorial plane implies the same for symmetric ($s=1$) and anti-symmetric ($s=-1$) modes with respect to that plane.
Thus, for a given $m$ and symmetry $s$, we expand the linear perturbations in spherical coordinates $(r,\theta,\phi)$ as
\begin{equation}
	\left [ \boldsymbol{u}, \Theta, \xi \right ] (\boldsymbol{r},t) = \left [ \widehat{\boldsymbol{u}}, \widehat{\Theta}, \widehat{\xi} \right ] (r,\theta) \exp \left [ \mathrm{i}m \phi + \lambda t \right ],
    \label{eq:eigenvalexpand}
\end{equation}
where $\lambda = \sigma + \mathrm{i} \omega$ is the complex eigenvalue with the growth rate $\Re_e (\lambda) = \sigma$ and the angular frequency $\Im_m (\lambda) = \omega$.
Substituting expansions (\ref{eq:eigenvalexpand}) into equations (\ref{eq:dimensionlessEQN}) yields the generalised eigenvalue problem (in symbolic form)
\begin{equation}
	\mathcal{A} \boldsymbol{X} = \lambda \mathcal{B} \boldsymbol{X},
	\label{eq:GEP_RDDC}
\end{equation}
with $\boldsymbol{X} = [ \widehat{\boldsymbol{u}}, \widehat{\Theta}, \widehat{\xi} ]$ the state vector and $(\mathcal{A},\mathcal{B})$ two linear operators, associated with the left and right hand sides of equations (\ref{eq:dimensionlessEQN}) and taking into account boundary conditions (\ref{eq:nsBC})-(\ref{eq:nofluxBC}). Problem (\ref{eq:GEP_RDDC}) is a boundary value problem, giving the dispersion relation for the complex eigenvalue
\begin{equation}
	\lambda = \lambda(m, s, Ra_T, Ra_C, Pr, Sc, Ek).
    \label{eq:dispGEP}
\end{equation}
From relation (\ref{eq:dispGEP}), the linear onset of convection is defined by the marginal state $\sigma=0$, realized for a set of Rayleigh numbers $(Ra_T, Ra_C)$ for given values of $(m, s, Ek, Pr, Sc)$.

We use the SINGE code \citep{vidal2015quasi} to solve the generalised eigenvalue problem (\ref{eq:GEP_RDDC}), by using an efficient sparse eigenvalue solver provided by the SLEPC library \citep{hernandez2005slepc}.
At the parameters of our study, we found that the onset of RDDC is systematically governed by equatorially symmetric ($s=1$) perturbations (i.e. they have a lower onset than anti-symmetric perturbations).
This is similar to purely thermal convection in spheres \citep[e.g.][]{busse1970thermal,jones2000onset} and RDDC with an inner-core at similar parameters \citep{net2012numerical}. Nevertheless, the antisymmetric modes may still play a role \citep{landeau2011,net2012numerical}, see here \S\ref{sec:nonlinear}.

We survey dispersion relation (\ref{eq:dispGEP}) by fixing all parameters except one of the Rayleigh numbers $Ra_X$ (where $X$ can be $T$ or $C$), that we vary until the growth rate $\sigma=0$ is bracketed within a small tolerance.
This is done automatically by the SINGE code using an optimization procedure based on Brent's method.
Having computed a collection of Rayleigh numbers $Ra_X$ at the onset for various azimuthal wave numbers $m$, we can usually define the critical number $Ra_X^c$ obtained for the critical wave number $m^c$, yielding the minimum Rayleigh number over all computed azimuthal numbers. 

\subsection{Marginal stability}
\subsubsection{Convection for unstably stratified fluids}
\begin{table}
 \centering
\begin{tabular}{cccc}
 \hline
     $Ek$ & $m^{c}$ & $Ra_C^{c}$ & $\omega$ \\ [0.5ex]
 \hline
 $10^{-4}$ & 10 & $1.59 \times 10^6$ & $-7.74 \times 10^1$ \\[0.5ex]
 $10^{-5}$ & 20 & $3.40 \times 10^7$ & $-4.63 \times 10^2$ \\[0.5ex]
 $10^{-6}$ & 55 & $7.60 \times 10^8$ & $-2.32 \times 10^3$ \\
\hline
\end{tabular}

\caption{Critical azimuthal wave number $m^{c}$, compositional Rayleigh number $Ra_C^{c}$ and angular frequency $\omega^c$ at the linear onset ($\sigma=0$) of compositional overturning convection (i.e. for $Ra_T=0$). Computations with SINGE at $Sc=3$.} 
\label{table:onset}
\end{table}

\begin{figure*}
	\centering
	\includegraphics[width=0.99\linewidth]{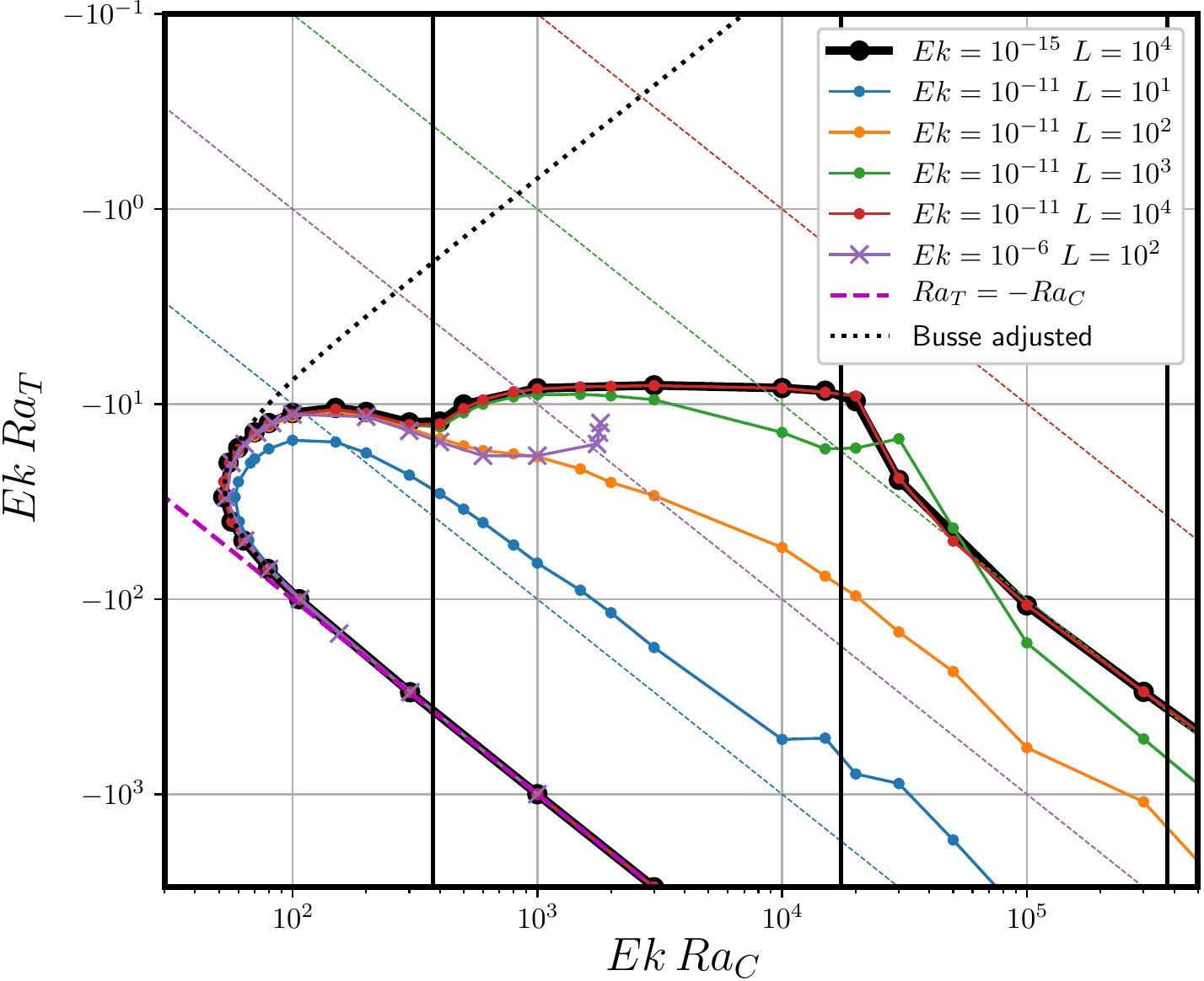}
    \caption{Influence of the Lewis and Ekman numbers on the linear onset of RDDC for the $m=1$ mode computed with SINGE. The curves were computed using $Sc=3$, but the exact same curves are obtained with $Sc=30$ or $Sc=100$.
    The dotted (black) line is the marginal curve from the theory of \citet{busse2002low} with $\Gamma = 2.4 $ (see appendix \ref{appendix:Busse}).
	The thin (colored) dashed lines are the neutral stability curve $N^2_0=0$ for the corresponding Lewis numbers.
	The thick (magenta) dashed line is the upper bound (\ref{eq:fingeringNRDDC}) for non-rotating finger convection, i.e. $Ra_T = -Ra_C$.
	Vertical lines mark the theoretical onset of convection for $Ra_T=0$ for $Ek=10^{-6}$, $10^{-11}$ and $10^{-15}$ from left to right.}
    \label{fig:inviscid}
\end{figure*}

We set $Sc=3$ and $Pr=0.3$, giving a Lewis number $L=10$, and report in table \ref{table:onset} the critical parameters at the onset of pure compositional (overturning) convection for $Ra_T=0$. 
As already noticed for pure thermal convection \citep{zhang1992spiralling,jones2000onset}, we report only a broad agreement between our global numerical results and local predictions at the onset \citep[e.g.][]{busse1970thermal}. The critical Rayleigh numbers $Ra_C^c$ are typically under-estimated by a factor two in the local theories (compared to the numerics), whereas the critical wave number $m^c$ and the angular frequency $\omega$  are over-estimated (not shown).

Then, we investigate the stability in the presence of an additional stabilising thermal background, which we refer to as the finger regime ($Ra_C \geq 0, Ra_T \leq 0$).
For many fixed $Ra_T < 0$, we determine the critical value of the compositional Rayleigh number $Ra_C^c$, reported in figure \ref{fig:SINGE_Busse_bas} for three Ekman numbers $Ek = \{10^{-6}, 10^{-5}, 10^{-4}\}$.
When $|Ra_T| \ll |Ra_C|$, the preferred modes of convection are almost that of a pure compositional convection, with an onset almost unchanged. 
Indeed, double-diffusive effects become significant only when $|Ra_T| \sim |Ra_C|$.
This behaviour has also been observed in thick shells \citep{net2012numerical}.

\begin{figure*}
\centering
\begin{tabular}{cc}
   	\includegraphics[height=0.35\textwidth]{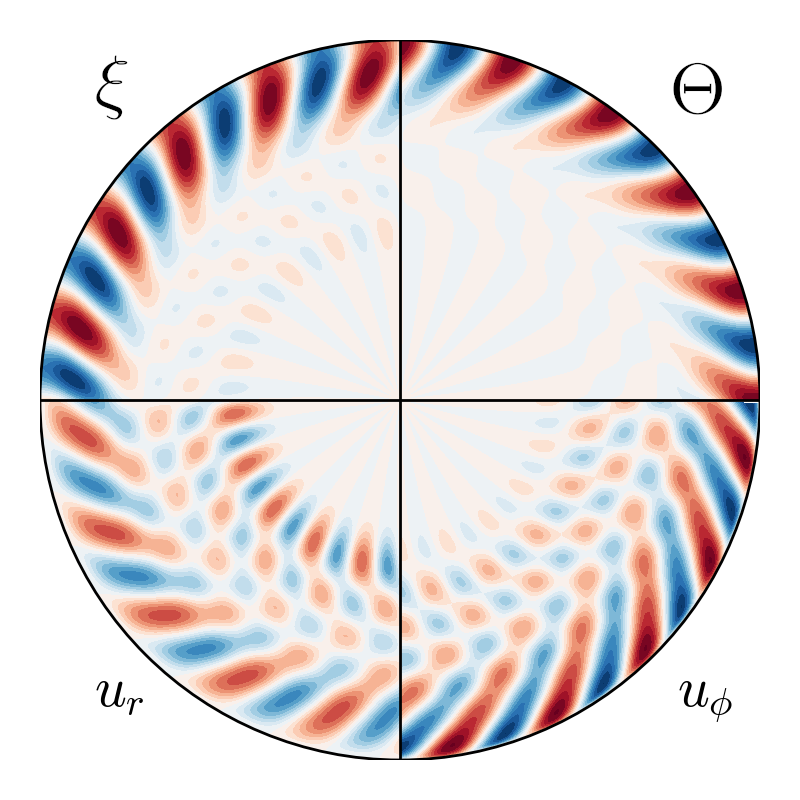} &
   	\includegraphics[height=0.35\textwidth]{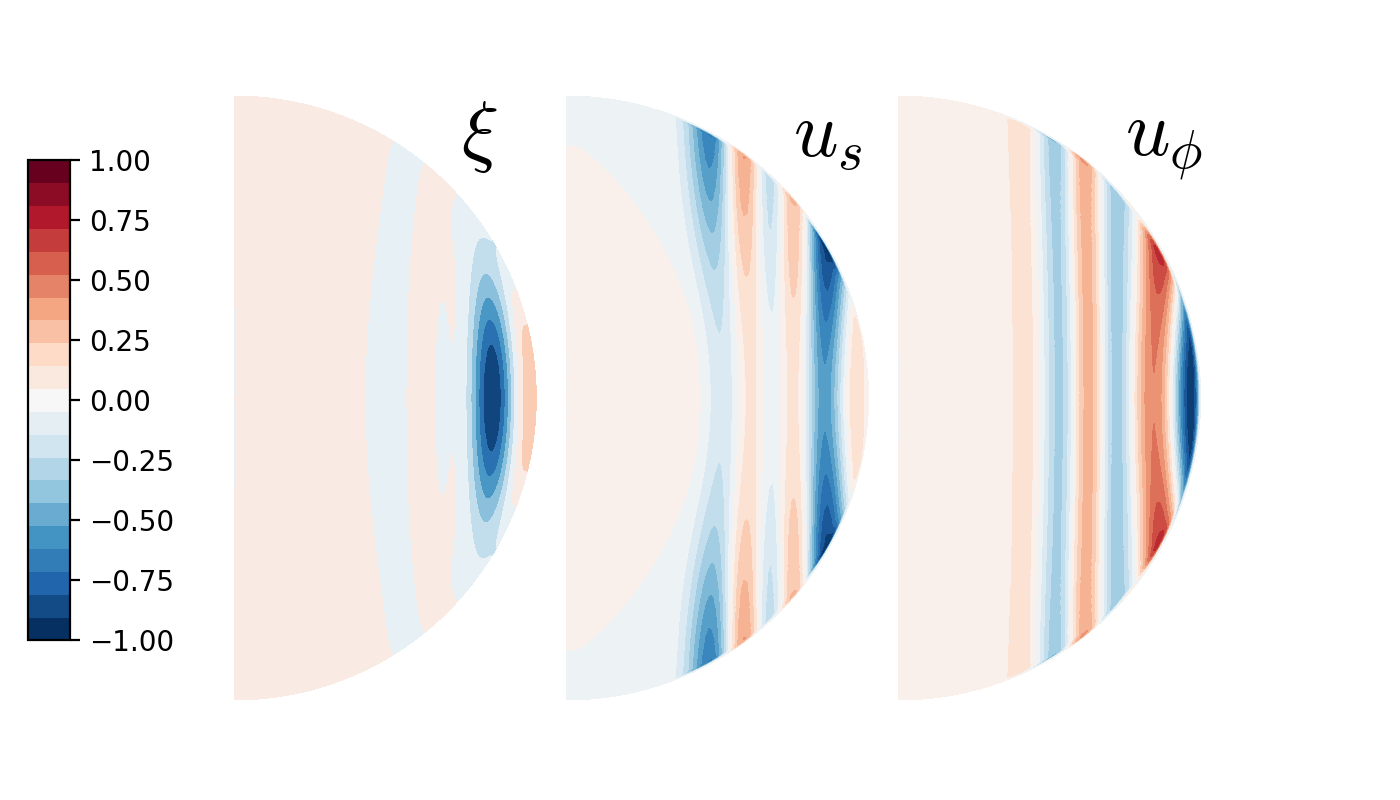} \\
    \multicolumn{2}{c}{(a) $m=20, Ra_T=0, Ra_C=3.4 \times 10^7, \omega / \Omega_s = -4.66 \times 10^{-3}$} \\
   \includegraphics[height=0.35\textwidth]{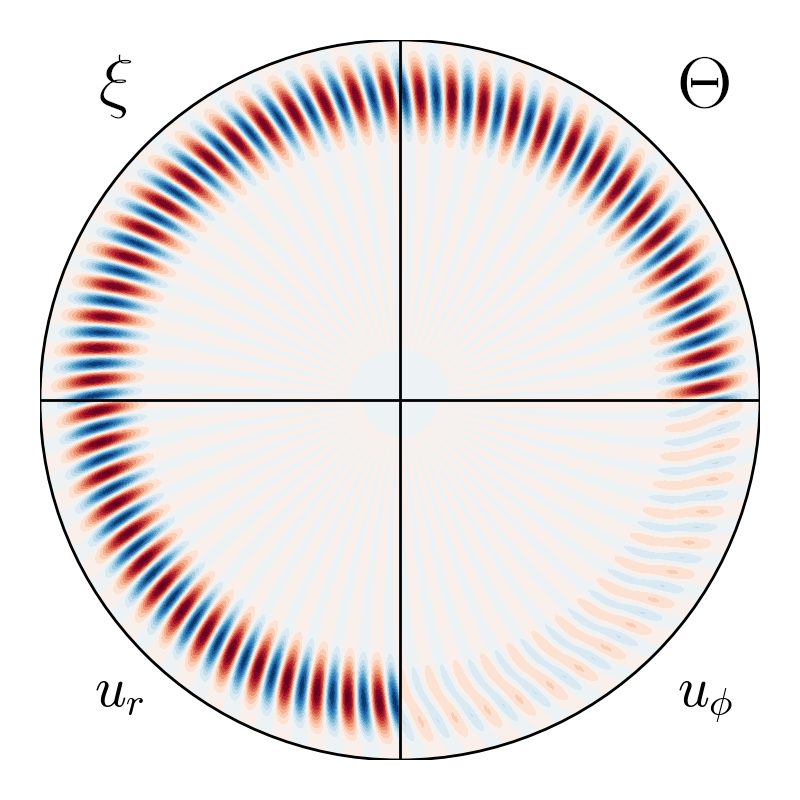} &
   \includegraphics[height=0.35\textwidth]{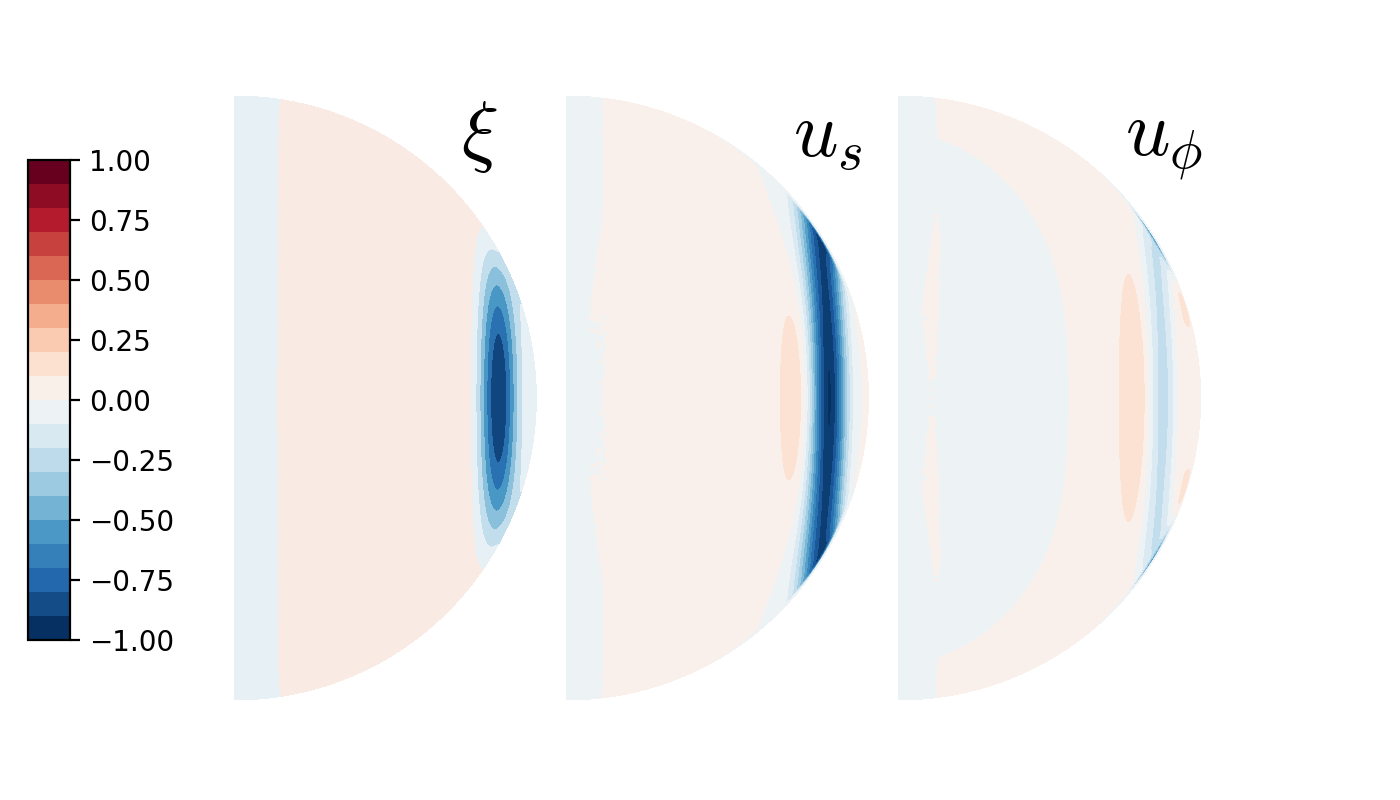} \\
    \multicolumn{2}{c}{(b) $m=60, Ra_C Ek=10^4=-Ra_T Ek, \omega / \Omega_s = 1.45 \times 10^{-5}$} \\
\end{tabular}
\caption{Eigenmodes at the onset of RDDC computed with SINGE. Linear computations at $Pr=0.3, Sc=3$ and $Ek={10}^{-5}$.
The full discs are equatorial cuts representing four different field components. The half discs are meridional cuts (taken at 3 o'clock in the equatorial planes) showing each a different component. The relative amplitude of $u_s$ (cylindrical radial velocity) and $u_\phi$ (azimuthal velocity) are preserved by using the same color maps. For the $m=60$ with $Ra_T=-Ra_C$, the shape and amplitude of the $\xi$ (composition) and $\Theta$ (temperature) fields are the same.}
\label{fig:flowonset}
\end{figure*}

\begin{figure*}
  \begin{tabular}{c}
   	\includegraphics[height=0.25\textwidth]{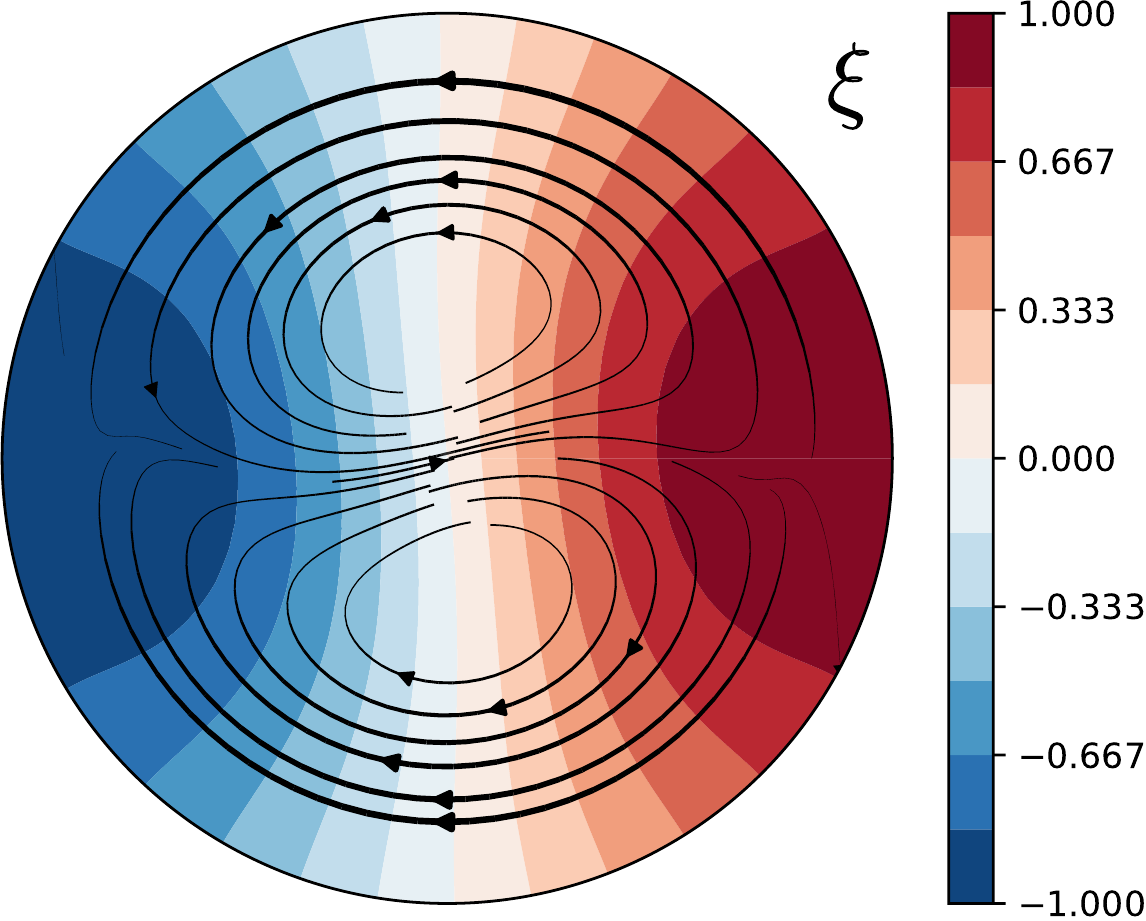}
   	\includegraphics[height=0.25\textwidth]{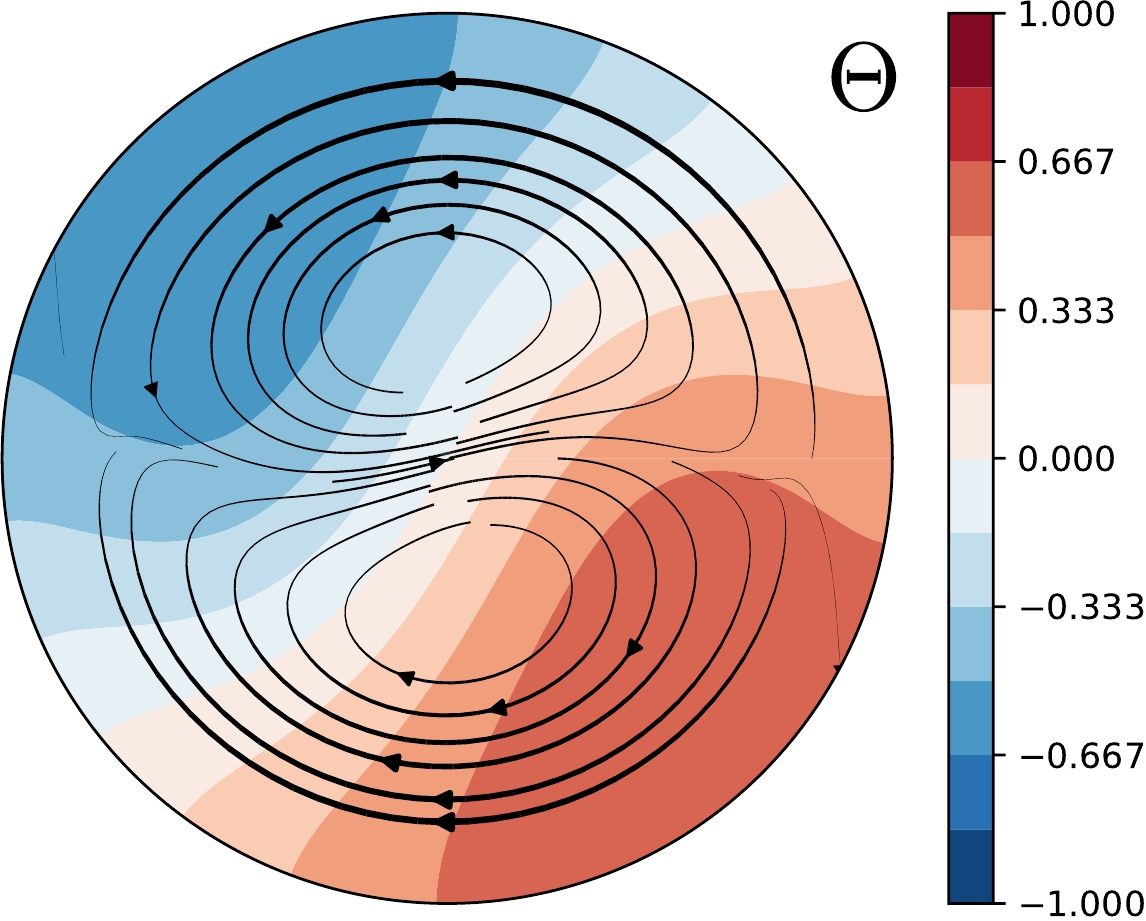} \quad
   	\includegraphics[height=0.25\textwidth]{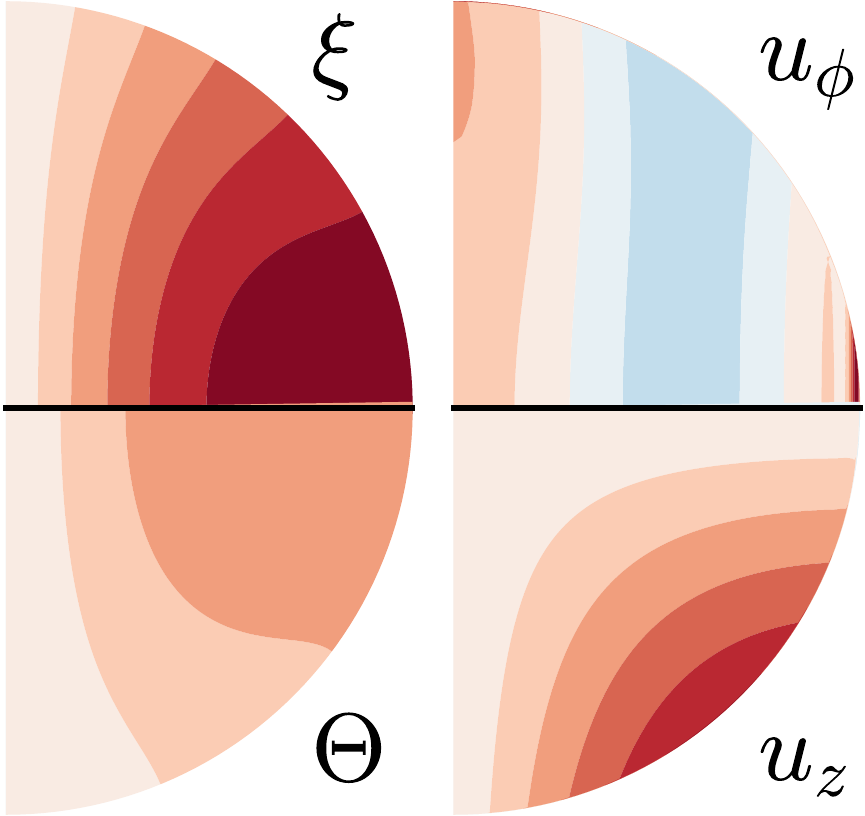} \\
  (a) $m=1, Ek\,Ra_T=-25, Ek\,Ra_C=52.5, \omega/\Omega_s = -0.16 \times 10^{-7}$ \\[1em]
   	\includegraphics[height=0.25\textwidth]{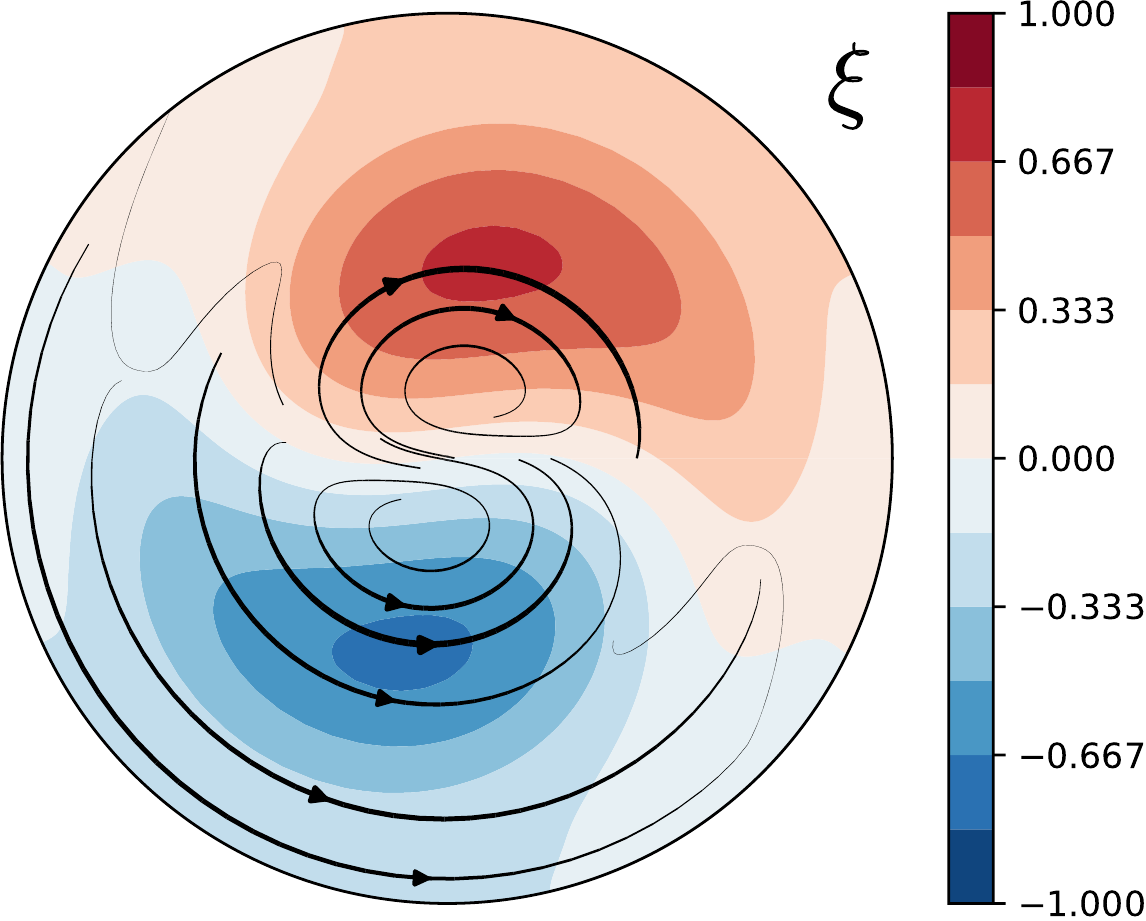}
   	\includegraphics[height=0.25\textwidth]{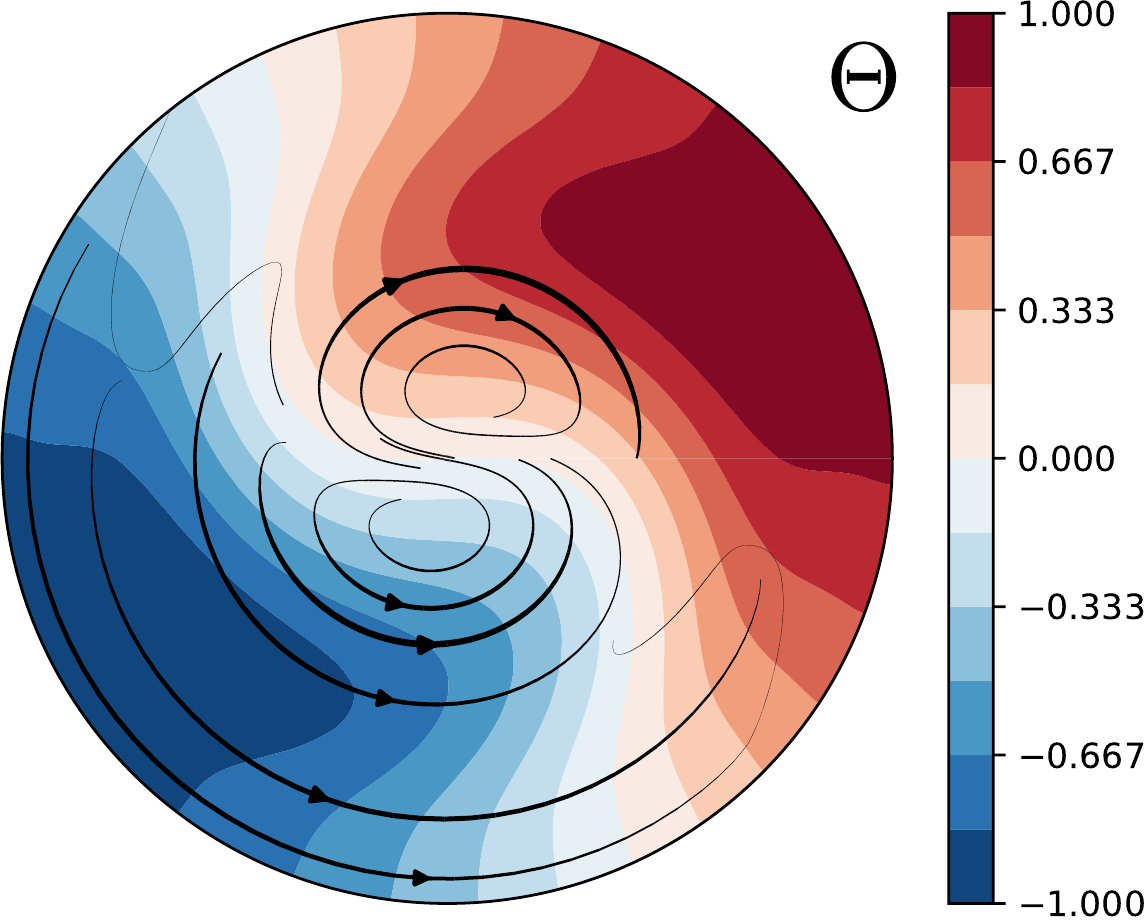} \quad
   	\includegraphics[height=0.25\textwidth]{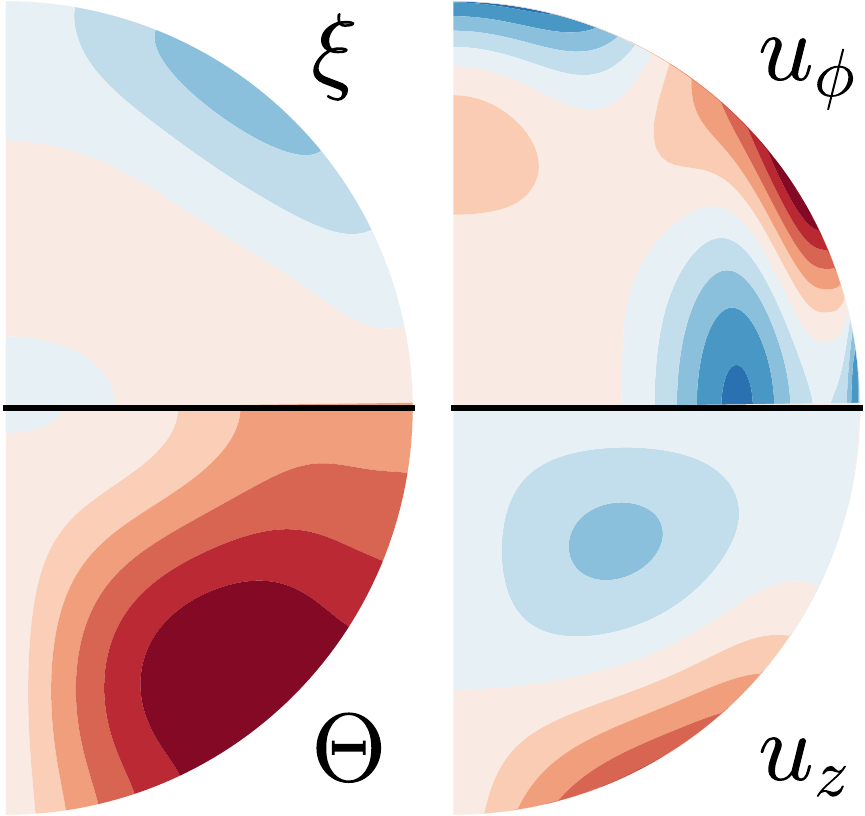}  \\
  (b) $m=1, Ek\,Ra_T=-9, Ek\,Ra_C=1000, \omega/ \Omega_s = -8.66 \times 10^{-7}$ \\[1em]
   	\includegraphics[height=0.25\textwidth]{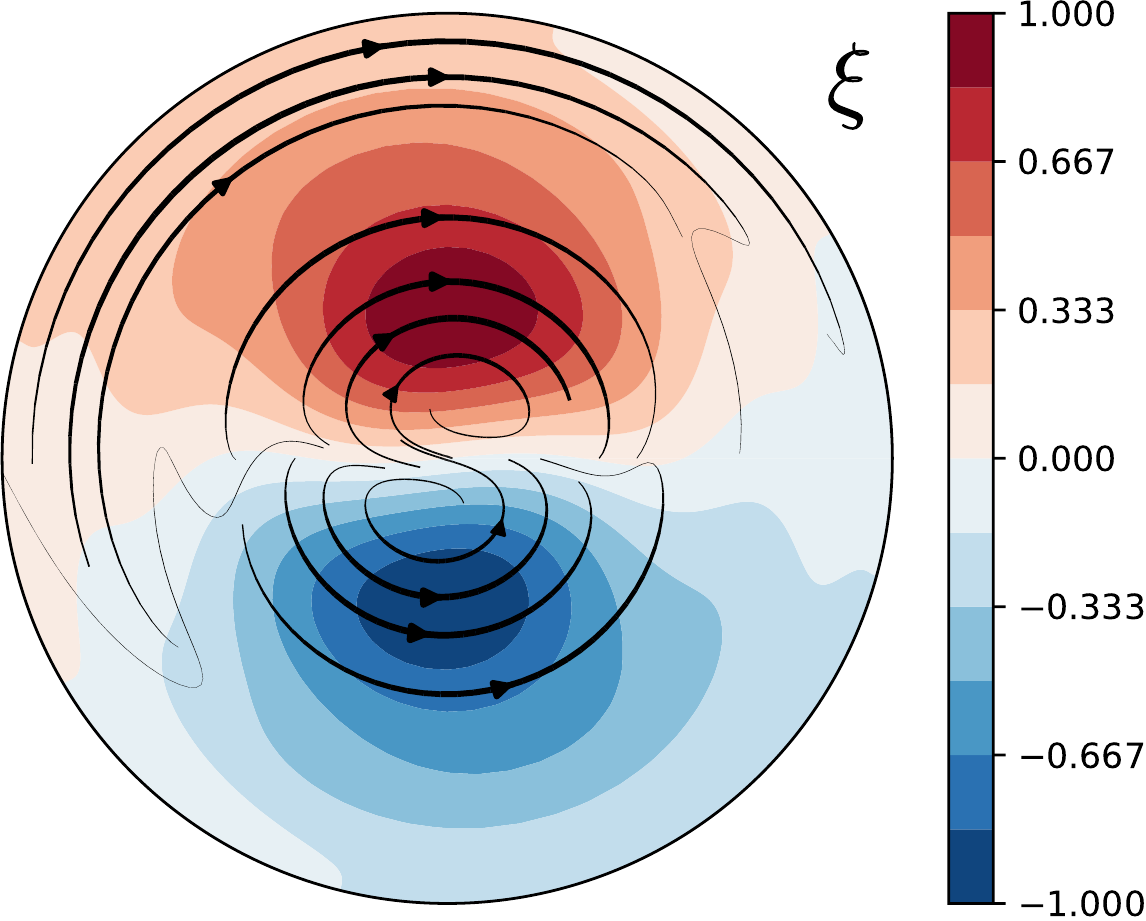}
   	\includegraphics[height=0.25\textwidth]{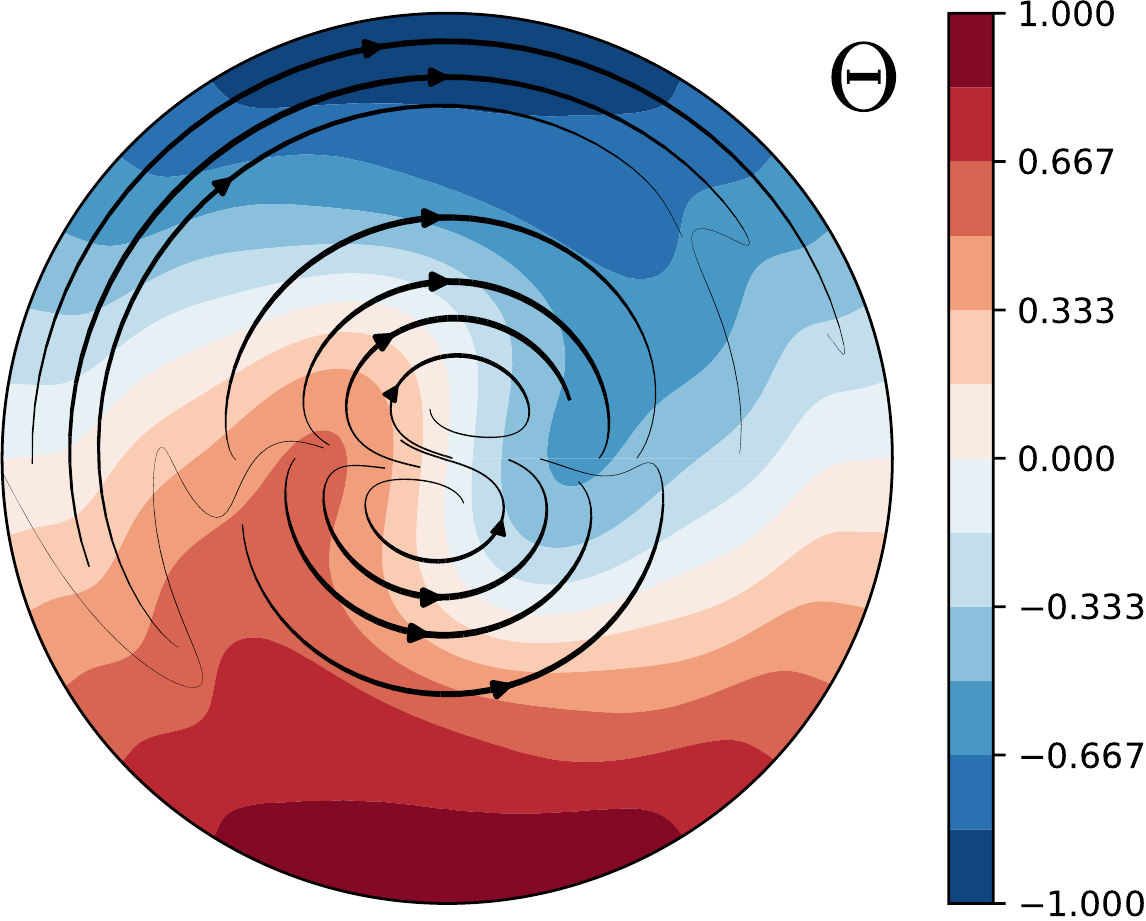} \quad
   	\includegraphics[height=0.25\textwidth]{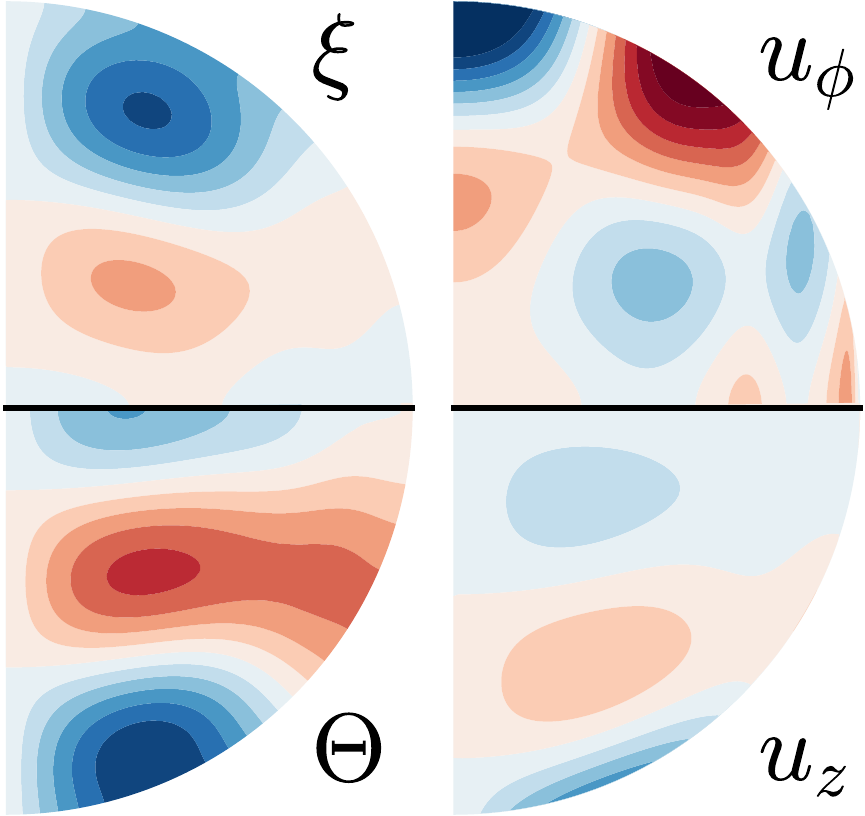} \\
  (c) $m=1, Ek\,Ra_T=-9, Ek\,Ra_C=3000, \omega/\Omega_s = -24.3 \times 10^{-7}$
  \end{tabular}
\caption{Eigenmodes at the onset of RDDC computed with SINGE at $Pr=0.03, Sc=30$ and $Ek={10}^{-7}$. The composition ($\xi$) and temperature ($\Theta$) perturbations are shown as color maps, superimposed with streamlines of the instantaneous velocity field.
The full discs are equatorial plane cuts, while the half discs are meridional cuts (taken at 3 o'clock in the equatorial planes). The relative amplitude of $\xi$ and $\Theta$ is preserved by using the same color map; likewise for $u_\phi$ and $u_z$.}
\label{fig:flowonset_m1}
\end{figure*}

\subsubsection{Inviscid convection for stably stratified fluids}
\label{sec:inviscid}
When $Ra_T<0$ is further decreased, double-diffusive effects start playing a significant role when the fluid is stably stratified in density.
For some values of $Ra_T$, there are now three values of $Ra_C$ that give $\sigma=0$, and $\sigma$ does not evolve monotonically with $Ra_C$.
The marginal stability curve $\sigma=0$ takes schematically the form of a tongue in the $Ra_c - Ra_T$ diagram (figure \ref{fig:SINGE_Busse_bas}), stretching towards lower $Ra_C$ within the stably stratified domain.
Within this tongue, convection occurs at $Ra_C$ and $m$ much lower than for $Ra_T=0$ (typically $m \leq 10$), down to $m=1$ near the edges.
This effect gets more important as $Ek$ is reduced, as observed in figure \ref{fig:SINGE_Busse_bas}.
At $Ek=10^{-6}$, $Ra_C$ in the tongue can go down to 10 times lower than the minimum $Ra_C^c$ of pure chemical convection. 
Furthermore, the smaller the $m$, the lower $Ra_C$ is at the onset.
Hence, the critical wave number $m^c$ severely drops, e.g. from $m^c = 20$ to $m^c = 1$ at $Ek=10^{-5}$.
This contradicts local stability analyses \citep[e.g.][]{sengupta2018effect}, which do not capture this puzzling double-diffusive behaviour.
Indeed, the existence of the double-diffusive tongue is due to the interplay between global rotation and the bounded geometry, as outlined by \citet{busse2002low}.
However, note that the onset of modes with large azimuthal wave numbers $m$ is almost unaffected by these effects, in agreement with the asymptotic limit of short-wavelength perturbations.

When $Ra_T$ is still further reduced, the critical $Ra_C^c$ increases again for all wave numbers. Ultimately, the stability curves for all azimuthal numbers $m$ collapse onto the asymptotic regime of non-rotating finger convection (\ref{eq:fingeringNRDDC}), i.e.
\begin{equation}
	Ra_C = -Ra_T \ \, \ \text{for} \ \, \ Ra_C, |Ra_T| \to \infty.
\end{equation}
However, we show in appendix \ref{appendix:BC} that limit (\ref{eq:fingeringNRDDC}) is not always valid in the sphere, depending on the thermal and compositional boundary conditions.

Because the edge of the tongue consists of a large-scale $m=1$ mode, we can expect being able to compute the onset with SINGE at the parameters of the Earth's core.
We remark that the tongue is stunningly invariant when plotted using inviscid dimensionless numbers, as shown in figure \ref{fig:inviscid}. 
We have also checked that $Pr$ and $Sc$ play only a role through the Lewis number $L$.
These two observations prove the inviscid nature of the low Rayleigh number double diffusive convection. To our knowledge, this behaviour has not been noticed by previous authors, although it can be inferred from the theory of \citet{busse2002low}, see appendix \ref{appendix:Busse}.

Furthermore, the tongue only weakly depends on the Lewis number when $L \gg 1$.
Hence, the black curve displayed in figure \ref{fig:inviscid}, computed at $Ek=10^{-15}$, fully characterises the convection onset within a stably stratified sphere, for any Ekman number $Ek \leq 10^{-4}$.
In particular, the lowest value of $Ra_C$ in this regime is given by $Ra_C \simeq 52 \, Ek^{-1}$ for $Ra_T \simeq -26 \, Ek^{-1}$.
Because the viscous convection onsets at $Ra_C \sim Ek^{-4/3}$ \citep{busse1970thermal,jones2000onset}, the Ekman number controls the transition between inviscid low-Rayleigh number convection and the standard viscous convection.
Thus, the domain of existence of inviscid convection increases as $Ek^{-1/3}$.

This behaviour supports the possibility of convection in planetary cores at low Rayleigh numbers (compared to the ones for pure compositional rotating convection).
However, unlike the suggestion of \citet{busse2002low} who mistakenly considered the non-rotating limit, the unstable Rayleigh numbers are not reduced down to non-rotating values, but rather to $Ra_C \simeq 52 \, Ek^{-1}$.
Note that the correct behaviour $Ra_C \sim Ek^{-1}$ is actually present in the annulus model (see appendix \ref{appendix:Busse}).

We also remark that these effects subsist with other boundary conditions, but the shape of the unstable tongue varies as shown in appendix \ref{appendix:BC}.
Interestingly, the asymptotic limit from local theory $Ra_T = -Ra_C$ is not always relevant (as pointed out above).
Finally, note that for the semi-convection quadrant ($Ra_C \leq 0, Ra_T \geq 0$ -- reported in appendix \ref{appendix:ODDC}), we find a similar behaviour with almost no effect of small stabilising compositional gradients.
However, for stably stratified fluids, the marginal curves $\sigma=0$ are significantly different, and should be studied in future work.

\subsection{Eigenmodes at the onset}
The rapid rotation does provide constraints on the velocity structure, not taken into account in local (unbounded) analyses. For instance in convective rotating spheres with the no-slip condition, flows approximately obey the Taylor-Proudman theorem \citep{greenspan1968theory}. This constraint yields quasi-geostrophic (QG) columnar motions, almost invariant along the rotation axis $\boldsymbol{1}_z$, as recovered numerically by SINGE \citep{kaplan2017subcritical}.
Then, we show in figure \ref{fig:flowonset} and \ref{fig:flowonset_m1} the spatial pattern of several eigenmodes at the onset of finger convection. They are representative of our linear numerical results, and do not depend much on the viscosity.

The eigenmode at the onset of almost pure compositional convection is shown in figure \ref{fig:flowonset}a. 
The flow is in the form of spiraling columnar rolls \citep{zhang1992spiralling}, extending spirally from near latitude $60^\circ$ to the equatorial region.
For this mode, the composition and temperature perturbations are phase-shifted by about $90^\circ$. 
Spiraling modes appear to be the preferred modes of convection for the moderate value $Sc=3$. 
However, in the limit $Sc\gg 1$, spiraling is expected to be small \citep{zhang1992spiralling,guervilly2010dynamos}.
In figure \ref{fig:flowonset}b, we show  a typical low-frequency mode ($m=60$) computed at $Ra_C = 10^9 = -Ra_T$.
For this mode, the composition and temperature perturbations are indistinguishable.
In that case, the critical Rayleigh number for all the modes are close, such that several modes are likely to be triggered in a slightly supercritical state.

Then, we show in figure \ref{fig:flowonset_m1} the $m=1$ mode at the onset within the double-diffusive tongue of figure \ref{fig:BasSINGE}.
At the tip of this tongue ($Ek\,Ra_C \simeq 52$ in figure \ref{fig:flowonset_m1}a) the mode is quite simple and spans the whole sphere and is almost stationary.
Remarkably, the composition and temperature perturbations are phase-shifted by about $45^\circ$.
The flow exhibits features reminiscent of quasi-geostrophy (columns aligned along the rotation axis).
For stronger forcing ($Ek\,Ra_C \gtrsim 1000$ in figures \ref{fig:flowonset_m1}b,c), the mode increases in complexity, with  several zeros in the direction parallel to the rotation axis.
There, it is no longer columnar and could not be captured by the quasi-geostrophic approach \citep{busse2002low,simitev2011double}.

\section{Nonlinear simulations of RDDC}
\label{sec:nonlinear}
\subsection{Nonlinear onset}
\begin{figure}              
  \centering
  \begin{tabular}{c}
  \includegraphics[width=0.7\textwidth]{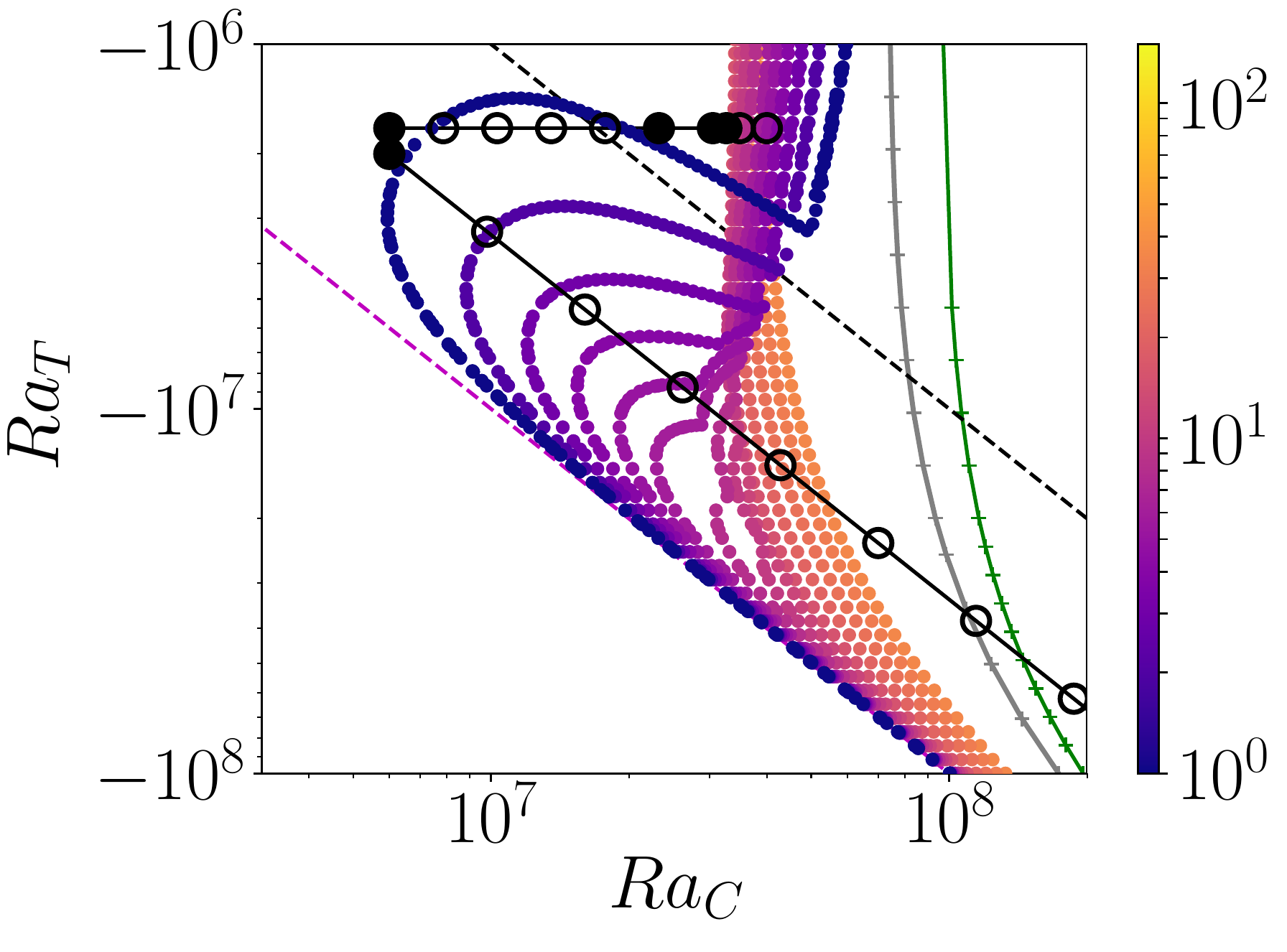} \\
  (a) \\
  \includegraphics[width=0.7\textwidth]{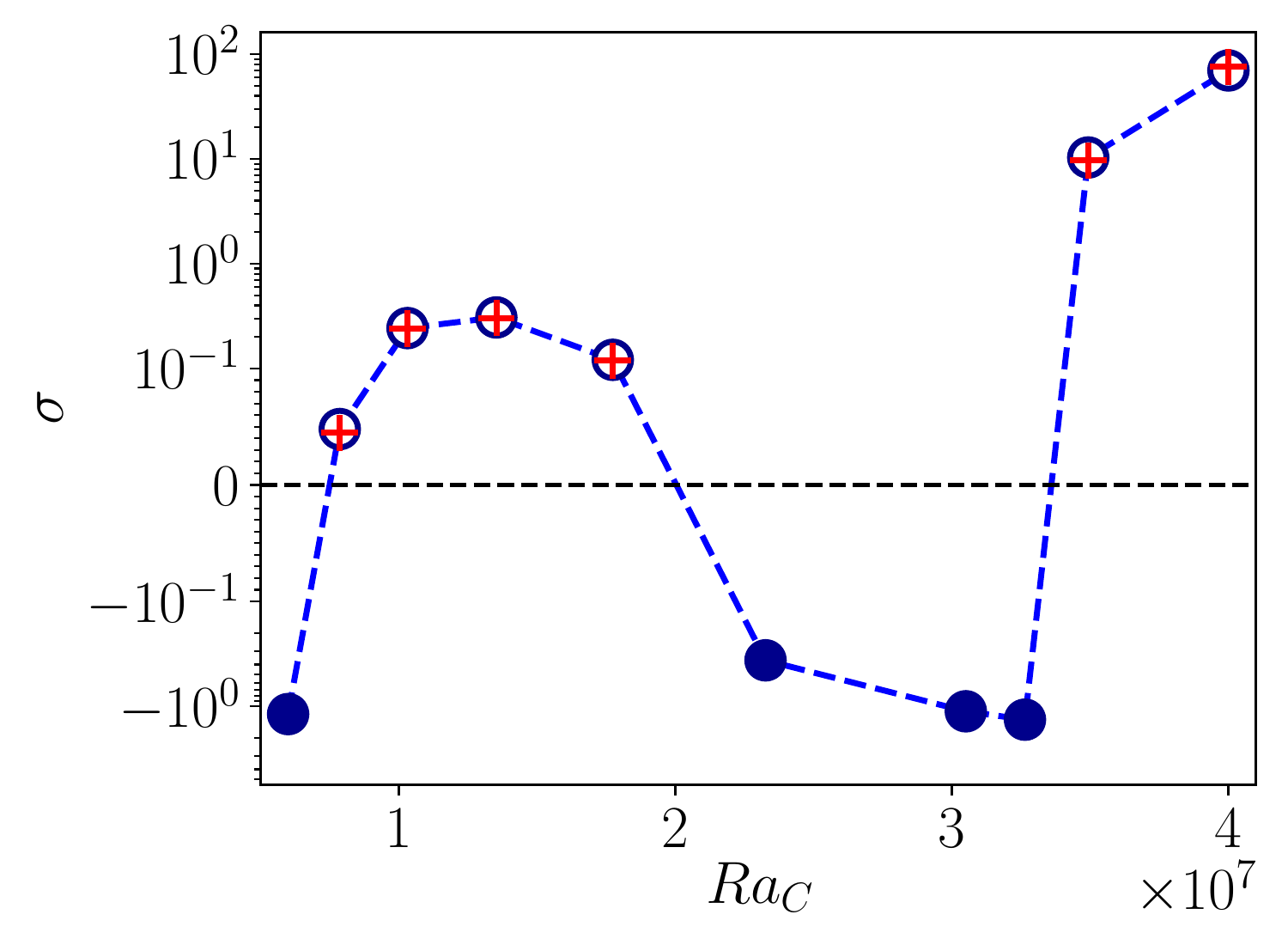} \\
  (b) \\
  \end{tabular}
  \caption{Onset of convection in the inviscid tongue at $Ek = 10^{-5}, Pr=0.3, Sc=3$. (a) Zoom in on figure \ref{fig:SINGE_Busse_bas}. The upper (black)  dashed line is the neutral curve $N^2_0=0$, i.e. $Ra_T = -Ra_C/L$. The lower (magenta) and dashed line is bound for non-rotating finger convection (\ref{eq:fingeringNRDDC}), here $Ra_T = -Ra_C$. Colour represents the azimuthal wave numbers $m$.
  Full black points: stable nonlinear simulations. Black circles: unstable nonlinear simulations. Green (respectively grey) solid lines with $+$ symbols are the onset for the anti-symmetric (respectively symmetric) $m=0$ mode.
  These nonlinear runs span the profile $Ra_T = -1.7 \times 10^6$ (horizontal line) and the profile $Ra_T = -Ra_C/3$, i.e. constant $R_0 = L/3$ (diagonal solid line).
  (b) Growth rate $\sigma$ along the profile $Ra_T = -1.7 \times 10^6$ shown in (a). Red crosses: computations with SINGE. Blue circles: computations with XSHELLS.
  The vertical scale is linear for $-0.1 \leq \sigma \leq 0.1$ and logarithmic for $|\sigma| > 0.1$.}
  \label{fig:BasSINGE}
\end{figure}  

As illustrated in figure \ref{fig:BasSINGE}a, the linear global analysis predicts the existence of alternating stabilising and destabilising double-diffusive effects when increasing $Ra_C$ for a fixed $Ra_T$ at the upper edge of the inviscid tongue.
We compare in figure \ref{fig:BasSINGE}b computations performed with SINGE and XSHELLS at $Ek=10^{-5}$, along the profile $Ra_T = -1.7 \times 10^6$ shown in figure \ref{fig:BasSINGE}a.
The growth rate computed with XSHELLS (during the exponential growth) is in perfect agreement with the eigenvalue computations.

Then, we aim at determining if this effect survives against finite-amplitude perturbations in nonlinear simulations. To do this, we have run the simulations sequentially for increasing value of $Ra_C$, and using the output of the previous simulation as initial state.
Starting from a linearly stable background state, increasing $Ra_C$ first destabilises the system, leading to RDDC within the unstable tongue.
Then, further increasing $Ra_C$ from a previous nonlinear state (at smaller $Ra_C$) abruptly inhibits the previously established RDDC when $Ra_C$ gets out of the tongue.
This is counter-intuitive as restabilisation occurs even though the compositional profile has \emph{a priori} a stronger destabilising gradient.
Finally, overturning convection sets up again in the system for larger values of $Ra_C > 3.4 \times 10^7$.
Similarly, we also find that the double-diffusive tongue subsists nonlinearly by varying $Ra_T$ at a fixed $Ra_C$ (not shown).

Thus, we have shown that this double-diffusive tongue is a linear mechanism, that persists against nonlinear perturbations of finite amplitude. We have found no evidence from the numerics that RDDC may onsets through a subcritical bifurcation, as recently obtained in pure thermal convection at much lower $Ek$ and $Pr$ \citep{kaplan2017subcritical}.

\subsection{Double-diffusive structures}
\label{subsec:surveyfingering}
In the following, we have conducted nonlinear simulations for stably stratified fluids along the profile $Ra_T = - Ra_C/3$ shown in figure \ref{fig:BasSINGE} as a diagonal line.
Along this profile, the density ratio (\ref{eq:criteriafinger}) is kept constant $R_0 = L/3$ but the background Brunt-V\"ais\"al\"a frequency increases according to formula (\ref{eq:N2/Ws2}). 
Note that we have also performed non-linear simulations in the semi-convection quadrant, as briefly discussed in appendix \ref{appendix:ODDC}.

\begin{figure}               
    \centering{
    \includegraphics[width=0.6\textwidth]{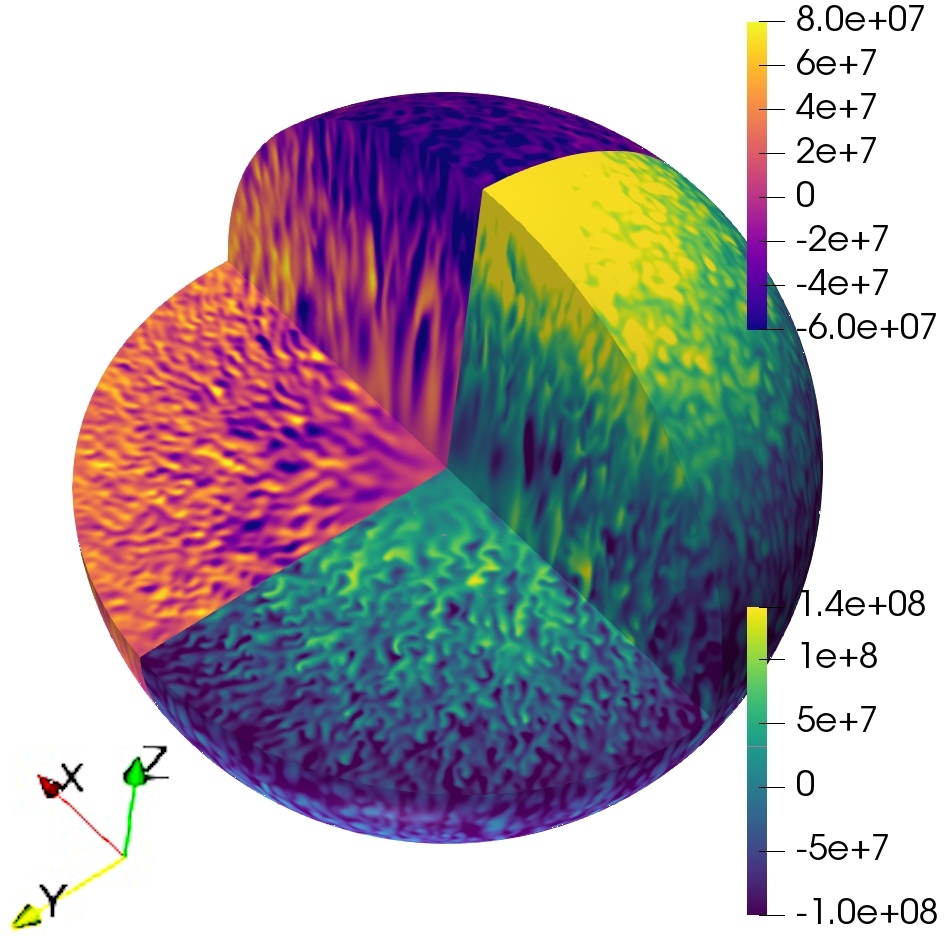} \\
    (a) \\
    \includegraphics[width=0.69\textwidth]{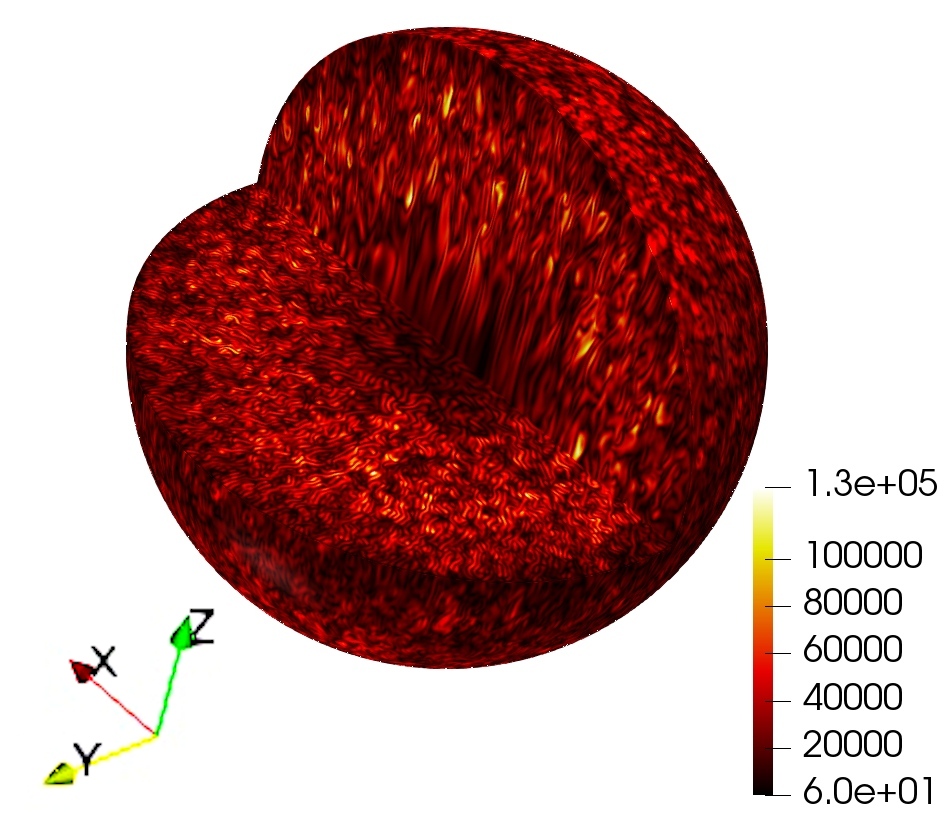} \\
    (b) \\
    }
    \caption{Nonlinear simulation of finger RDDC at $Pr=0.3, Sc=3, Ek=10^{-5}$, $Ra_C=10^{10}$ and $Ra_T=-Ra_C/3$ (before the saturation of the large-scale zonal flow). Rotation is along $\boldsymbol{1}_z$. (a) 3D snapshot of $Ra_T \Theta$ (upper colour bar corresponding to the left part) and compositional perturbation $Ra_C \xi$ (bottom colour bar corresponding to the right part). (b) 3D snapshot of the magnitude of the vorticity $|\boldsymbol{\nabla} \times \boldsymbol{u}|$.}
	\label{fig:fingers}                
\end{figure} 

Within the double-diffusive tongue, for typical compositional Rayleigh numbers $Ra_C \leq 10^8$, the nonlinear solutions are reminiscent of the eigenmodes at the linear onset (not shown). However, for higher Rayleigh number ($Ra_C \geq 10^8$), many high-wavenumber modes are unstable (see figure \ref{fig:BasSINGE}), leading to extremely thin convection fingers, elongated in the direction of the rotation axis due to the rapid rotation (figure \ref{fig:fingers}).

In non-rotating systems, finger DDC leads to spatial scales intrinsically governed by the fast (thermal) diffusion and viscosity \citep[e.g.][]{radko2013double}. Recently, \citet{bouffard2017double} proposed another empiric scaling law in the presence of rotation.
These two scaling laws predict the typical length of density structures in the equatorial plane $l_\perp$.
They read respectively in the non-rotating and rotating regimes (with our variables) 
\begin{subequations}
\label{eq:radkobouffard}
\begin{align}
  l_\perp &\propto |Ra_T|^{-1/4} \propto Ra_C^{-1/4}, \label{eq:radko}\\
  l_\perp &\propto \left ( Ek \, |Ra_T| \right )^{-1/2} \propto \left ( Ek \, Ra_C \right )^{-1/2},
    \label{eq:bouffard}
\end{align}
\end{subequations}
in which the rightmost forms involving $Ra_C$ are only valid for profiles characterised by $Ra_T \propto Ra_C$.
Note that scalings (\ref{eq:radko})-(\ref{eq:bouffard}) are expected for large enough values of the Lewis number.
In addition, the typical horizontal size of the fingers is reasonably well approximated by prediction (\ref{eq:radko}) in the non-rotating case, even for moderate values of $L$.
Indeed, relation (\ref{eq:radko}) holds for local computations at $L=3$ \citep[see figure 7a of][]{traxler2011dynamics}.

\begin{figure}
	\centering
    \includegraphics[width=0.7\textwidth]{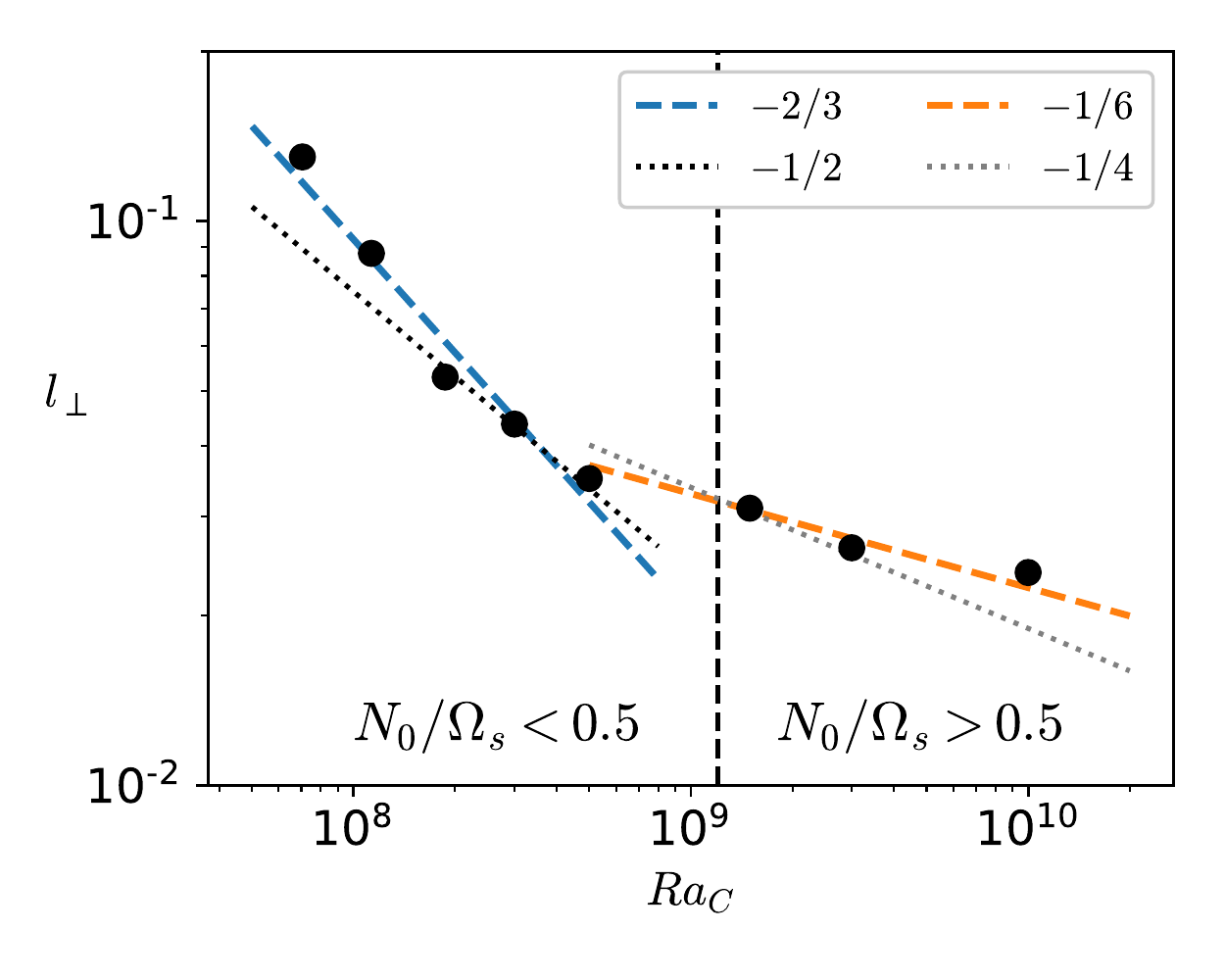}
    \caption{Typical dimensionless wavelength $l_\perp$ of fingers in the equatorial plane. Nonlinear simulations at $Pr=0.3,Sc=3, Ek=10^{-5}$ and $Ra_T=-Ra_C/3$.
    The following scalings are represented: $l_\perp = 20\,000 \: Ra_C^{-2/3}$ (blue dashed, left); $l_\perp = 750 \: Ra_C^{-1/2}$ (black dotted, left, eq. \ref{eq:bouffard}); $l_\perp = 1.04\: Ra_C^{-1/6}$ (orange dashed, right); $l_\perp = 6\: Ra_C^{-1/4}$ (gray dotted, right, eq. \ref{eq:radko}).
    The vertical dashed line separates heuristically the two regimes ($N_0/\Omega_s \lesssim 0.5$) of finger convection in the simulations, as determined from figure \ref{fig:ReRo} (see below).}
    \label{fig:Radko}
\end{figure}

We assess their relevance for RDDC against 3D simulations performed at the finite value of $L=10$ in figure \ref{fig:Radko}. We have determined the approximate number of fingers in the equatorial plane to estimate $l_\perp$.
We observe two regimes, with a transition between $Ra_C \simeq 5 \times 10^8$ and $Ra_C \simeq 1.5 \times 10^9$.
Our measurements do not seem to be in obvious agreement with the previous scaling laws, but the decrease of $l_\perp$ with increasing $Ra_C$ slows down at the transition, as predicted.
The transition occurs for the Brunt-V\"ais\"al\"a frequency $N_0/\Omega_s \simeq 0.5$, and will be seen in several other diagnostics in the following (see below).
We did not test the dependence of $l_\perp$ with the Ekman number $E$, which is predicted by eq. \ref{eq:bouffard}. This would require to reduce the Ekman number, and run several high-resolution simulations at the edge of what is feasible.

For the simulations in the strongly stratified regime at $Ra_C \geq 10^9$, we may look for density staircases \citep{stern1969salt}.
The latter are made of stacks of well-mixed convective layers, separated by stably stratified shells for the total density profile \citep[e.g.][]{stellmach2011dynamics}.
However, we have not found any evidence of density staircases in our simulations. 
In the non-rotating regime, \citet{brown2013chemical} found that local simulations performed at low values of the reduced density ratio $\widetilde{R}_0 \ll 0.01$ exhibit properties consistent with density layering, with
\begin{equation}
	\widetilde{R}_0 = \frac{R_0-1}{L-1}.
\end{equation}
The finger regime is mapped into $0\leq \widetilde{R}_0 \leq 1$. We have $\widetilde{R}_0 \sim 0.2$ in our simulations, such that the absence of staircases is expected even for non-rotating fluids. Thus, performing more turbulent simulations, at lower values of $\widetilde{R}_0 \ll 1$, appears necessary to investigate the interplay between rotational effects and density staircases.

\subsection{Turbulence and transport}
\begin{figure}               
    \centering
	\includegraphics[width=0.7\textwidth]{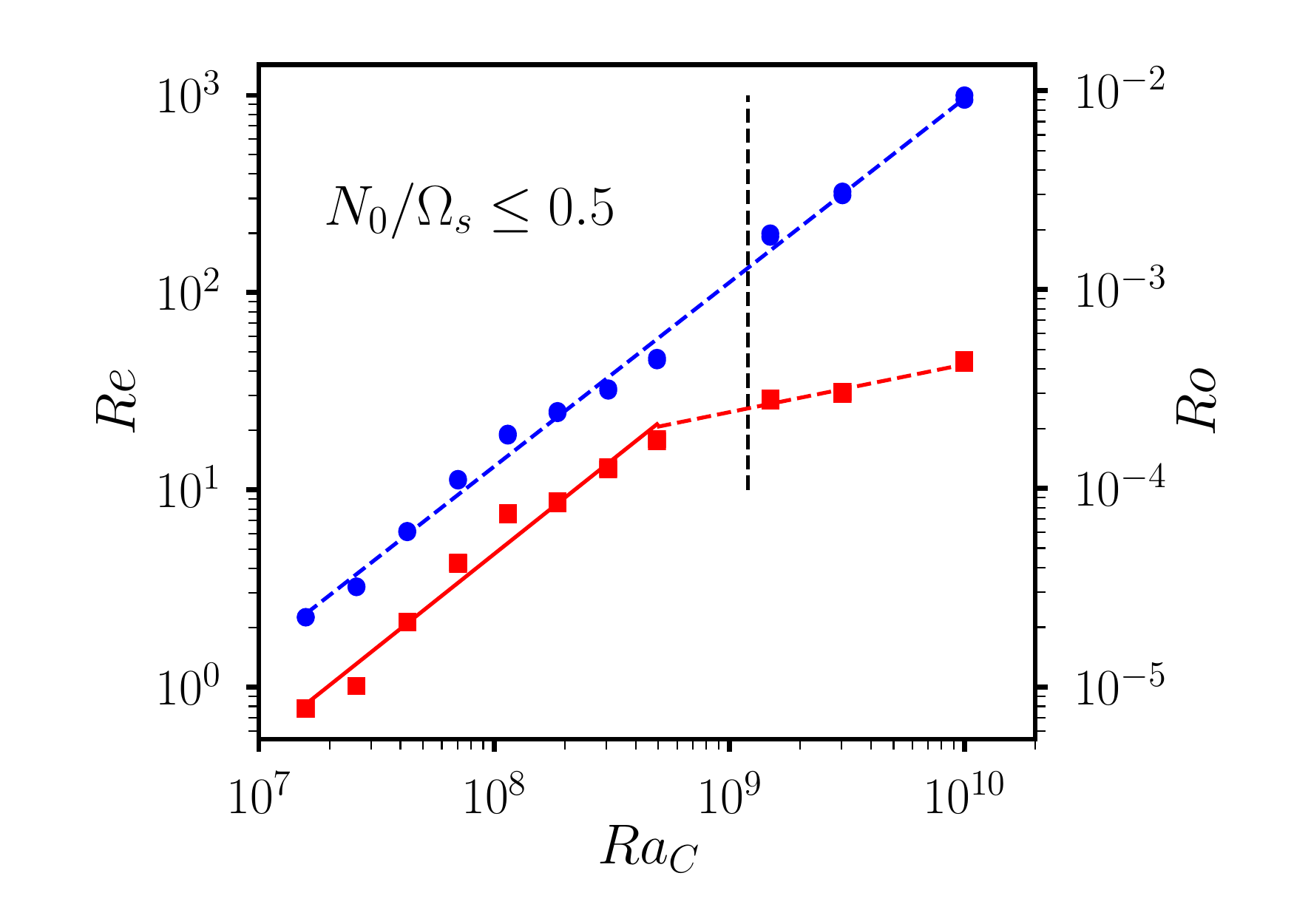} \\
	(a) \\
	\includegraphics[width=0.7\textwidth]{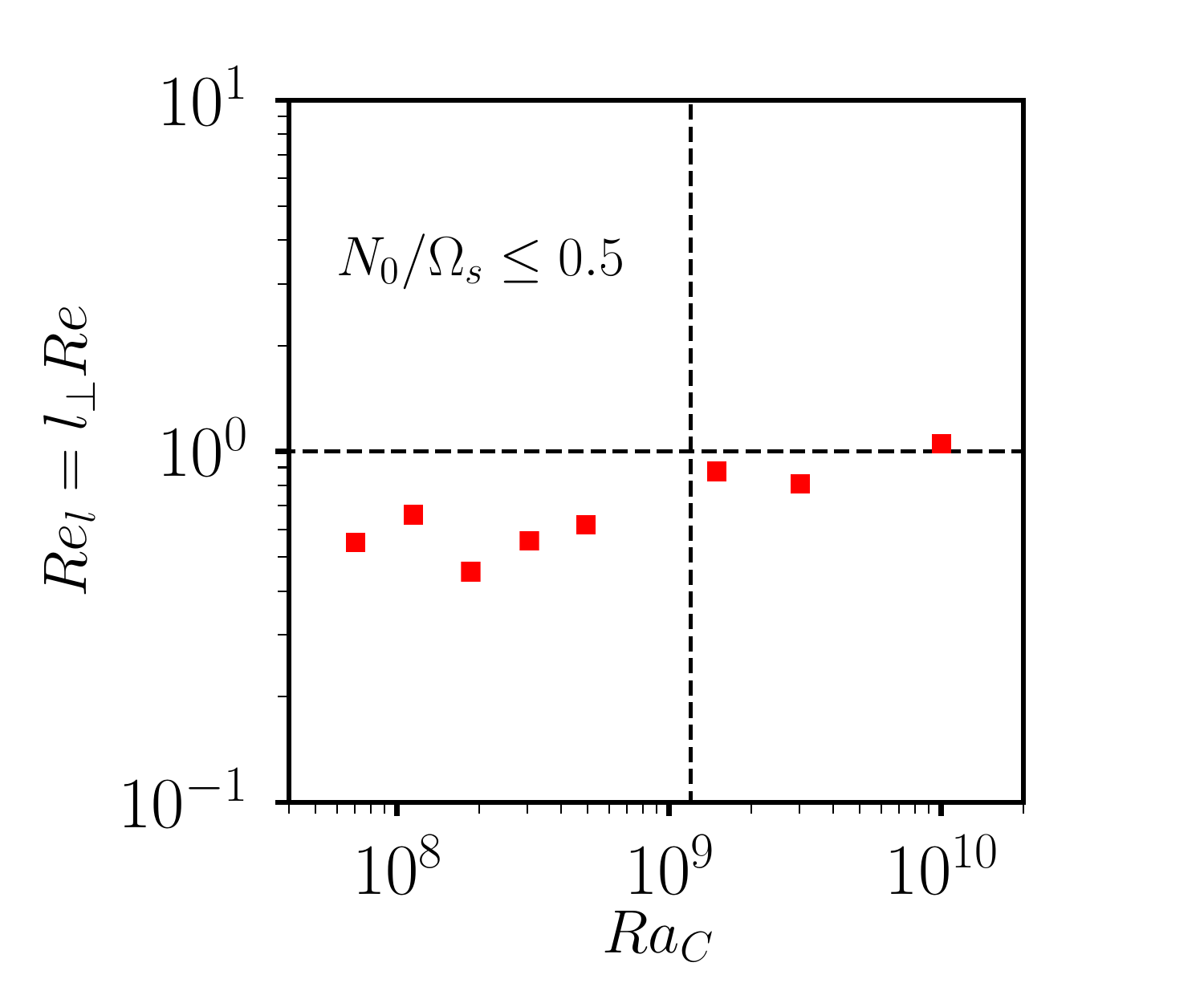} \\
	(b) \\
    \caption{(a) Reynolds and Rossby numbers $[Re, Ro = Re \, Ek]$ and (b) local Reynolds number $Re_l = l_\perp \, Re$ (with $l_\perp$ given in figure \ref{fig:Radko}), as a function of $Ra_C$ for nonlinear simulations along the diagonal profile shown in figure \ref{fig:BasSINGE}, i.e. for a constant buoyancy number $R_0$. Computations at $Pr=0.3, Sc=3, Ek={10}^{-5}$ and $Ra_T=-Ra_C/3$, for which the double-diffusive inviscid tongue exists for $Ra_C \geq 6.37 \times 10^6$. Blue circles: rms velocity based on total energy. Red squares: rms velocity based on non-zonal, poloidal energy (proxy of the radial velocity). In (a), the tilted lines are the best least-square fit to the data, yielding the scalings $Re \propto Ra_C^{0.93}$ (blue dashed),  $Re \propto Ra_C^{0.95}$ (red thin for $N_0/\Omega_s \lesssim 0.5$) and $Re \propto Ra_C^{0.24}$ (red thin for $N_0/\Omega_s \gtrsim 0.5$).  }
	\label{fig:ReRo}                
\end{figure} 

We now focus on specific features of finger convection in the turbulent regime. To quantify the nonlinear outcome, we compute in figure \ref{fig:ReRo} the root mean square (rms) Reynolds and Rossby numbers
\begin{equation}
	Re = \sqrt{\frac{2}{V} E_u}, \ \, \ Ro = Re \, Ek,
\end{equation}
with $V = 4\pi/3$ the dimensionless spherical volume and $E_u$ the kinetic energy defined by formula (\ref{eq:KETECE}). We have used the time average of $E_u$ in the saturated regime to determine the rms velocity. 
We have also separated $Re$ and $Ro$ based on total and non-zonal poloidal energies, to illustrate several regimes of finger convection. 

First, when $N_0/\Omega_s \lesssim 0.5$, the Reynolds numbers based on total and radial velocities both exhibit the same scaling $Re \propto Ra_C^{0.93}$. However, when $N_0/\Omega_s \gtrsim 0.5$, another regime appears. Although $Re$ based on the total velocity is still nearly proportional to $Ra_C$, the scaling of $Re$ based on the poloidal energy is suddenly altered for $Ra_C \geq 10^9$, yielding $Re \propto Ra_C^{0.24}$.
Hence, for strong stratification, radial (poloidal) motions are inhibited, while toroidal ones are not. 
This behaviour is consistent with scaling arguments and simulations of sustained stratified turbulence \citep{billant2001self,brethouwer2007scaling}. Indeed, a transition is expected between two turbulent regimes, characterised by strong and weak radial (here poloidal) motions. Such a dichotomy has been also evidenced in pioneering global simulations of tidally driven stratified flows \citep{vidal2018magnetic}.

We can compare our results with the unbounded RDDC recently studied by \citet{sengupta2018effect}.
They find the local Reynolds number $Re_{l}$, based on the convective velocity (analog to our non-zonal poloidal energy), to scale as
\begin{equation}
    Re_{l} \propto [Pr\, (R_0 - 1)]^{-1/2}.
    \label{eq:scalingReGaraud}
\end{equation}
In our case, this formula gives a constant value of $\sim 1$, in apparent contrast with the evolution of $Re$ shown in figure \ref{fig:ReRo}a.
However, $Re$ is based on the (global) radius of our sphere, which is not relevant to estimate a local Reynolds number. Using rather the local length $l_{\perp}$, we estimate $Re_{l}= l_{\perp} \, Re$ in our simulations.
As shown in figure \ref{fig:ReRo}b, we then obtain a constant $Re_{l} \simeq 1$ in agreement with formula (\ref{eq:scalingReGaraud}). In both regimes $N_0/\Omega_s \leq 0.5$ and $N_0/\Omega_s \geq 0.5$, we thus recover the behaviour observed in unbounded RDDC.
This can be understood with the following physical argument.
Finger convection works because during the motion of a fluid particle, the temperature can be exchanged with its surroundings.
Hence, the thermal diffusion time-scale $l_\perp^2/\kappa_T$ must not be smaller than the advection time-scale $l_\perp/u$.
This leads to the condition $u\, l_\perp /\kappa_T \lesssim 1$, that is the Péclet number is of order one.
This also translates into $Re_l \lesssim 1/Pr$, which is consistent with our findings (fig. \ref{fig:ReRo}b, $Re_l \simeq 1$ independent of $Ra_C$).
Note that, because we have set $Pr=0.3$, the two predictions cannot be distinguished.

\begin{figure}
	\centering
    \includegraphics[width=0.6\textwidth]{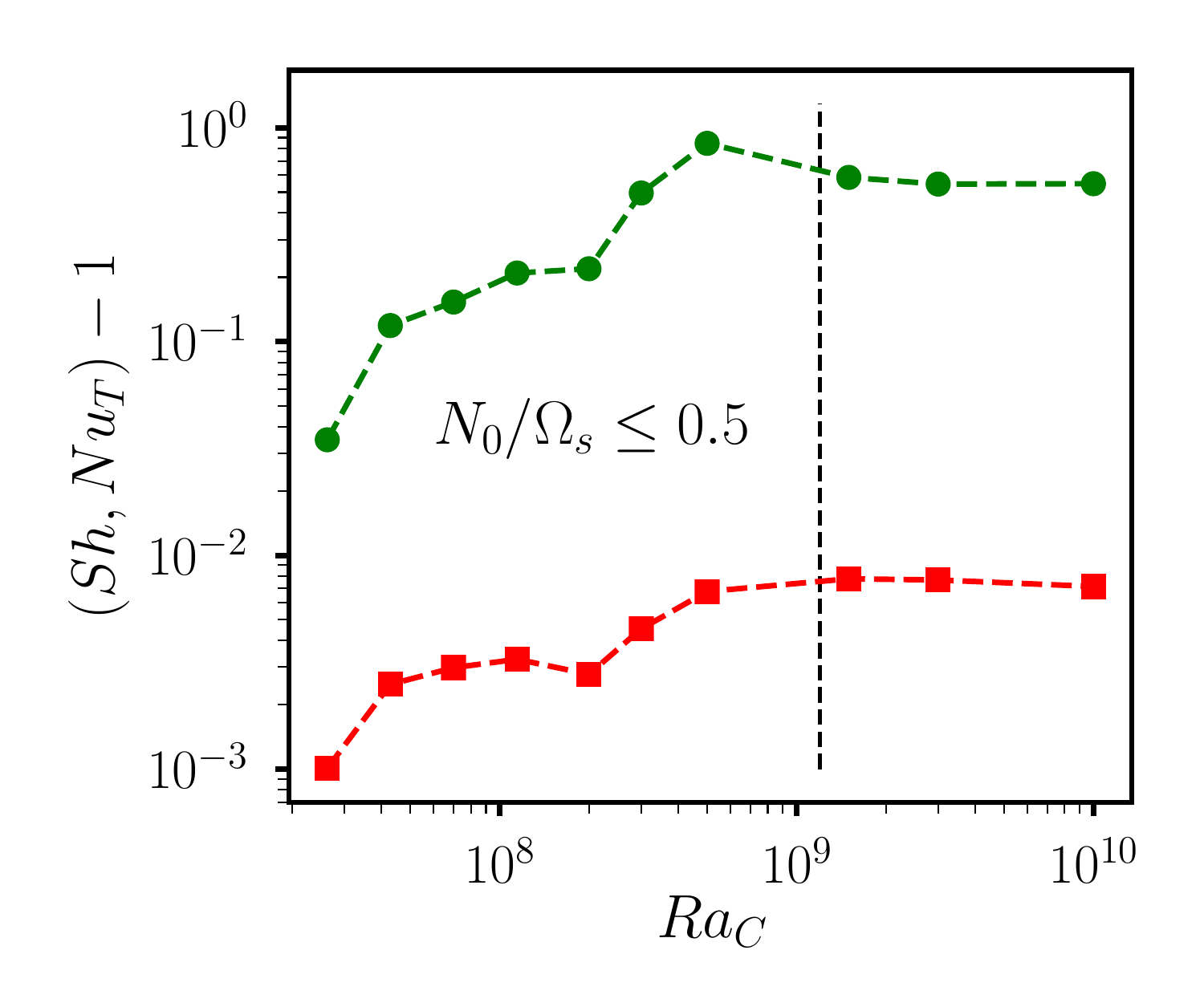} \\
    \caption{Turbulent thermal (red squares) and compositional (green circles)  Nusselt numbers $[Nu_T, Sh]$, defined by formulas (\ref{eq:nusselt}), as a function of the compositional Rayleigh number $Ra_C$ for rotating finger convection. Nonlinear simulations at $Pr=0.3, Sc=3, Ek=10^{-5}$ and $Ra_T=-Ra_C/3$ (i.e. constant $R_0 = L/3$). The vertical dashed line separates the two rotating regimes in the simulations.}
    \label{fig:Nusselt}
\end{figure}

We now turn to the efficiency of convective transport of temperature and composition, which are quantified by the Nusselt $Nu_T$ and Sherwood $Sh$ numbers respectively.
Their value is 1 for pure diffusion, and increase with increasing convection strength.
In a convective sphere with internal sources and fixed flux at the outer boundary, they are given by
\begin{subequations}
\label{eq:nusselt}
\begin{align}
	Nu_T &= \frac{T_0(0) - T_0(1)}{T_0(0)-T_0(1) + \Theta_{rms}(0) - \Theta_{rms}(1)}, \\
    Sh &= \frac{C_0(0)-C_0(1)}{C_0(0)-C_0(1) + \xi_{rms}(0) - \xi_{rms}(1)},
\end{align}
\end{subequations}
with $[T_0, C_0](r)$ the dimensionless background profiles (\ref{eq:prof-tc}) and the rms values of temperature and compositional perturbations $[\Theta_{rms},\xi_{rms}](r)$, defined from thermal and compositional energies at the radius $r$.
In figure \ref{fig:Nusselt}, we observe that $Nu_T$ is only weakly affected by varying $Ra_C$, yielding $Nu_T-1 \leq 10^{-2}$. This is in agreement with local models of non-rotating finger convection. Indeed, \citet{brown2013chemical} showed that $Nu_T$ is always low and drops to 1 as $L$ (or $R_0$) is increased.
This shows that the significant thermal diffusion necessary for finger patterns to develop always dominates the heat transport.

The compositional Nusselt number exhibits more significant variations. When $Ra_C$ increases in the regime $N_0 /\Omega_s \lesssim 0.5$ defined above, $Sh$ increases up to $Sh \sim 2$.
Thus, the turbulent compositional flux is enhanced, for a fixed $R_0$ and an increasing strength of the background stratification along the profile.
Then, in the second regime ($N_0 /\Omega_s \gtrsim 0.5$), increasing further $Ra_C$ does not yield significant changes in $Sh$. 

Note that the scaling of the Nusselt and Sherwood numbers in figure \ref{fig:Nusselt}, are in agreement with the laws of non-rotating finger convection \citep[e.g.][]{garaud2018double}.
Indeed, they predict constant Nusselt and Sherwood numbers for constant buoyancy ratio $R_0$ and Prandtl number $Pr$.

By contrast, the regime $N_0/\Omega_s \lesssim 0.5$ is more puzzling.
One one hand, no clear scaling was observed for $Sh$, $Nu_T$ or $l_\perp$, but on the other hand the $Re \sim Ra_C$ scaling found in this regime
has also been put forward in rotating thermal convection \citep{guervilly2019}, also obtained at low $Pr$ and low $Nu$.

\begin{figure}
	\centering
    \includegraphics[width=0.6\textwidth]{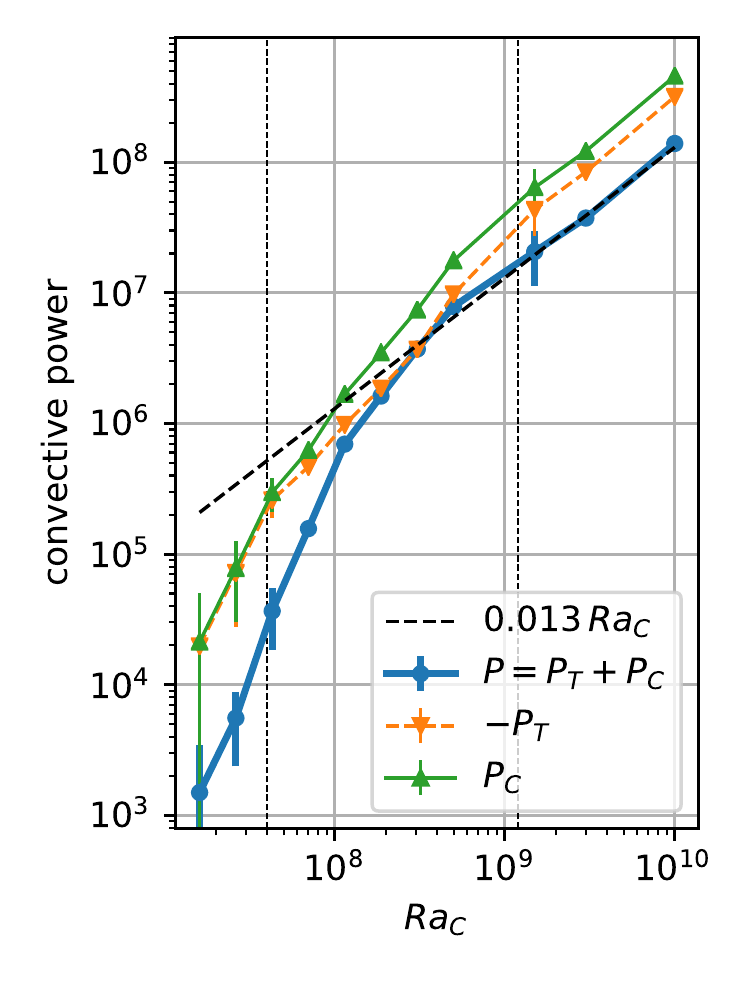} \\
    \caption{Convective power as a function of chemical Rayleigh number $Ra_C$, for a stably stratified setup at $Pr=0.3, Sc=3, Ek=10^{-5}$ and $Ra_T=-Ra_C/3$ (i.e. constant $R_0 = L/3$).
    The thermal and solutal convective powers, respectively $P_T$ and $P_C$, measure the work done by the thermal and compositional buoyancy forces respectively.
    Here, $P_T < 0$ opposes fluid motion, while $P_C > 0$ drives the flow.
    For large $Ra_C$, the net convective power $P=P_T+P_C$ scales as $P \simeq 0.013 Ra_C$.
    Error bars indicate the temporal fluctuations of the quantities.
    The vertical dashed lines mark the limit of the anomalous inviscid tongue ($Ra_C \simeq 4 \times 10^7$) and the limit $N_0/\Omega_s = 0.5$ ($Ra_C \simeq 1.2 \times 10^9$).}
    \label{fig:convpower}
\end{figure}

Figure \ref{fig:convpower} shows the evolution of the convective power for the same set of simulations, as a function of $Ra_C = -3 Ra_T$.
The convective power -- the work done by buoyancy forces -- is the sum of the thermal buoyancy power $P_T = \alpha_T (T_0^*+\Theta) \boldsymbol{g} \boldsymbol{\cdot} \boldsymbol{u}$ and the solutal buoyancy power $P_C = \alpha_C (C_0^*+\xi) \boldsymbol{g} \boldsymbol{\cdot} \boldsymbol{u}$.
Three regimes may be distinguished here. Within the anomalous inviscid tongue ($Ra_C \lesssim 4 \times 10^{-7}$, see Fig. \ref{fig:BasSINGE}a), the compositional buoyancy almost balances the thermal buoyancy. Only a small amount of net power drives convection.
For larger $Ra_C$, the solutal buoyancy overcomes more easily the thermal stabilizing gradient, leading to more efficient convection.
This coincides with the identification of small-scales fingers (see Fig. \ref{fig:Radko}).
Interestingly, for the largest forcings, we find that the net convective power $P$ evolves like $P \simeq 0.013 Ra_C$.
A scaling of $P$ proportional to $Ra_C$ is also reported in standard convection and convective dynamos \citep[e.g.][]{christensen2006scaling}, but with a proportionality constant close to one.

\subsection{Zonal flows}

For strong forcings, two different symmetries of zonal flow emerge which are discussed below.

\begin{figure}                
  \centering
  \includegraphics[width=0.8\textwidth]{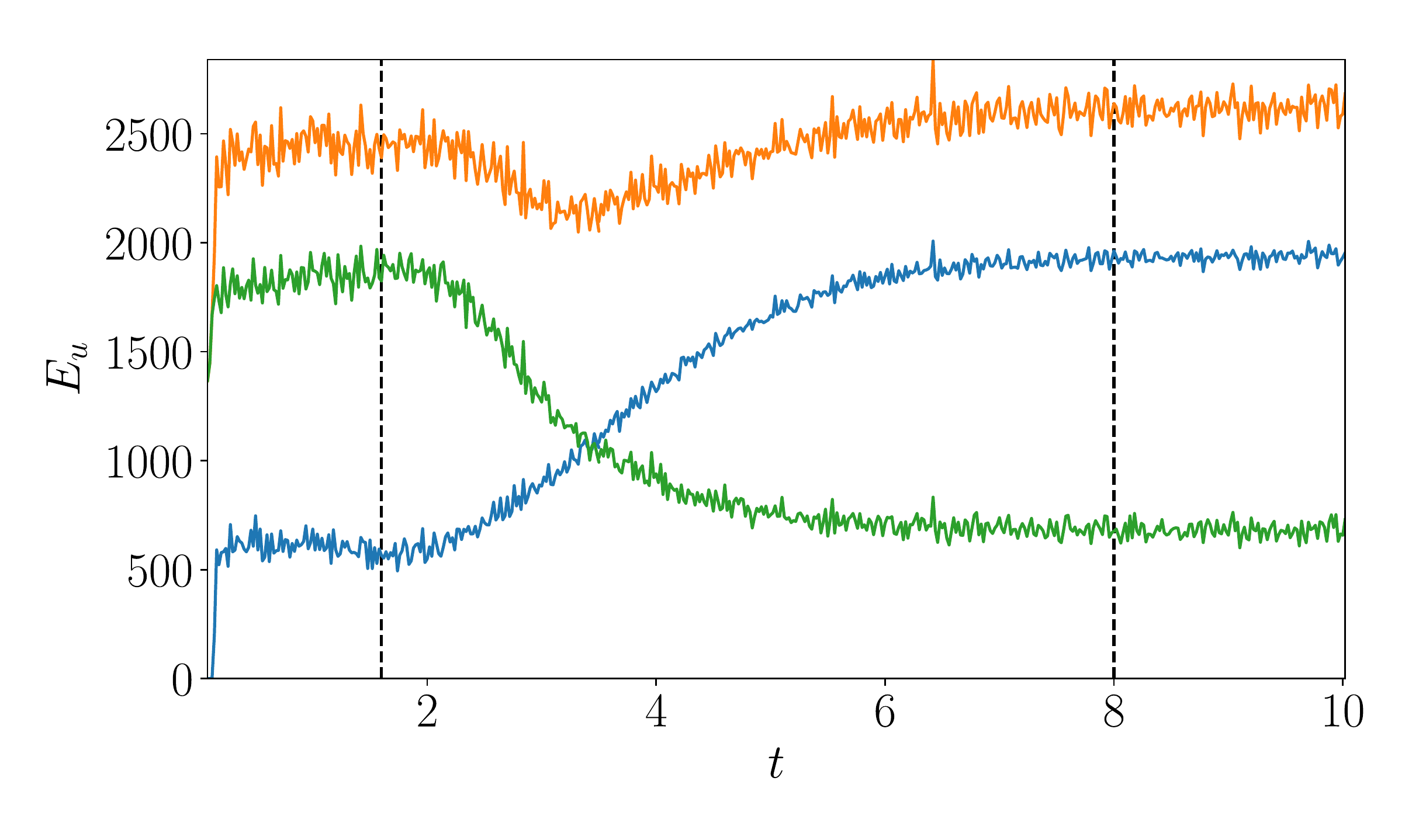} \\
  (a) \\[2em]
  \begin{tabular}{cc}
  \includegraphics[width=0.4\textwidth]{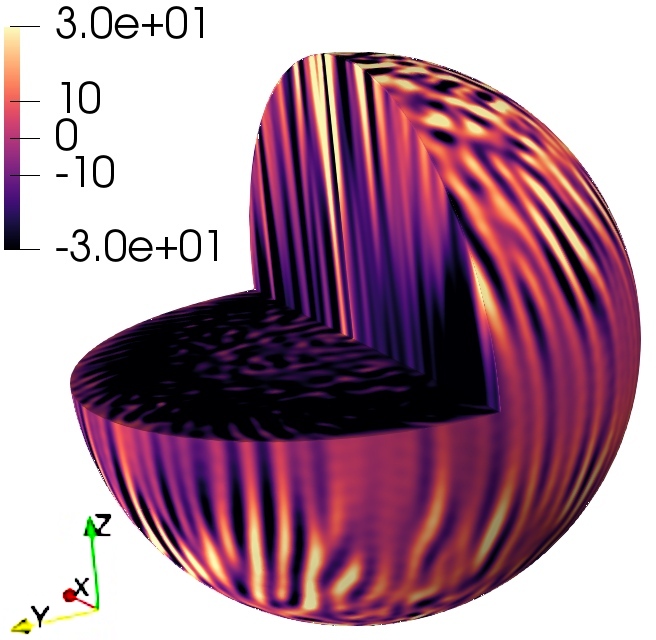} &
  \includegraphics[width=0.4\textwidth]{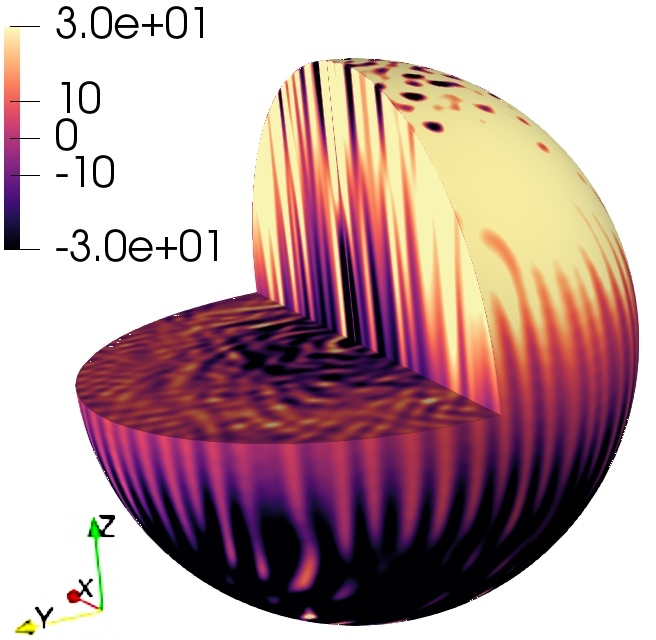} \\
  (b) $t=1.6$ &
  (c) $t=8$
  \end{tabular}
  \caption{Nonlinear simulation of finger RDDC for stably stratified fluids at $Pr=0.3, Sc=3, Ek=10^{-5}$ and $Ra_C= 3 \times 10^8$ and $Ra_T = -Ra_C/3$.
  (a) Equatorially anti-symmetric (blue, i.e. the lowest one at $t=1$), symmetric (green, i.e. the lowest one at $t=8$) and total (orange, i.e. the uppermost one at any time) kinetic energies  $E_u$ as a function of the dimensionless time $t$.
  (b) \& (c) Dimensionless azimuthal velocity $u_\phi$ at the outer boundary ($r=1$), in a meridional slice and in the equatorial plane ($z=0$) at the times shown by the dashed vertical lines in (a) ($t=1.6$ and $t=8$). The rotation axis is along $\boldsymbol{1}_z$.}
  \label{fig:snapnrj}                
\end{figure}

\begin{figure}
    \centering
    \includegraphics[width=0.6\textwidth]{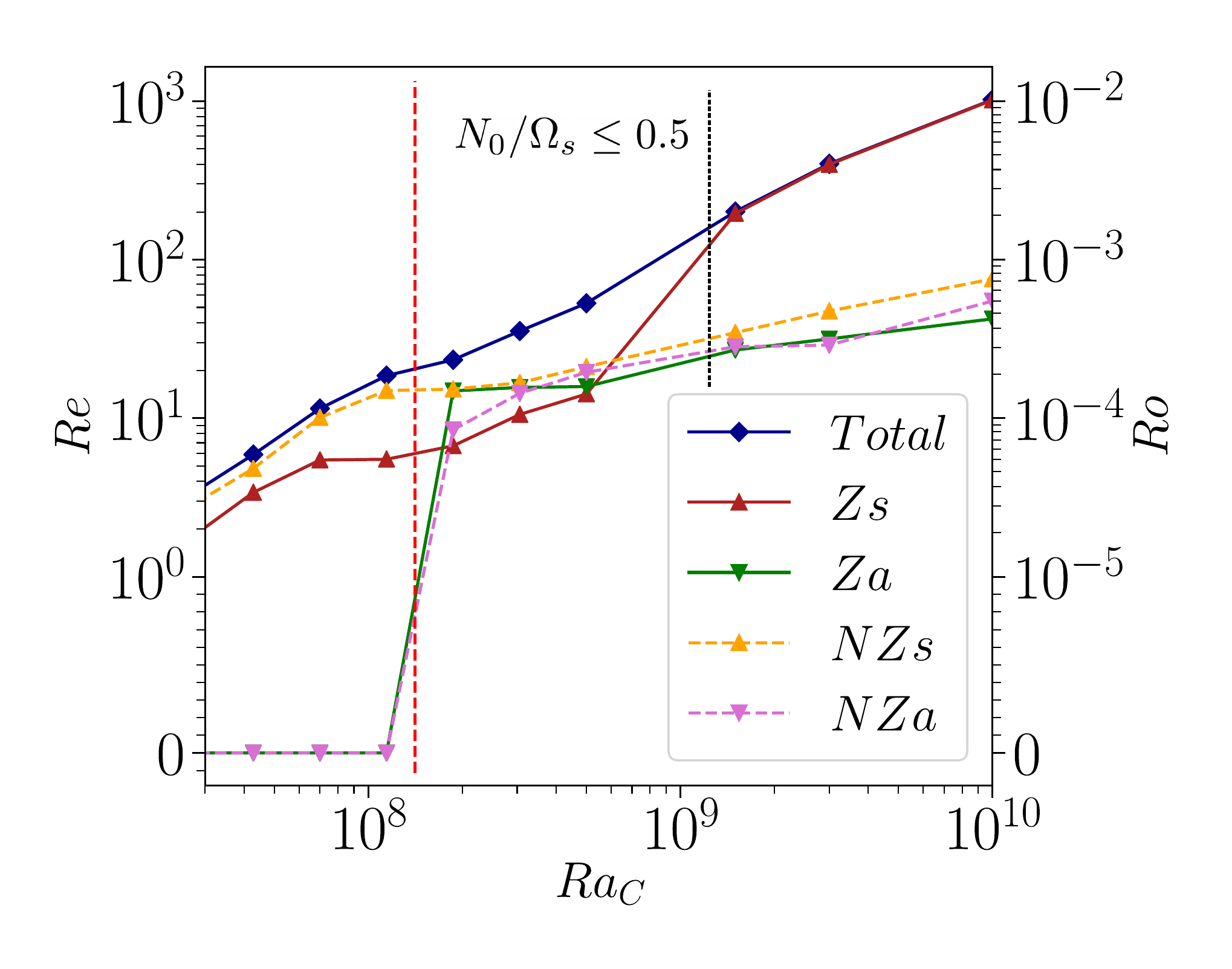}
    \caption{Reynolds and Rossby numbers as a function of $Ra_C$ for nonlinear simulations along the diagonal profile $Ra_T=-Ra_C/3$ (shown in figure \ref{fig:BasSINGE}) at $Pr=0.3, Sc=3, Ek={10}^{-5}$. $Zs$ and $Za$ are respectively the zonal equatorially symmetric and anti-symmetric components. $NZs$ and $NZa$ are their non-zonal analogues. The vertical red dashed line is the anti-symmetric linear onset for the mode $m=0$ while the black one at $Ra_C \simeq 1.2 \cdot 10^9$ corresponds to $N_0/\Omega_s \simeq 0.5$. As expected, the non-zonal symmetric perturbation dominates near the RDDC onset.} 
    \label{fig:NRJ_diag}
\end{figure}
\subsubsection{Equatorial anti-symmetry for moderate forcing}
Within the regime $N_0/\Omega_s \lesssim 0.5$, an equatorially anti-symmetric, differential rotation emerges from saturated quasi-geostrophic motions.
A typical temporal evolution is summarised by figure \ref{fig:snapnrj}a. Initially, the saturated nonlinear flow is dominated by quasi-geostrophic vortices (see figure \ref{fig:snapnrj}b). They are associated with density fingers (thicker than the ones illustrated in figure \ref{fig:fingers}, obtained at larger $Ra_C$). Columnar motions are predominant as long as $t\leq2$ in the simulation (figure \ref{fig:snapnrj}a).
Then, an anti-symmetric flow grows on few viscous time units (figure \ref{fig:snapnrj}a).
Meanwhile, the energy of the equatorially symmetric flow is significantly reduced, such that the total energy of the fluid remains roughly constant.
This anti-symmetric flow is mainly toroidal, consisting of a strong differential rotation: the flow is prograde in the Northern hemisphere and retrograde in the the Southern hemisphere (see figure \ref{fig:snapnrj}c), and is associated with a segregation of both compositional and temperature anomalies in one hemisphere (not shown).
This is the first report of such flows in finger convection.
Actually, as shown in figure \ref{fig:NRJ_diag}, they appear just above the linear onset for the equatorially anti-symmetric and axisymmetric (EAA) mode.
By contrast, \citet{landeau2011} found the appearance of the EAA mode in purely thermal convection much further above its onset.
In our case, the EAA flow appearing in the nonlinear regime is clearly linked to the crossing of the associated linear threshold. It is quite unexpected that, far from the instability onset, a purely linear mechanism can explain the symmetry breaking of a nonlinear flow at saturation.
This highlights the potential importance of linear modes even far from the global stability threshold in this systems. Furthermore, it emphasizes the importance of long simulations spanning several diffusion times.

\subsubsection{Equatorial symmetry for strong forcing}
\begin{figure}                
	\centering
    \begin{tabular}{c}
	    \includegraphics[width=0.55\textwidth]{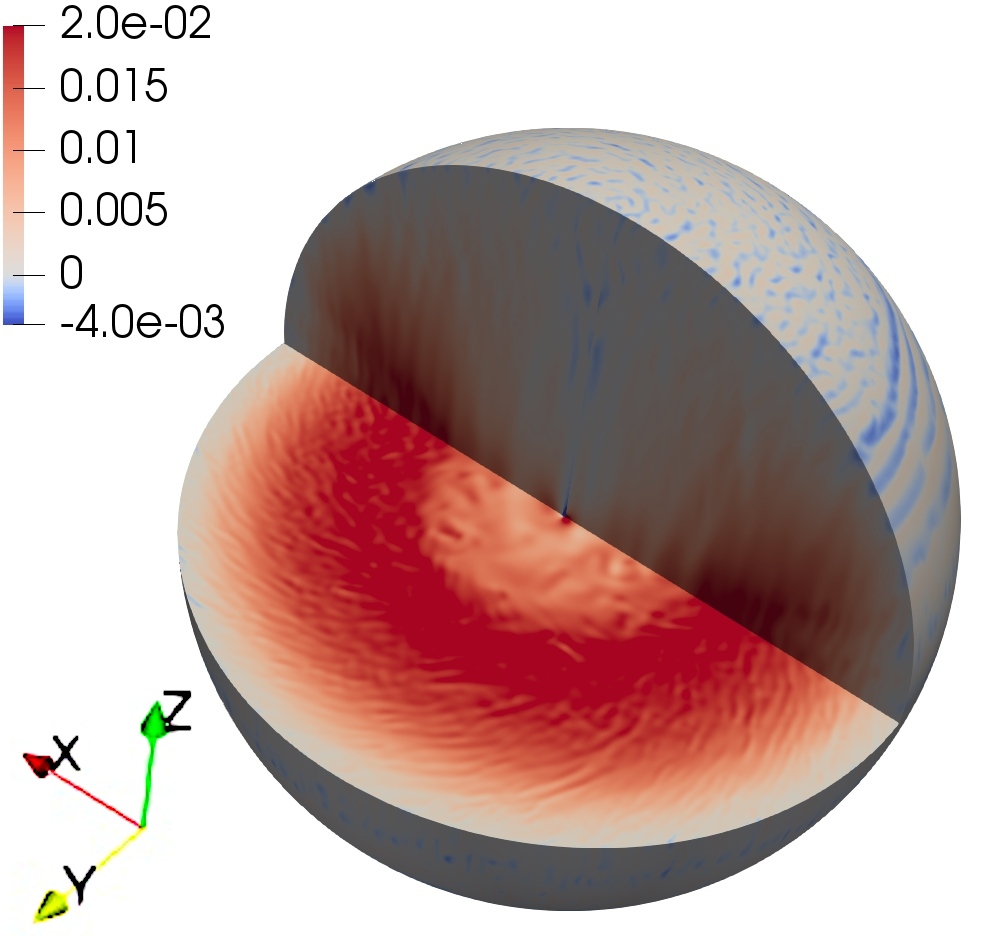} \\
        (a) \\
		\includegraphics[width=0.7\textwidth]{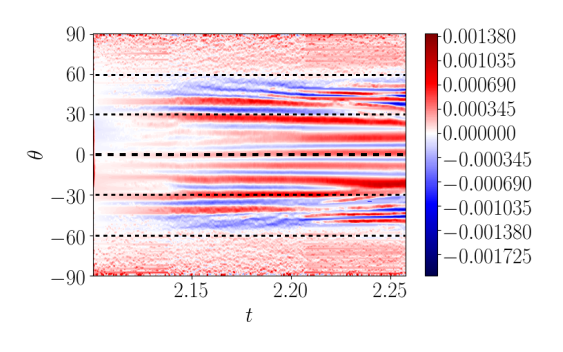} \\
        (b) \\
    \end{tabular}
    \caption{Equatorially symmetric zonal jet in rotating finger convection.
    Azimuthal average of the dimensionless rotation of the fluid $\Omega = Ek \, u_\phi/r$. Simulations at $Pr=0.3, Sc=3, Ek=10^{-5}$, $Ra_C=10^{10}$ and $Ra_T=-Ra_C/3$. (a) Instantaneous 3D snapshot of $\Omega$ up to radius $r=0.995$. The rotation axis is along $\boldsymbol{1}_z$. (b) $\Omega$ as a function of time $t$ and colatitude $\theta$ (in degrees) at the radius $r=0.995$, below  the Ekman boundary layer.}
    \label{fig:Uphizon}                
\end{figure} 

Figure \ref{fig:NRJ_diag} shows that the EAA is overtaken by equatorially symmetric zonal flows for $N_0/\Omega_s \gtrsim 0.5$.
This contrasts with \citet{landeau2011}, where the EAA mode increasingly dominates with forcing.
Being mainly toroidal, the zonal flow does not affect the Reynolds number based on poloidal non-zonal energy shown in figure \ref{fig:ReRo}.
The more $Ra_C$ increases, the larger the amplitude of this zonal flow, which quickly dominates all other components, as shown in figure \ref{fig:NRJ_diag}.
A typical kinetic energy spectrum is shown as a function of $m$ in figure \ref{fig:spectre}b. The zonal $m=0$ component has an amplitude up to several orders of magnitude larger than the non-zonal components.

This zonal flow has a strong radial dependence, as illustrated in figure \ref{fig:Uphizon}a. In the bulk (here at $r=0.5$), the zonal flow is prograde. However, it naturally exhibits multiple alternating prograde and retrograde jets at the outer spherical boundary, with rich dynamics (figure \ref{fig:Uphizon}b).

These zonal flows may be seen as the manifestation in spheres of large-scale vortices (LSV) found in (unbounded) local simulations. LSV are conspicuous in local simulations of rotating finger convection in the polar regions \citep{sengupta2018effect}. They also appear in rotating semi-convection \citep{moll2017double} and in rotating pure-thermal convection \citep[e.g.][]{guervilly2014large,guervilly2017jets}. \citet{julien2018impact} argued that the formation of zonal flows and jets is a robust feature resulting from an inverse energy cascade, provided that the flow is strongly anisotropic.
In our simulations it is the zonal flows that allow rapid velocities ($Re \sim 1000$) to be reached by convection in stably stratified fluids.

\section{Towards planetary core conditions}
\label{sec:applications}
\subsection{Linear onset in the early Earth}
\begin{table}
	\centering
      \begin{tabular}{cccc}
      	\hline
        & \multicolumn{2}{c}{\textsc{Thermal}} & \textsc{Compositional} \\ [1ex] 
     	$Pr$ & $0.01$ & $0.1$ &  \\
        $Sc$ &  &  & $10^2$ \\
        $Ra_T^c$ & $6.00\times 10^{19}$ & $1.42\times 10^{20}$ & $0$ \\
        $Ra_C^c$ & $0$ & $0$ & $1.08 \times 10^{21}$ \\
        $m^c$ & $10755$ & $22301$ & $67570$ \\
      	\hline
      \end{tabular}
    \caption{Parameters for the onset of convection at core conditions ($Ek=10^{-15}$): critical thermal $Ra_T^c$ or compositional $Ra_C^c$ Rayleigh numbers, critical wave number $m^c$ computed from table 2 of \citet{jones2000onset}. Note that our dimensionless numbers $Ek$ and $Ra$ differ from theirs.}
    \label{table:jones2000}
\end{table}

\begin{figure}
	\centering
    \begin{tabular}{c}
    	\includegraphics[width=0.65\textwidth]{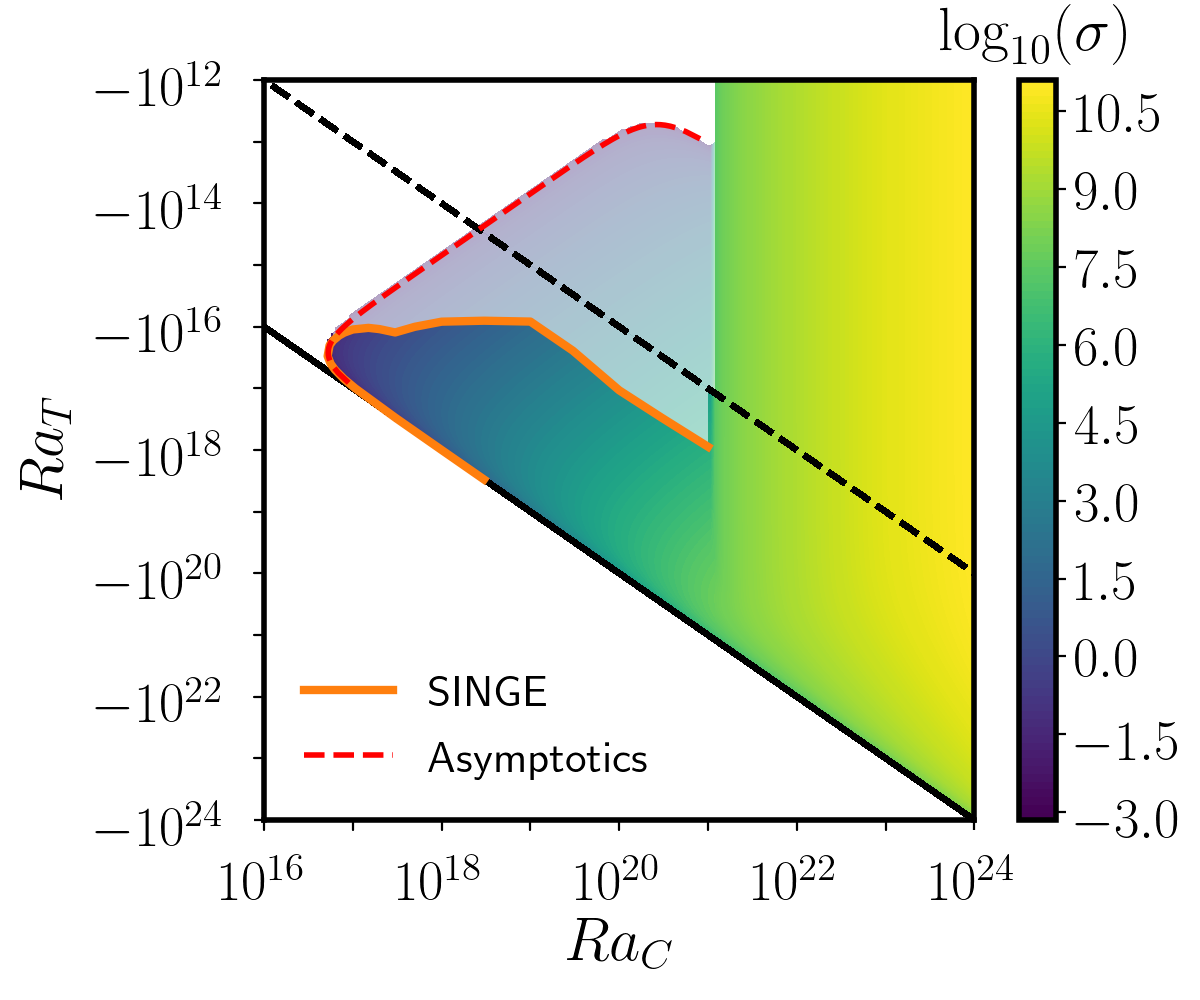} \\
        (a) Growth rate \\
        \includegraphics[width=0.65\textwidth]{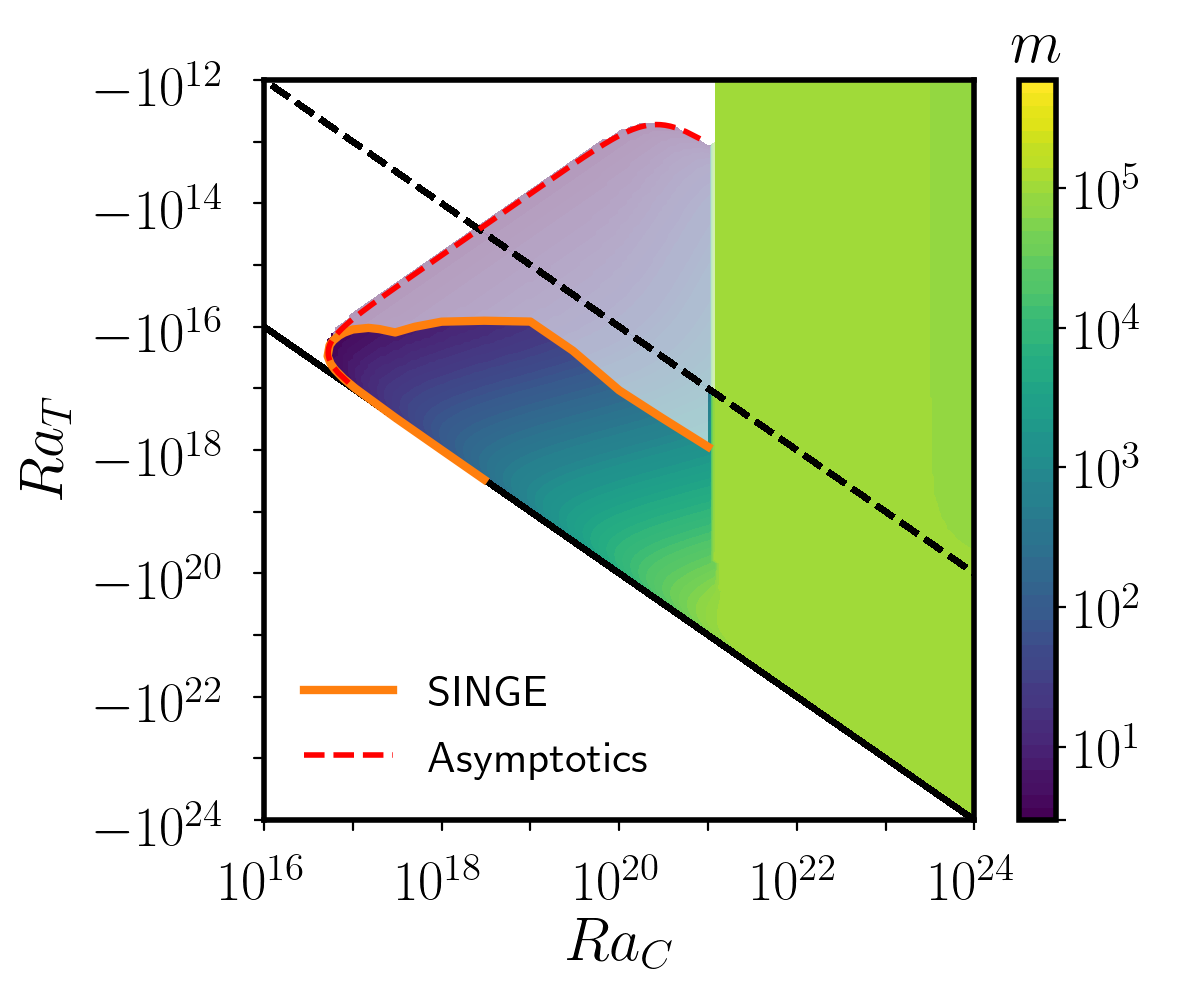} \\
        (b) Most unstable azimuthal wave number \\
    \end{tabular}
    \caption{Linear onset of RDDC for Early Earth core conditions, i.e. $Ek=10^{-15}$, $L=10^4$ in the finger quadrant. Orange thick line is the onset for $m=1$, computed with SINGE for these parameters (and setting $Pr=3.10^{-4}$ for numerical convenience, the results being independent of $Pr$, see \S\ref{sec:inviscid}). Colour maps are obtained from the approximate annulus theory of \citet{busse2002low}, adjusted to fit the SINGE data, for the dimensionless growth rate (a) and the most unstable wave number (b). In this model, the RDDC tongue is given by formula (\ref{eq:parametric}), shown as red dashed lines. Dashed oblique line is the neutral curve $N_0^2 = 0$, i.e. $Ra_T = -Ra_C/L$. By contrast with the annulus theory, faded colored zones are stable according to SINGE (see appendix \ref{appendix:Busse} for details).}
    \label{fig:BusseEarlyEarth}
\end{figure}

We have highlighted, using linear and nonlinear simulations, that rotation has surprising effects for rotating finger convection. Now, a complete quantitative picture of the onset of rotating finger convection is emerging for typical core conditions ($Ek=10^{-15}$ and $L=10^4$). 
Indeed, we remind the reader that double-diffusive effect are negligible except for $|Ra_T| \sim |Ra_C|$. Hence, we gather in table \ref{table:jones2000} the parameters of pure thermal or compositional convection, as predicted by the global theory of \citet{jones2000onset}.
Moreover, thanks to the inviscid nature of the instability in the stably stratified region ($N_0^2>0$), we have already determined the onset of finger convection at Earth's core conditions (see figure \ref{fig:inviscid}).

The scenario is illustrated in figure \ref{fig:BusseEarlyEarth}. 
For core conditions, we clearly observe the possibility of convection at reduced Rayleigh number, immensely facilitated by the stably stratified thermal profile.
First, the wave number at the onset is strongly reduced within the tongue, yielding typical values $m \leq 10$.
The growth rate increases with $Ra_C$ from a few to several thousands per viscous time-scale within the tongue.

Then, as expected, the critical Rayleigh number at the onset of pure compositional convection $Ra_C^c \sim 10^{21}$ (table \ref{table:jones2000}) is orders of magnitude larger than the critical value at the upper edge of the double-diffusive tongue, i.e. $Ra_C^c \simeq 10^{17}$.
The composition Rayleigh number $Ra_C$ is thus reduced by four decades for the early Earth by adding a stabilising temperature gradient.

\subsection{Speculative estimates for the early Earth}
To investigate the relevance of RDDC in the early Earth, we can use orders of magnitude arguments. They are presently highly speculative, due e.g. to the large modeling uncertainties. They will be certainly revised by future additional constraints, provided by mineral physics and thermal models of the Earth. 

A typical estimate of the compositional Rayleigh number is 
\begin{equation}
	\frac{Ra_C}{Sc} \sim \frac{g_0 R^4}{\nu^2} \frac{\Delta \rho_C}{\rho_m},
    \label{eq:RaCearlyEarth}
\end{equation}
where $\Delta \rho_C$ is the typical density yielding the compositional buoyancy (due to light elements)  and $\rho_m$ the typical density of the core. 
Following \citet{jones2015tog}, typical values are $R = 3500$ km for the radius of the core, $Rg_0 = 10 \, \text{m.s}^{-2}$ for the gravitational acceleration, $\nu=10^{-6}$ m$^2$.s$^{-1}$for the (molecular) kinematic viscosity and $Sc=10^2$ for the Schmidt number. 
The amount of light elements attributable to compositional sources is highly debated. The compositional gradient was likely destabilising in the early Earth, due to the exsolution or crystallisation of light elements in the core. 
The equilibration at high temperatures in the aftermath of giant impacts would be responsible for a small amount of magnesium to partition into the core, yielding the exsolution of light magnesium oxides in the core \citep{badro2016early,o2016powering}. 
This mechanism is energetically efficient, since precipitating a layer of magnesium-bearing material with a typical thickness of 10 km above the CMB would be equivalent to crystallising the entire inner core \citep{o2016powering}. 
Instead of invoking such singular events, \citet{hirose2017crystallization} advocated the crystallization of silicon dioxide. 
Nonetheless, in the two scenarios, roughly the same mass of light elements is precipitated/crystallised. 
Thus, typical (speculative) upper bounds are $0.2$ wt \% of precipitated magnesium-bearing minerals \citep[see figure 2 of][]{o2016powering} or $0.4$ wt \% of crystallised silicon dioxide \citep{hirose2017crystallization}. Based on these two scenarios, we may consider the typical value $\Delta \rho_C / \rho_m = 10^{-3}$ as an upper bound. Then, formula (\ref{eq:RaCearlyEarth}) yields the estimate $Ra_C/Sc \leq 10^{29}$ in the early Earth. 
This upper estimate is much larger than the critical values required at the onset (figure \ref{fig:BusseEarlyEarth}), typically $Ra_C^c \sim 10^{17} - 10^{18}$ in the inviscid tongue and $Ra_C^c \sim 10^{21}$ for compositional convection (even without stabilising thermal effects).
This suggest that the early Earth did undergo highly supercritical RDDC, either for unstably or stably stratified fluids.

The properties of convection would be certainly different in the two regimes. Hence, to argue in favour of one regime, we have to estimate the square of the total background Brunt-V\"ais\"al\"a frequency $N_0^2$. On the one hand, the compositional part is
\begin{equation}
{N_{0,C}^2}/{\Omega_s^2} \sim - Ek^2 \, {Ra_C}/{Sc}.
\end{equation}
This gives the speculative estimate $-10^{-1} \leq {N_{0,C}^2}/{\Omega_s^2} \leq 0$. 
On the other hand, the presence of a thick, thermally stratified layer seems probable prior to the formation of the inner core if there was a sub-adiabatic heat flux at the top of the core $Q_{\text{cmb}}$. 
To our knowledge, there is no reliable agreement between thermal models of the Earth \citep[e.g.][]{labrosse1997cooling,nimmo2015tog,nakagawa2018thermo}.
Therefore, we estimate a speculative upper bound from the difference between the total heat flux $Q_{\text{cmb}}$ and the adiabatic flux $Q_a$ at the CMB 
\begin{equation}
	\frac{N_{0,T}^2}{\Omega_s^2} \sim \frac{\alpha_T g_0 R}{k \mathcal{S} \Omega_s^2} \left [Q_a - Q_{\text{cmb}} \right],
\end{equation}
with $\mathcal{S}$ the surface of the outer core, $\alpha_T \sim 10^{-5}$ K$^{-1}$ the thermal expansion coefficient \citep{nimmo2015tog} and $k$ the thermal conductivity. The latter quantity is badly constrained \citep{williams2018thermal}, so does the thermal history of the Earth. Possible values are $40-160$ W.m$^{-1}$.K$^{-1}$. 
A broad range of values appears possible for $Q_a - Q_{\text{cmb}}$. Upper-bound estimates are presently a few TW, yielding the (highly) speculative bounds for a thermal stratification $0 \leq N_{0,T}^2 /\Omega_s^2 \ll 1-10$. The upper bound values are very close to the plausible geophysical estimates of the thermal Brunt-V\"ais\"al\"a frequency at the top of the Earth in the present time \citep[e.g.][]{labrosse1997cooling,buffett2014geomagnetic,helffrich2013causes}. Consequently, $N_{0,T}^2 + N_{0,C}^2$ may have been either positive or negative. 

To sum up, the Early Earth may have been prone to either overturning convection (for unstably stratified fluids) or finger RDDC (for stratified fluids). 

\section{Conclusion}
\label{sec:ccl}
\subsection{Summary}
We have revisited rotating double-diffusive convection (RDDC) in planetary cores, by considering flows driven by buoyancy forces of thermal and compositional origins. We have studied RDDC with a Boussinesq model in a full sphere, with internal source and sink terms. We have separated thermal and compositional effects, to go beyond the codensity approach  \citep{braginsky1995equations,lister1995strength} commonly used in planetary simulations. 
We have mainly focused on the finger regime ($Ra_C \geq 0, Ra_T < 0$), by considering stabilising thermal effects and destabilising compositional effects.

First, we have performed the linear stability analysis of background diffusive state (\ref{eq:T0C0sol}) in the finger quadrant, by using a global (spherical) method.
A global picture is now emerging. 
A quantitative proxy of the strength of rotational and stratified effects is the absolute value of the square of the dimensionless background Brunt-V\"ais\"al\"a frequency, i.e. the ratio $|N_0^2/\Omega_s^2|$. Overturning convection occurs for unstably stratified fluids ($N_0^2/\Omega_s^2 \leq 0$). When overturning convection is controlled by rotational effects ($-1 \leq N_0^2/\Omega_s^2 \leq 0$), the onset is largely unaffected by double-diffusive effects when $|Ra_T| \ll Ra_C$ in the finger regime. 
Then, the linear spherical analysis recovers asymptotically the onset of non-rotating DDC in strongly stratified regime ($N_0^2/\Omega_s^2 \gg 1$). On the other hand, it strongly differs in the other regime $N_0^2/\Omega_s^2 \ll 1$ with local analyses. Indeed, local analyses predict that rotation has a simple stabilising effect, merely increasing the critical Rayleigh numbers at the onset. 
However, rotational effects are more subtle in the presence of double diffusion. 
Indeed, the global analysis shows that the linear onset of RDDC can occur for lower Rayleigh numbers for stably stratified fluids than for unstably stratified fluids. 
This phenomenon, first outlined by \citet{busse2002low}, is intrinsically due to rotational effects in the bounded spherical geometry. Therefore, they are filtered out by local models. 
The associated flows at the linear onset do not always take the form of quasi-geostrophic motions (aligned with the rotation axis), unlike in standard rotating convection \citep[e.g.][]{zhang2007asymptotic,kaplan2017subcritical}. 
In addition, for a specific combination of boundary conditions (namely fixed temperature and imposed composition flux), rotating double-diffusive convection surprisingly occurs for density ratios $R_0 > L$, which is beyond the limit of non-rotating double-diffusive convection.
In the finger regime, double-diffusive effects become preponderant only for stably stratified fluids ($N_0^2/\Omega_s^2 \geq 0$). On the contrary, as discussed in appendix \ref{appendix:ODDC}, double-diffusive effects start playing a role even for unstably stratified fluids ($N_0^2/\Omega_s^2 \leq 0$) in the semi-convection quadrant ($Ra_C \leq 0, Ra_T > 0$).

Second, we have conducted high-resolution, nonlinear simulations for rotating stratified fluids ($N_0^2/\Omega_s^2 \geq 0$) in the finger regime. Several nonlinear features have been obtained. 
Outside the DD tongue for large enough $Ra_C$, the flow structures (fingers) strongly differ from the linearly unstable tongue modes at the upper edge of the DD tongue. Moreover, we have identified a sharp transition outside the tongue in the rapidly rotating finger regime. This transition empirically occurs at $N_0/\Omega_s \simeq 0.5$ in the simulations, for the fixed value $Ek=10^{-5}$. In the first regime, the nonlinear flows exhibit equatorially anti-symmetric, large-scale zonal flows, which appears when the associated linear onset is crossed. In the second regime, strong equatorially symmetric zonal flows are sustained. 
The latter flows are reminiscent of the large-scale vortices found in local models of finger convection \citep[e.g.][]{sengupta2018effect}. The turbulent properties, e.g. the output Reynolds or Nusselt numbers, are also significantly different in the two regimes. Notably, we have found scalings for the second regime that appear in broad agreement with the scalings proposed for local DDC. 

Finally, we have succeeded in predicting the onset of RDDC numerically at core conditions, after noticing the inviscid nature of finger convection in the weakly stratified regime.
We have shown that the combination of rotation and double-diffusive effects is strongly destabilising in the inviscid tongue for stably stratified fluids. The critical Rayleigh number is reduced by four decades for realistic core conditions.
Then, we have crudely estimated the thermal and compositional stratification in the Early Earth. We support that it may have undergone highly turbulent RDDC, either in the overturning compositional convection (unstably stratified) or in the finger regime associated with strong zonal flows.

\subsection{Perspectives}
\subsubsection{Discussion and improvements}
A considerable amount of work remains to be done, e.g. to expand the surveyed parameter space and to refine the model. 
Further simulations are required to understand the nonlinear saturation of finger convection (figure \ref{fig:ReRo}), e.g. by varying $L$, $Ra_T$ and $Ra_C$. On the one hand, we have found that local scalings of non-rotating finger convection \citep{garaud2018double} may qualitatively hold in the second rotating regime. Nonetheless, a more exhaustive numerical survey of the parameter space is required to assess their quantitative validity. Moreover, it remains an open question whether regimes of rotating thermal convection \citep{gastine2016scaling} apply for RDDC, both for destabilising and stabilising density profiles. 
Therefore, this calls for assessing and possibly improving the scaling laws describing rotating convection in the presence of significant double-diffusive effects.

For numerical reasons, we have considered moderate values for the Lewis $L=10$ and Ekman $Ek=10^{-5}$ numbers in the nonlinear simulations. The value of $L$ is about two orders of magnitude smaller than the expected values in planetary cores. 
Larger values of $L$ may facilitate the generation double-diffusive structures. In particular, we have not found any density staircases \citep{stern1969salt}, resulting from secondary instabilities. Several theories have been proposed in the non-rotating case \citep{stern1969salt,radko2013double}. For the moderate values of $Pr$ characterising planetary cores, their generation may rely on the mixing by nonlinear internal waves \citep{garaud2015excitation}. Yet, these mechanisms remain to be confirmed in the presence of rapid rotation. 
Their existence may strongly affect the turbulent regime. Indeed, it has been shown that density staircases can increase the turbulent heat and compositional fluxes by several orders of magnitude \citep[e.g. in oceanography][]{schmitt2005enhanced}. Thus, the conditions of existence for density staircases in rotating finger convection remain unanswered and studying them deserves future work.

We have outlined that we cannot rule out RDDC in the Early Earth. Now, investigating the dynamo capability is necessary to assess the validity of the proposed mechanisms for the origin of the Early geodynamo \citep{badro2016early,o2016powering,o2017thermal,hirose2017crystallization}.
The dynamo capability of rotating finger convection remains an open question.
Typically, dynamo action requires $Rm > 100$, where $Rm = Re Pm$ is the magnetic Reynolds number with $Pm = \nu/\eta$ the magnetic Prandtl number ($Pm \ll 1$ for cores) and $\eta$ the magnetic diffusivity.
With this first study, we cannot establish scaling laws that would allow us to infer $Rm$ at core conditions.
However, Fig. \ref{fig:ReRo} shows that $Re$ can be large, possibly allowing large $Rm$ too.
For large $Ra_C$, the flow organizes itself into strong large-scale zonal shears and weak small-scale fingers.
Even though the radial velocity of the small-scale finger is small, the large-scale zonal shear is large.
This situation could in principle sustain an $\alpha\omega$ dynamo, in which the large-scale shear is responsible for a so-called $\omega$-effect while the small-scale convection produces an $\alpha$-effect \citep[e.g.][]{roberts1972}.
We have also checked that our flow displays a significant amount of helicity, an ingredient thought to be important to obtain an important $\alpha$-effect.
From a numerical point of view, we reach $Re \sim 10^3$ in our simulations.
In an $\alpha\omega$ dynamo context, the relevant magnetic Reynolds number would be the geometric mean $Rm^*$ of the $Rm$ based on the large-scale zonal flow and the $Rm$ based on the small-scale one \citep[e.g.][]{roberts1972}.
According to Fig. \ref{fig:ReRo}a, this leads to $Rm^* \sim 200 Pm$, potentially allowing dynamos for $Pm \gtrsim 1$.

Beyond the question of the dynamo capability, we can wonder about the strength of the generated magnetic field.
In the case of simple convective dynamos (ie without double-diffusive effects), the field strength scales as $P^{1/3}$, where $P$ is the convective power \cite[e.g.][]{christensen2006scaling}.
Despite the small values of $Nu$ and $Sh$ (see Fig. \ref{fig:Nusselt}), we find significant buoyancy power in our simulations, scaling like $P \simeq 0.013 Ra_C$ (see Fig. \ref{fig:convpower}).
This scaling is similar to the one found in standard convective dynamos \cite[see][]{christensen2006scaling}, differing only by the constant factor which is about 100 times smaller here.
Assuming this scaling holds, we can expect strong magnetic fields to be generated, provided that $Ra_C$ is large enough.
Nevertheless, the saturation of a dynamo driven by double-diffusive convection may behave differently.
In addition, the large-scale zonal flows we have found in these simulations, which may persist for core conditions, are known to be important for the dynamo process in stratified interiors \citep[e.g.][]{spruit2002dynamo} whereas it does not change much the radial transport (and thus the Nusselt and Sherwood numbers).
Indeed, such zonal flows can sustain various hydrodynamic and magnetic instabilities \citep[e.g][]{knobloch1982nonlinear,jouve2015three}. 
Hence, dynamo onset, field strength at saturation, and extrapolation to core conditions all require a future study of dynamo driven by double-diffusive convection in the turbulent rotating regime.

Recently, \citet{guervilly2016subcritical} and \citet{kaplan2017subcritical} found that the smooth (linear) onset of rapidly rotating thermal convection is replaced by (nonlinear) hysteresis cycles and subcritical behaviours, at small enough Ekman numbers. These effects may survive with double-diffusive effects in the overturning regime. 
Finger convection may also occur though a subcritical bifurcation when $L \gg 1$, as proposed for non-rotating stratified fluids in planar models \citep{veronis1965finite,proctor1981steady}. This mathematical observation has not been confirmed yet numerically. Notably, we have not found evidence supporting this behaviour in the numerics. However, these nonlinear effects may only appear for $L$ larger than in our simulations. Therefore, studying finite-amplitude perturbations appears of special interest to investigate the transition towards turbulence in RDDC when $L \gg 1$.

Finally, we have neglected so far several double-diffusive effects occurring in a binary mixture. More relevant compositional boundary conditions may be implemented, e.g. the intricate boundary condition proposed by \citet{braginsky1995equations,glatzmaier1996anelastic}. Investigating additional binary effects in the thermal and heat fluxes is also worthy of interest (still in the Boussinesq approximation). They are only responsible for second order effects at the linear onset \citep[e.g.][]{hort1992onset,net2012numerical}, when a background state state is imposed. 
However, they may play a dynamical role in nonlinear simulations. For instance, barodiffusion is the tendency of light material to migrate down the pressure gradient. Barodiffusion sustains the accumulation of light elements at the top of the core \citep{gubbins2013stratified}, to naturally increase the Brunt-V\"ais\"al\"a frequency. Handling barodiffusion is not demanding numerically, e.g. in shells by considering a system forced by the boundaries (i.e. no background state) but with an additional mass sink \citep[e.g.][]{davies2011buoyancy,bouffard2017double}. These effects should be considered for consistent future nonlinear simulations.

\subsubsection{Towards planetary applications and beyond}
Beyond the origin of the early geodynamo, the (possible) outermost stable stratification in the Earth's core is another long standing geophysical issue \citep[e.g.][]{loper1981study,si1993mac,lister1998stratification}.
The existence of such a layer has been outlined by seismological \citep{helffrich2010outer,helffrich2013causes,irving2018seismically}, geodetic \citep{buffett2010stratification} and geomagnetic \citep{gubbins2007geomagnetic,buffett2014geomagnetic} data. The density stratification may have a thermal and/or compositional origin \citep[e.g.][]{buffett2010stratification,davies2018partitioning,nakagawa2018thermo,bouffard2019chemical}.
Indeed, the thermal conductivity has been revised upward by ab-inito calculations \citep{pozzo2012thermal,de2012electrical,pozzo2013transport} and experiments \citep{gomi2013high,ohta2016experimental,konopkova2016direct}.
This may favour an outer sub-adiabatic thermal stratification, but large thermodynamical uncertainties remain \citep{williams2018thermal}. 
Moreover, \citet{mound2019regional} pointed out that this outermost stratification may be regional (rather than global), being generated by the lateral variations in heat flux at the core–mantle boundary .
Stratification may be also sustained by the accumulation of light elements \citep[e.g.][]{loper1981study}.
This stratified layer may affect the geodynamo \citep[e.g.][]{olson2017dynamo,christensen2018geodynamo}, e.g. by filtering small-scale internal convective motions \citep{vidal2015quasi} or trapping waves \citep{knezek2018influence}. 
However, this hypothetical layer may be prone to either rotating finger convection or semi-convection \citep{braginsky2006formation}, making the internal core dynamics more complex.
In particular, intense zonal flows could develop, as we have found in this work.
Partially stratified core layers may also exist in other planets, e.g. Mercury \citep{manglik2010dynamo,takahashi2019mercury} or Venus \citep{jacobson2017formation}. 
Therefore, it is of special interest to determine whether thermally and/or compositionally stably stratified layers can survive dynamically against RDDC. 

In addition, double-diffusive effects are also relevant for giant planets \citep{stevenson1982formation}, such as Saturn \citep{stevenson1977dynamics,leconte2013layered} and Jupiter \citep{moll2017double}. 
Stellar interiors may also undergo DDC \citep{garaud2018double}, e.g. low-mass hosting exoplanets \citep{vauclair2004metallic} or massive stars \citep[e.g.][]{merryfield1995hydrodynamics,woosley2002evolution}. 
Even though they were largely neglected, rotational effects may be significant in these objects, e.g. for the giant planets of our Solar system which are rapidly rotating (9.9 hr for Jupiter and 10.7 hr for Saturn) or for some radiative stars \citep[e.g.][]{jouve2015three}. 

The validity of the Boussinesq model for compressible interiors should be assessed.
The scalings for the typical length scale of density structures, applied to planetary Earth-like parameters, yield \citep{bouffard2017double} $l_\perp \sim 20$ cm for rapid rotations  cm and $l_\perp = 40$ cm in the non-rotating case. \citet{spiegel1960boussinesq} showed that the Boussinesq approximation is relevant for dynamical scales smaller than the pressure scale height, typically one-tenth of the radius of stars. Therefore, the compressible dynamics may be surprisingly well described by using the Boussinesq approximation, as advocated in the non-rotating regime \citep{radko2016thermohaline}. A comparison between Boussinesq and anelastic models of RDDC \citep[e.g.][]{glatzmaier1996anelastic} is certainly worthy of interest for astrophysical objects. 

In addition, gaseous planets would require to consider stress-free conditions for the flow.
Our results show that, in the limit $Ek \ll 1$, stress-free conditions do not affect the onset of inviscid RDDC, which remains symmetric with respect to the equatorial plane. 
However, these bodies are characterised by much smaller values of $Pr \ll 1$ (compared to planetary cores).
In this regime, flows at the onset can be equatorially anti-symmetric torsional modes. They sometimes appear as the preferred unstable modes of (pure) thermal convection in spheres in the limit $Pr\ll1$ (e.g. at $Pr/Ek=10$), but only for stress-free conditions \citep{sanchez2016critical,zhang2017asymptotic} as commonly used for giant planets and stars.
Moreover, polar anti-symmetric modes have also been found at the onset when $Pr \ll 1$, for (pure) thermal convection in thick \citep{garcia2008antisymmetric} and thin \citep{garcia2018onset} spherical shells.
The nonlinear regime in the low-$Pr$ regime is expected to differ from the high-$Pr$ regime \citep[e.g. in the non-rotating regime][]{garaud2018double}.
Therefore, studying RDDC in the low-$Pr$ regime with stress-free conditions may lead to different double-diffusive effects than those previously obtained in shells \citep[e.g.][]{net2012numerical}.

Finally, we remark that the large-scale inviscid mode in the stably stratified regime is always $m=1$, with a net flow at the center within the equatorial plane.
Such a mode could constrain the translation direction of a freshly-nucleated inner core to be perpendicular to the rotation axis, in agreement with seismological observation of the hemispherical dichotomy of the inner core \citep[see e.g.][]{deguen2012}.

\section*{acknowledgments}
This project was funded by ANR-14-CE33-0012 (MagLune). JV acknowledges the support of STFC Grant ST/R00059X/1. We thank the geodynamo team (Univ. Grenoble Alpes), R. Deguen and T. Gastine for fruitful discussions and comments. Computations were performed on the Froggy platform of CIMENT (\url{https://ciment.ujf-grenoble.fr}), supported by the Rh\^one-Alpes region (CPER07${}_{}$13 CIRA), OSUG\@2020 LabEx (ANR10 LABX56) and Equip\@Meso (ANR10 EQPX-29-01). 
ISTerre is part of Labex OSUG\@2020 (ANR10 LABX56). All figures were produced using matplotlib (\url{http://matplotlib.org/}) or paraview (\url{http://www.paraview.org/}).


\appendix

\section{Other boundary conditions at the linear onset}
\label{appendix:BC}

\begin{figure}
	\centering
    \includegraphics[width=0.6\textwidth]{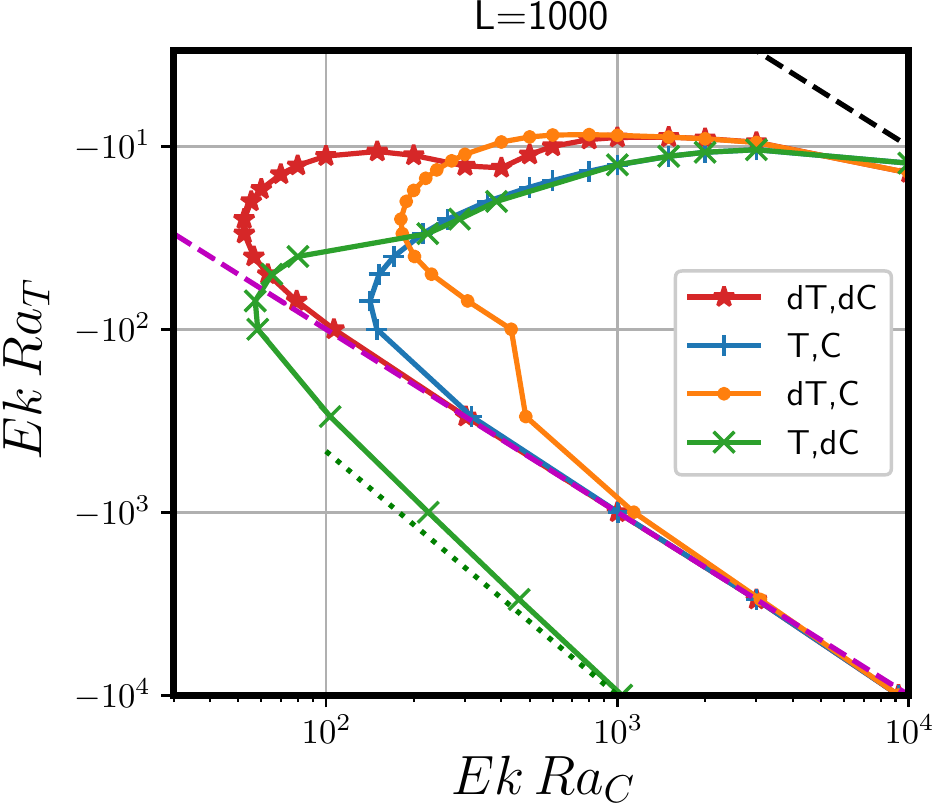}
    \caption{Linear onset for the rotating finger regime for various boundary conditions, encoded as T or C for fixed temperature or composition, and dT or dC for fixed flux of temperature or composition. Computations with SINGE at $Pr=0.003, Sc=3, Ek=10^{-11}$ for azimuthal wave number $m=1$ and no-slip boundary condition.
    The dotted line is $Ek Ra_T = -(Ek\,Ra_C)^{4/3}$, while the dashed lines are $N_0=0$ (upper, black) and $Ra_T=-Ra_C$ (lower, magenta).
    }
    \label{fig:BC}
\end{figure}

We investigate the effects of different mechanical, thermal and compositional boundary conditions (BC) on RDDC in spheres. We substitute no-slip conditions (\ref{eq:nsBC}) by stress-free conditions for the velocity field
\begin{equation}
	\boldsymbol{1}_r \boldsymbol{\cdot} \boldsymbol{u} = 0, \ \, \ \boldsymbol{1}_r \times [ \boldsymbol{\mathcal{E}} \boldsymbol{1}_r ] = \boldsymbol{0} \, \ \, \text{at} \ \, \ r=1
    \label{eqappendix:SF}
\end{equation}
with $\boldsymbol{\mathcal{E}} =  \left [ \boldsymbol{\nabla} \boldsymbol{u} + (\boldsymbol{\nabla} \boldsymbol{u})^\top \right ]/2$ the strain rate tensor (incompressible Newtonian fluid). Instead of fixed flux conditions (\ref{eq:nofluxBC}), we consider fixed temperature or composition at the boundary
\begin{equation}
	\Theta = \xi = 0 \, \ \, \text{at} \ \, \ r=1.
        \label{eqappendix:fixedTC}
\end{equation}
Numerical results, computed with SINGE, have been performed for $m=1$ and $m=6$ at $Ek=10^{-4}$ and $m=1$ at $Ek=10^{-11}$. Given that the results lead to the same conclusions, we only show the results for $m=1$ and $Ek=10^{-11}$ in figure \ref{fig:BC}.
Within the stable double-diffusive tongue given by the Ledoux criterion (\ref{eq:ledoux}), the linear onset is independent of the mechanical conditions. For the low Ekman number considered here, using stress-free (\ref{eqappendix:SF}) or no-slip condition (\ref{eq:nsBC}) leads to the same marginal stability curve (not shown).
However, changing the boundary condition on the temperature or composition field has important effects on the shape of the marginal stability curve, but the latter still remains independent of viscosity.
Surprisingly, with a fixed temperature and imposed buoyancy flux, the double-diffusive convection extends to $Ra_T < - Ra_C$, which corresponds to density ratios $R_0 > L$. This linear instability, located beyond the expected range of finger convection, has been confirmed by time-stepping nonlinear simulations with XSHELLS (at $Ek\,Ra_T = -10^3$, $Ek \, Ra_C=6 \times 10^2$ and $Ek=10^{-5}$).

\section{Semi-convection}
\label{appendix:ODDC}

\begin{figure}
	\centering
    \includegraphics[width=0.6\textwidth]{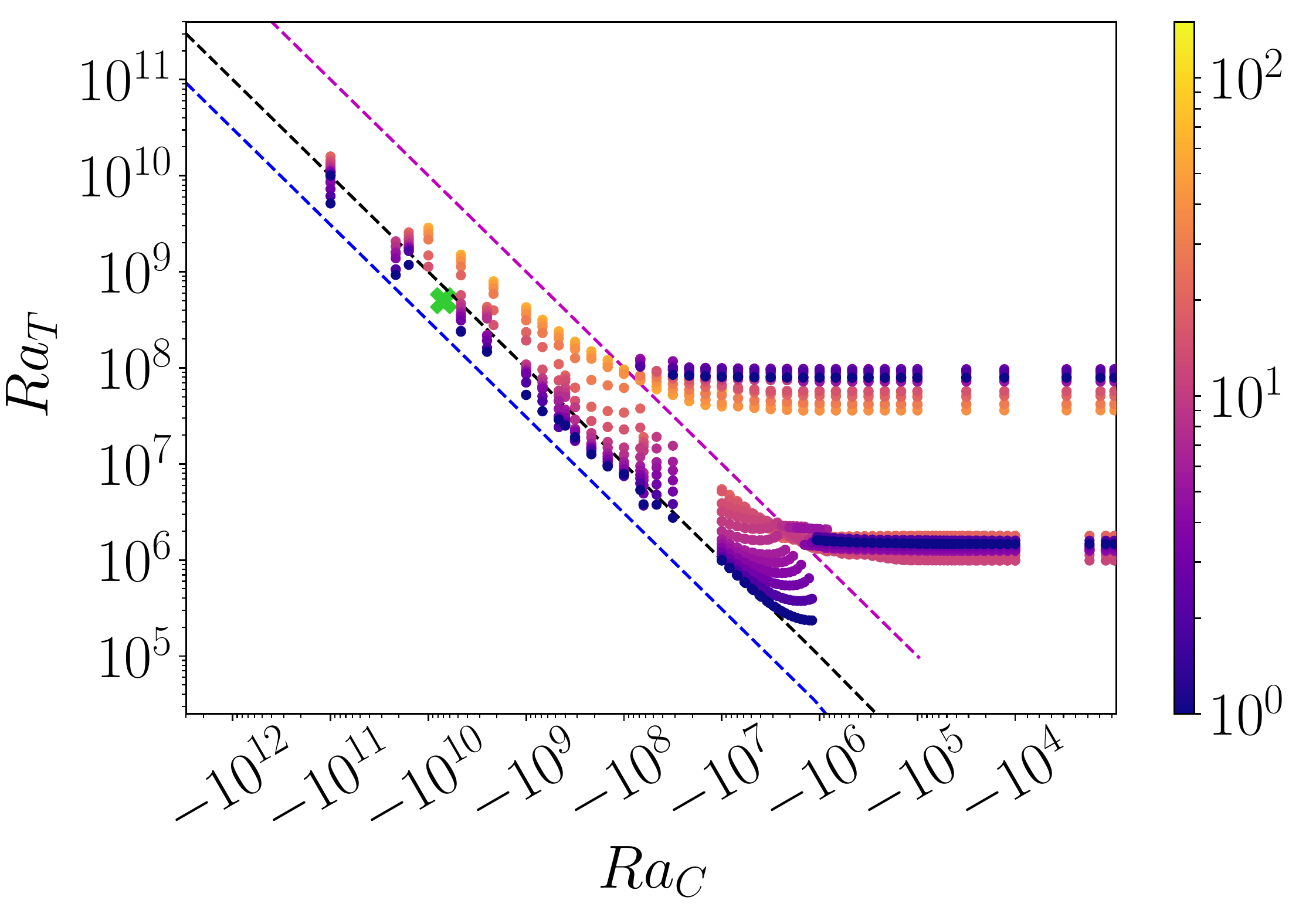}
    \caption{Linear onset of rotating semi-convection ($Ra_T > 0, Ra_C < 0$ at $Pr=0.3$ and $Sc=3$)  computed with SINGE at $Ek=10^{-4}$ (bottom points) and $Ek=10^{-5}$ (top points).
    The colour shows the azimuthal wave number $m$ at the onset. The (middle) black dashed line is the neutral curve $N_0^2 = 0$ (i.e. $Ra_C=- L Ra_T$), the (upper) magenta dashed one is $Ra_C=-Ra_T$ and the (lower) blue dashed one is $Ra_T = (-Ra_C/L) (Pr+ 1/L)/(Pr+1)$. The green cross locates the nonlinear simulation shown in figure \ref{fig:ODDC_NL}.}
    \label{fig:ODDC_lin}
\end{figure}

\begin{figure}
	\centering
    \includegraphics[width=0.6\textwidth]{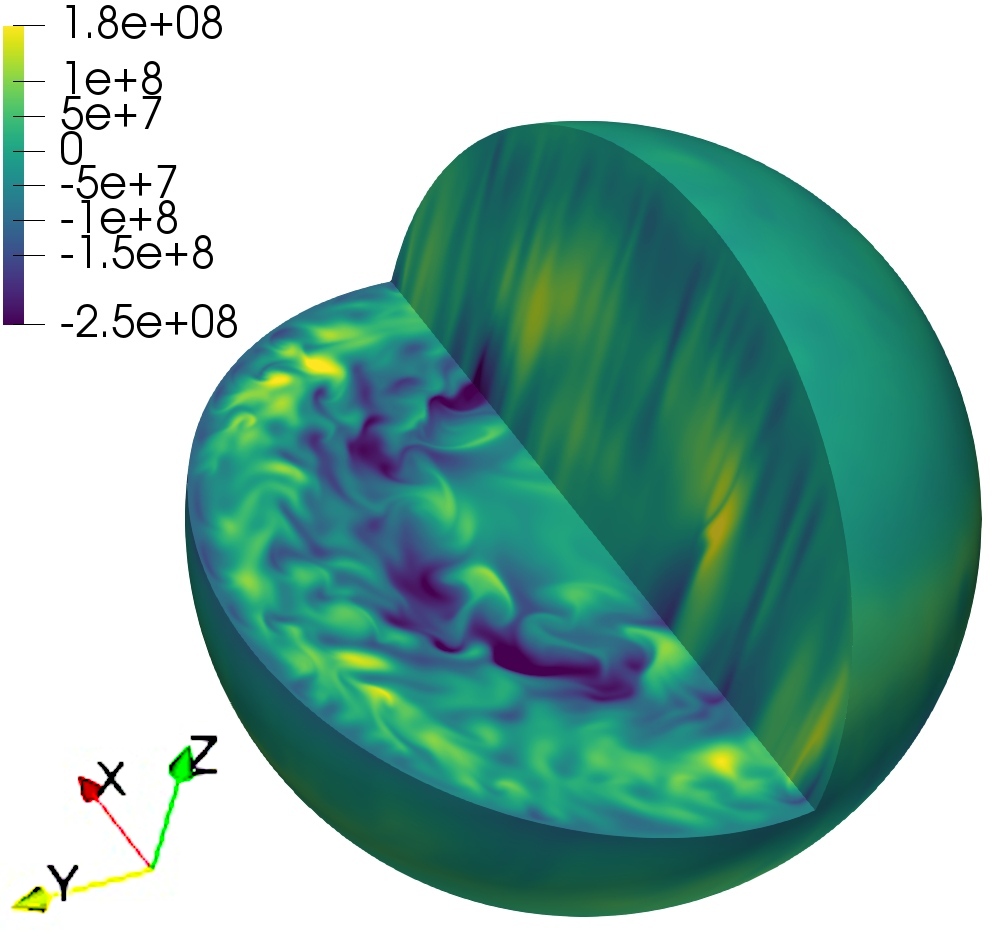}
    \caption{3D snapshot of a nonlinear simulation of rotating semi-convection ($Ra_T > 0, Ra_C < 0$ at $Pr=0.3$ and $Sc=3$), showing the chemical buoyancy $Ra_C \xi$ for a stably stratified background state at $Ek=10^{-5}$, $Ra_C=-7 \times 10^9$ and $Ra_T = 5 \times 10^8$. Rotation axis is along $\boldsymbol{1}_z$.}
    \label{fig:ODDC_NL}
\end{figure}

The onset of RDDC in the semi-convection quadrant ($Ra_T > 0, Ra_C < 0$) is represented in figure \ref{fig:ODDC_lin} the linear computations at the onset computed with SINGE, for two values of $Ek$. The critical parameters at the onset of pure thermal convection are given in table \ref{table:onsetT}, for completeness with table \ref{table:onset} for pure compositional convection. 
The onset of convection is largely insensitive to double-diffusive effects as long as $|Ra_C| \ll Ra_T$. This refers to the overturning regime of thermal convection. For higher $|Ra_C|$, double-diffusive effects start to be important when $|Ra_C| \geq Ra_T$. As in the finger regime, the marginal stability curve $\sigma$ takes the form of a tongue in the $Ra_c - Ra_T$ diagram (figure \ref{fig:SINGE_Busse_bas}). However, double-diffusive effects become significant even for unstably stratified fluids ($N_0^2 < 0$), as opposed to the finger quadrant in which only stably stratified fluids ($N_0^2 \geq 0$) are strongly affected.
Within this tongue, modes with small azimuthal wave number are triggered at the onset, which also occurs for smaller thermal Rayleigh number than in the overturning regime. In the limit $|Ra_C| \to \infty$, RDDC reaches asymptotically the non-rotating regime predicted by formula (\ref{eq:criteriaODDC}). Then, we show in \ref{fig:ODDC_NL}b an illustrative nonlinear simulation of semi-convection at $Ek=10^{-5}$ and $Ra_T=10^8$. Density structures exhibit larger spatial scales than the ones obtained in simulations within the finger regime (for similar absolute values of the Rayleigh numbers).

\begin{table}
 \centering
\begin{tabular}{cccc}
 \hline
     $Ek$ & $m^{c}$ & $Ra_T^{c}$ & $\omega$ \\ [0.5ex]
 \hline
$10^{-4}$ & 12 & $9.86 \times 10^5$ & $-5.48 \times 10^2$ \\[0.5ex]
$10^{-5}$ & 40 & $3.60 \times 10^7$ & $-2.08 \times 10^3$ \\[0.5ex]
\hline
\end{tabular}
 
\caption{Critical wave number $m^{c}$, thermal Rayleigh number $Ra_T^{c}$ and angular frequency $\omega^c$ at the marginal onset ($\sigma=0$) of thermal overturning convection (i.e. for $Ra_C=0$). Computations at $Sc=3$ and $Pr=0.3$. The first row is obtained at $Ek=10^{-4}$ and the second one at $Ek=10^{-5}$.}
\label{table:onsetT}
\end{table}

\section{Revisiting the annulus geometry}
\label{appendix:Busse}
\subsection{Mathematical formulation}
\begin{figure}                
  \centering
  \includegraphics[width=0.4\textwidth]{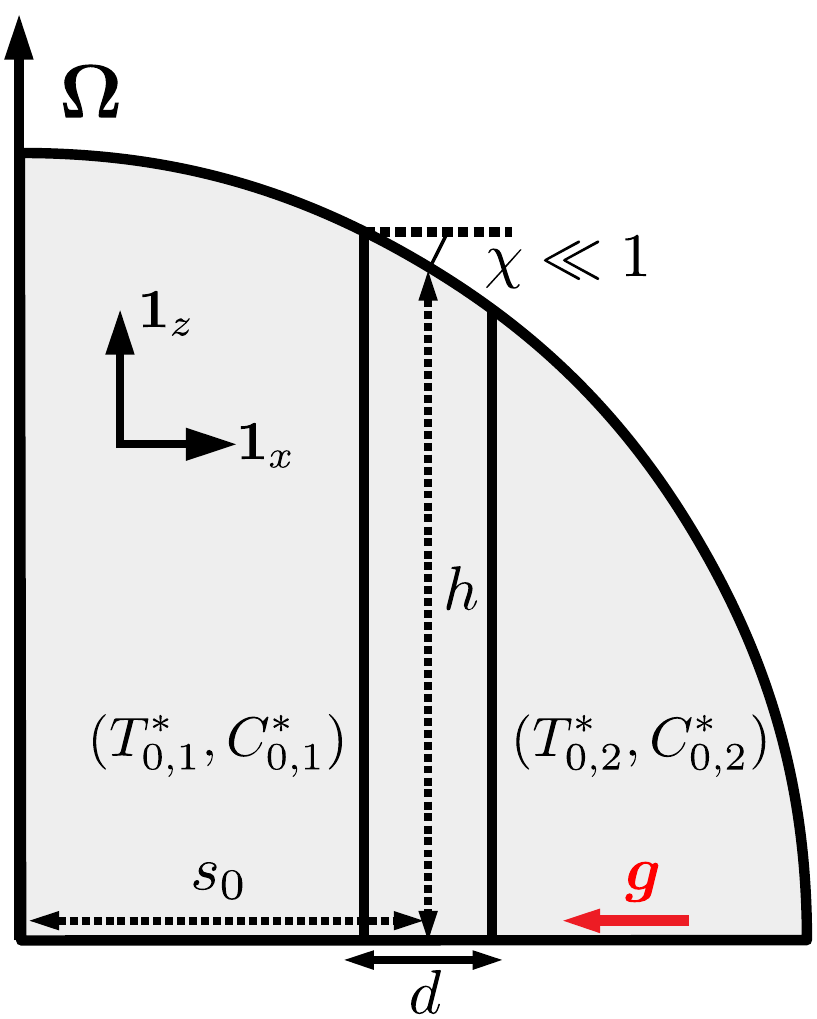}
  \caption{Configuration of the rotating cylindrical annulus. The thin gap of the annulus at the cylindrical radius $s_0$ is $d$, $h$ the spatially varying height of the annulus, $\chi$ the angle of the conical upper and lower (not shown) caps. $\boldsymbol{\Omega}$ the planetary angular velocity and $\boldsymbol{g}$ is the radially directed acceleration of gravity. The inner cylindrical wall at $x=-d/2$ (respectively outer wall at $x=d/2$) is kept at the (dimensional) temperature $T_{0,1}^*$ and composition $C_{0,1}^*$ (respectively $T_{0,2}^*$ and $C_{0,2}^*$) in the background state. }
  \label{fig:confAn}                
\end{figure}

We revisit the model of RDDC in a cylindrical annulus.
A few misprints are present in \citet{busse2002low}, which also used other dimensionless variables.
Furthermore, \citet{busse2002low} made wrong assumptions when drawing his conclusions, mistakenly considering the non-rotating limit.
Before taking the annulus model further, we clearly explain the theory, going through the derivation of the equations in our formalism.

For the sake of tractable analytical developments, \citet{busse1970thermal} pointed out that a simplified model of QG convection in spheres should consider a thin cylindrical annulus, with sloping top and bottom boundaries. Using this asymptotic model, he investigated the onset of thermal convection with $Ra_C=0$ \citep{busse1986asymptotic}, and extended it to RDDC \citep{busse2002low}. This model considers a thin-gap geometry centered on the QG columns at the onset. Moreover, this asymptotic theory can embrace core conditions in the limit $Ek \ll 1$ and $L \gg 1$.
The annulus geometry is illustrated in figure \ref{fig:confAn}. We consider the cylindrical annulus region, located at the cylindrical radius $s_0$ in a full sphere rotating at the angular velocity $\Omega_s$. We use the small-gap approximation, by assuming $d/s_0 \ll 1$. Thus, the effects of the spherical curvature can be neglected and we use the Cartesian coordinate system of unit vectors $(\boldsymbol{1}_x,\boldsymbol{1}_y,\boldsymbol{1}_z)$ centered at $s_0$. The annular channel is bounded at top and bottom by rigid conical caps with the angle of inclination $\chi$. We denote $h$ the half-height of the cylindrical annulus (with respect to the equatorial plane). In the background state, the fluid is stratified in temperature and composition under the inward gravity field $\boldsymbol{g} = -g_0 s_0 \boldsymbol{1}_x$, which is constant at the scale of the annulus. The inner wall (respectively the outer one) is kept at the constant temperature $T_{0,1}^*$ and composition $C_{0,1}^*$ (respectively $T_{0,2}^*$ and $C_{0,2}^*$).

We choose the gap $d$ as length scale, $d^2/\nu$ as time scale, $\Delta T^* \, Pr = (T_{0,1}^* - T_{0,2}^*) \, Pr$ as thermal scale and $\Delta C^* \, Sc = (C_{0,1}^* - C_{0,2}^*) \, Sc$ as compositional scale. These thermal and compositional scales are the local analogues of the global scales chosen in the main text. Dimensionless variables are denoted in the following without asterisk.
We assume that the slope $\chi$ of the upper and lower caps shown in figure \ref{fig:confAn} is small ($\chi \ll 1$), such that the local conductive background state is close to the one in the annulus of uniform depth \citep[e.g.][]{busse1970thermal}. Hence, the dimensionless background state is
\begin{equation}
	\nabla T_0 = -\frac{1}{Pr}\boldsymbol{1}_x, \ \, \ \nabla C_0 = -\frac{1}{Sc}\boldsymbol{1}_x.
\end{equation}
Then, the local form of equations (\ref{eq:dimensionlessEQN}) for the dimensionless perturbations $(\boldsymbol{u}, \Theta, \xi)$ takes the form
\begin{subequations}
\label{eqappend:governBusselike}
\begin{align}
\frac{\partial \boldsymbol{u}}{\partial t} + \frac{2}{\widetilde{Ek}}\boldsymbol{1}_z \times \boldsymbol{u} &= -\nabla p + \boldsymbol{\nabla}^2 \boldsymbol{u} \label{eqappend:governBusselikeU} \\ 
&+ \left( \widetilde{Ra}_{T} \, \Theta + \widetilde{Ra}_{C}\,  \xi \right) \boldsymbol{1}_x, \nonumber \\
\frac{\partial \Theta}{\partial t} &= \frac{1}{Pr} \left (\boldsymbol{1}_x \cdot\boldsymbol{u} + \nabla^2 \Theta\right ), \label{eqappend:governBusselikeT} \\
\frac{\partial \xi}{\partial t} &= \frac{1}{Sc} \left ( \boldsymbol{1}_x \cdot\boldsymbol{u} + \nabla^2 \xi \right ). \label{eqappend:governBusselikeC}
\end{align}
\end{subequations}
We have introduced in equations (\ref{eqappend:governBusselike}) the local Ekman and Rayleigh numbers
\begin{subequations}
\label{eqappend:RayleighBusse}
\begin{align}
	\widetilde{Ek} &= \frac{\nu}{\Omega_s d^2}, \\
	\widetilde{Ra}_{T} &= \frac{\alpha g_0 s_0 \Delta T^* d^3}{\nu^2 \kappa_T}, \\
    \widetilde{Ra}_{C} &= \frac{\alpha g_0 s_0 \Delta C^* d^3}{\nu^2 \kappa_C}. 
\end{align}
\end{subequations}
Note that Rayleigh numbers (\ref{eqappend:RayleighBusse}) are the local versions of the spherical Rayleigh numbers (\ref{eq:RaTRaC}) introduced in the main text. 

We seek velocity solutions of equations (\ref{eqappend:governBusselike}) with small variations along the rotation axis $\boldsymbol{1}_z$. Hence, the velocity takes the form of QG flows
\begin{equation}
	\boldsymbol{u} \sim  \nabla \times \left ( \Psi \,  \boldsymbol{1}_z \right ) + u_z \boldsymbol{1}_z,
    \label{eq:QGJault}
\end{equation}
with $u_z$ the small vertical velocity (at the order $\chi$) and $\Psi$ the velocity stream function in the equatorial plane $(z=0)$. 
The linear onset given by equations (\ref{eqappend:governBusselike}) can be solved by considering stress-free, iso-thermal and iso-compositional boundaries
\begin{equation}
	\Psi = \frac{\partial^2 \Psi}{\partial^2 x} = \Theta = \xi = 0 \, \ \, \text{at} \ \, \ x=\pm \frac{1}{2}.
    \label{eqappend:BCBusse}
\end{equation}
We also assume that the upper and lower conical caps at $z=\pm h/(2d)$ are rigid, with fixed vertical thermal and compositional fluxes \citep{busse1986asymptotic}. This yields
\begin{equation}
	\boldsymbol{u} \boldsymbol{\cdot} \boldsymbol{1}_z = \pm \tan \chi \, (\boldsymbol{u} \boldsymbol{\cdot} \boldsymbol{1}_x), \ \, \ \frac{\partial}{\partial z} [ \Theta, \xi] = 0.
\end{equation}
Other conditions are irrelevant in the analysis. In particular, the neglected viscous boundary layer vanishes in the limit $Ek \ll 1$ in the annular geometry \citep{hunter1967axisymmetric}. This is in agreement with the observation that the viscous boundary condition is of second order importance for the onset of convection in spheres \citep{zhang1993influence,jones2000onset}, at least for not too small values of $Pr$ at fixed $Ek$ \citep{zhang2017asymptotic}.
Although these boundary conditions are not physically realistic \citep{braginsky1995equations}, they do not hinder from investigating the leading order double-diffusive effects. 
Then, following \citet{busse2002low}, we take the $z$-component of the curl of momentum equation (\ref{eqappend:governBusselikeU}) and average it over $z$ (from bottom to upper caps). This yields at first order in $\chi$ \citep[see][for the derivation]{busse1986asymptotic}
\begin{equation}
	\left ( \frac{\partial}{\partial t} - \Delta_\perp \right ) \Delta_\perp \Psi - \beta \, \frac{\partial \Psi}{\partial y} - \widetilde{Ra}_T  \frac{\partial \Theta}{\partial y} - \widetilde{Ra}_C  \frac{\partial \xi}{\partial y} = 0,
    \label{eaeppend:vortz}
\end{equation}	
with the two-dimensional horizontal Laplacian $\Delta_\perp = \partial^2/\partial x^2 + \partial^2 /\partial y^2$ and the parameter
\begin{equation}
	\beta = \frac{4 \Omega_s d^3}{h \nu} \tan \chi.
    \label{eqappend:etaBusse}
\end{equation}
In the rapidly rotating limit $\widetilde{Ek} \ll 1$, $\beta$ is a leading order parameter containing the effects of the boundary curvature (the so-called $\beta$-effect).

We assume periodicity in the $\boldsymbol{1}_x$ direction to satisfy boundary conditions (\ref{eqappend:BCBusse}), yielding the form of the solutions \citep[e.g.][]{busse1986asymptotic}
\begin{multline}
	[\Psi, \Theta, \xi ] (x,y,t) = \left [ \widehat{\Psi}, \widehat{\Theta}, \widehat{\xi} \right ] \exp(\mathrm{i} m y + \lambda t) \\ \cos \left [ n \pi \left ( x+\frac{1}{2} \right ) \right ]), 
    \label{eq:QGPsi}
\end{multline}
where $\left [ \widehat{\Psi}, \widehat{\Theta}, \widehat{\xi} \right ]$ are complex-valued amplitudes, $\lambda$ is the complex eigenvalue with $\Re_e(\lambda) = \sigma$ the growth rate, $m$ is the azimuthal wave number and $n$ is the degree of spatial complexity along the horizontal direction.  We substitute solutions (\ref{eaeppend:vortz}) into equations (\ref{eqappend:governBusselikeT})-(\ref{eqappend:governBusselikeC}) and (\ref{eaeppend:vortz}). They can be recast into a single equation for $\widehat{\Psi}$. 
This equation can be recast into the original form introduced by \citet{busse2002low}, i.e.
\begin{multline}
(\lambda Pr + a^2 )\left(\lambda Pr + \frac{a^2}{L}\right)[(\lambda +a^2)a^2 +\mathrm{i} m \beta] \\ - m^2 \left [ R_T \left(\lambda Pr + \frac{a^2}{L} \right) + R_C \, (\lambda Pr + a^2) \right ] = 0
\label{eqappendix:polBusse2}
\end{multline}
with $a^2 = m^2 + n^2 \pi^2$ and by introducing the thermal and compositional Rayleigh numbers in the Busse's notation
\begin{equation}
	R_T = \widetilde{Ra}_T \ \, \ \text{and} \ \, \ R_C = \widetilde{Ra}_C/L.
\end{equation}
Because all azimuthal wave numbers $m$ are separated, the marginal stability curve $\sigma = 0$ is obtained by minimising the critical Rayleigh number over all values of $m$.
In the following, we will survey the properties of RDDC in the annulus geometry by varying Busse's parameters $(R_T,R_C)$.

In our notations, the growth rate $\sigma = \Re_e(\lambda)$ of RDDC is predicted by the following polynomial equation \citep{busse2002low}
\begin{multline}
(\lambda Pr + a^2 ) (\lambda Sc + a^2 ) [(\lambda +a^2) \, a^2 + \mathrm{i} m \beta] \\ - m^2 \left [Ra_T \left(\lambda Sc + a^2 \right) + Ra_C  \, (\lambda Pr + a^2) \right ] = 0,
\label{eq:polBusse}
\end{multline}
with $a^2 = m^2 + \pi^2$, $[Ra_C,Ra_T]$ the Rayleigh numbers (\ref{eq:RaTRaC}) and $\beta$ a geometrical parameter in the annulus geometry. 
When double-diffusive effects are negligible, the onsets of pure thermal or compositional rotating convection are naturally recovered, given by the critical values
\begin{equation}
Ra_T^c = g(Pr) \ \, \ \text{and} \ \, \ Ra_C^c = g(L Pr),
\label{eq:criticalpureconv}
\end{equation}
with the function
\begin{equation}
g(x) = \frac{a^6}{m^2} + \left( \frac{\beta x}{1+x} \right) ^2 a^{-2}. 
\label{eq:gx}
\end{equation}

\subsection{New asymptotic predictions}
In the limit $Ek \ll 1$, we have obtained an analytical expression for the double-diffusive onset from formula (\ref{eq:polBusse}).
This contradicts the prediction \citet{busse2002low}, made by mistakenly considering the non-rotating limit.
Within the double-diffusive tongue, the onset is given by (see details in the supplementary material)
\begin{align}
Ra_T^c &= \pm \frac{\beta a^2 (L+1) }{m \sqrt{(K-1)(L^2-K)}}  \ \, \ \text{with} \ \, \ \frac{Ra_C^c}{Ra_T^c} =-K ,
\label{eq:parametric}
\end{align}
with negative (respectively positive) values of $Ra_T^c$ in the finger (respectively semi-convection) regime. Predictions (\ref{eq:parametric}) agree very well with our numerical simulations in the sphere (see figure \ref{fig:inviscid}). In particular, we recover that the results do not depend on $Sc$ and $Pr$, but only on $L$ and $Ek$. Moreover, for each $m$, the minimum $|Ra_C^c|$ is located along the line
\begin{align}
Ra_T^c &= - \frac{1+L^2}{2L^2} Ra_C^c  \ \, \simeq Ra_C^c/2 \ \, \ \text{for} \ \, \ L\gg 1
\end{align}
and is given by
\begin{align}
\min_{Ra_T} |Ra_C^c| &=   \frac{2 \beta a^2 L }{m (L-1)}  \ \, \simeq \frac{2 a^2 \beta}{m} \ \, \ \text{for} \ \, \ L\gg 1 . \label{eq:cebron_law}
\end{align}
This corrected expression of the reduced onset agrees with our numerical results in the sphere (figure \ref{fig:inviscid}). Note that we also recover that, near this point, the onset is independent on $L$, and thus only depends on $Ek$.

\subsubsection{Matching the annulus to the sphere}
\citet{simitev2011double} showed numerically that $n=1$ is always the most unstable radial wave number in the annulus geometry. So, we have fixed $n=1$ in the following, as originally considered by \citet{busse2002low}.
Then, parameters (\ref{eqappend:RayleighBusse})-(\ref{eqappend:etaBusse}) are local parameters. Moreover, the latter parameter $\beta$ is constant in the thin-gap approximation. However, the spherical curvature, here measured by $\chi$, is spatially varying in the sphere.
For a matching to the sphere, these local parameters should be adjusted at the location of the QG structure at the onset, as schematically illustrated in figure \ref{fig:confAn}.
Indeed, $\beta$  strongly depends on the critical cylindrical radius $s_0$ at which columnar QG motions first appear, which is known to vary in spheres \citep{jones2000onset}.
Similarly, $(\widetilde{Ra}_T,\widetilde{Ra}_C)$ depend not only on the global Rayleigh numbers $(Ra_T, Ra_C)$ introduced in the main text (\ref{eq:RaTRaC}), but also on the local position $s_0$.

Therefore, $(\widetilde{Ra}_T, \widetilde{Ra}_C, \beta)$ are free parameters in the model.
To heuristically link the local and global parameters, we introduce one adjustable parameters $\Gamma$ such that
\begin{equation}
   \beta = \Gamma \, Ek^{-1}, \label{eqappendix:etagamma}
\end{equation}
$\Gamma$ should depend on the dimensionless parameters at the onset, i.e. $\Gamma = \Gamma (Ek,Pr,Sc,Ra_T,Ra_C)$. Thus, this parameter is not \emph{a priori} uniquely determined.

\begin{figure}                
  \centering
  \includegraphics[width=0.6\textwidth]{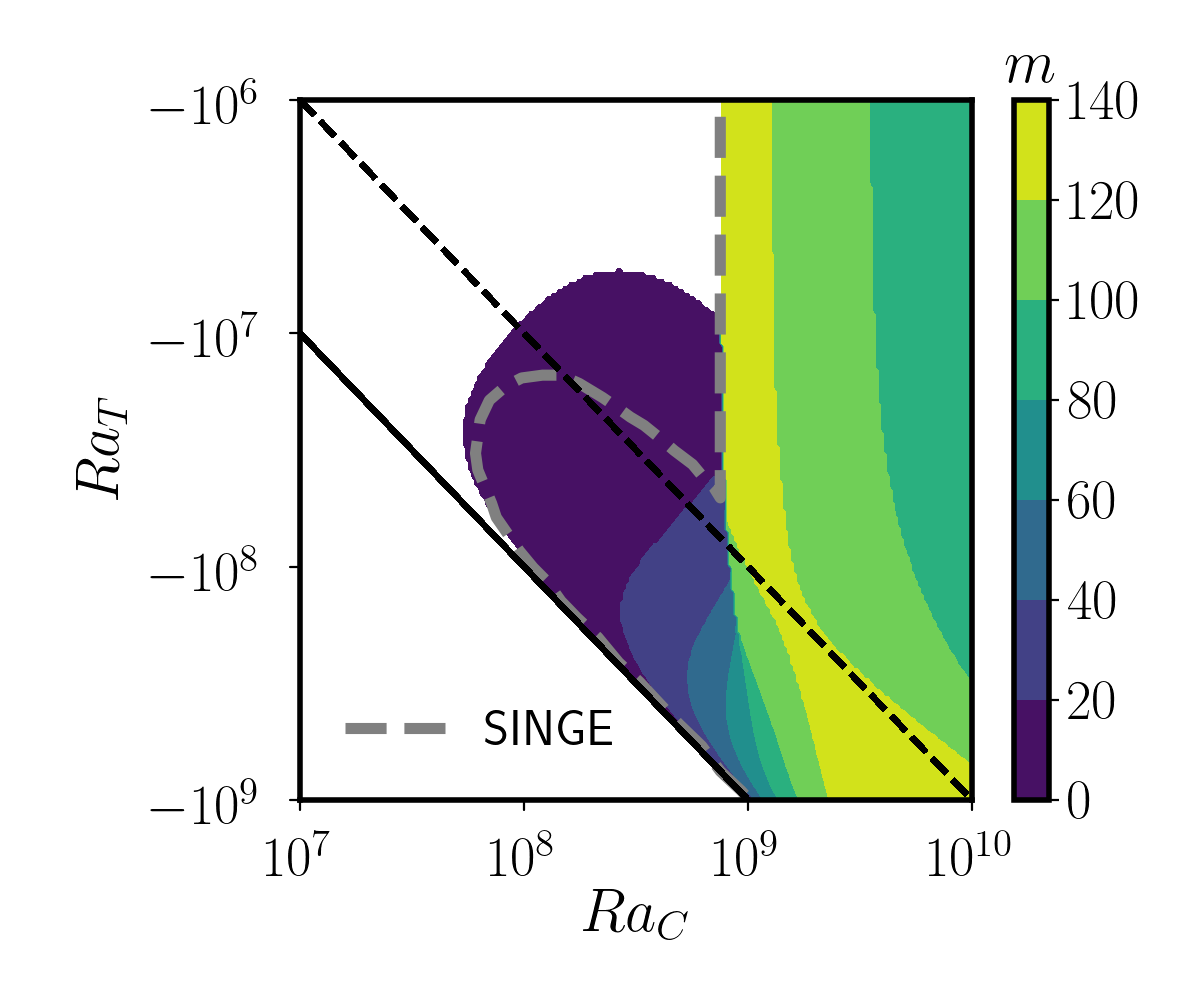}
  \caption{Comparison between the annulus asymptotic theory and SINGE computations at $Pr=0.3$, $Sc=3$, $Ek=10^{-6}$.
  Colour bar shows the most unstable azimuthal wavenumber. At the upper edge of the double-diffusive tongue, the critical number is $m^c = 3$ in (a) and (b). The dashed (thick) gray line is the marginal stability curve $\sigma$ computed with SINGE. The dashed tilted line is the neutral curve $N_0^2 = 0$, i.e. $Ra_T = -Ra_C/L$. The tilted solid line is the bound for non-rotating finger convection (\ref{eq:fingeringNRDDC}), i.e. the curve $Ra_T = -Ra_C$.
  For the annulus theory, we set $\Gamma = 3.786$ to match the pure compositional onset (at $Ra_T=0$) given by SINGE.}
  \label{fig:SINGE_Busse}                
\end{figure}

\subsubsection{Benchmark with SINGE}

We now compare the prediction of the previous model with the actual data given by SINGE.
To do so, we have adjusted $\Gamma$ such that the marginal stability curve $\sigma = 0$, predicted by (\ref{eqappendix:polBusse2}), coincides with the critical Rayleigh numbers at the onset of pure compositional convection ($Ra_T=0$) as computed by SINGE. 
We show in figure \ref{fig:SINGE_Busse} the superposition of the marginal stability curve $\sigma=0$ determined by SINGE and the stability map predicted by equation (\ref{eqappendix:polBusse2}) in the finger quadrant. 

Several points are worthy of comment.
First, the critical wave number $m^c$ in the theory is over-estimated compared to the numerical values in table \ref{table:onset}, roughly by a factor three. This confirms that local theories can only predict the order of magnitude of the wave number at the onset \citep[e.g.][]{busse1970thermal}.
On the marginal stability curve within the double-diffusive tongue, SINGE always find an $m=1$ mode.

Second, the reduced model recovers the non-rotating limit of finger convection. Indeed, the non-rotating limit (\ref{eq:fingeringNRDDC}), i.e. $Ra_T = -Ra_C$, is asymptomatically reached for large enough Rayleigh numbers.
Note however that we found convective motion beyond this limit with SINGE for some boundary conditions (see \S\ref{appendix:BC}).

Finally, double-diffusive effects are over-estimated in the reduced model for unstably stratified fluids (above the dashed-line in figure \ref{fig:SINGE_Busse}), predicting unstable regions where the system is in fact stable.
In addition, in the reduced model, the unstable double-diffusive tongue widens without bound when increasing $L$, whereas it reaches a limit for $L \gtrsim 10^3$ in our numerical computations (see figure \ref{fig:inviscid}).
Quantitatively, these discrepancies increase when $Ek$ decreases.

 
 \setlength{\bibsep}{0pt plus 0.3ex}

\end{document}